%% file: ms.tex
\documentclass[12pt,preprint]{aastex}

\shorttitle{Turbulence Cascade}
\shortauthors{Jiang et al.}

\begin{document}

\title{Cascade and Damping of Alfv\'{e}n-Cyclotron Fluctuations: Application
to Solar Wind Turbulence}

\author{Yan Wei Jiang,\altaffilmark{1,2}, Siming Liu,\altaffilmark{3} and Vah\'e
Petrosian,\altaffilmark{1,2,4}}

\altaffiltext{1}{Center for Space Science and Astrophysics, Department of
Physics, Stanford University, Stanford, CA 94305; arjiang@stanford.edu,
vahep@astronomy.edu}
\altaffiltext{2}{Kavli Institute of Particle Physics and Cosmology,
Stanford University}
\altaffiltext{3}{Department of Physics and Astronomy, University of
Glasgow, Glasgow, G12 8QQ, UK; sliu@astro.gla.ac.uk}
\altaffiltext{4}{Also Department of Applied Physics}

\begin{abstract}

It is well-recognized that the presence of magnetic fields will lead to
anisotropic energy cascade and dissipation of astrophysical turbulence.
With the diffusion approximation and linear dissipation rates,
we study the cascade and damping of Alfv\'{e}n-cyclotron fluctuations
in solar plasmas numerically for two diagonal  diffusion tensors, one
(isotropic) with identical components for the parallel and perpendicular
directions (with respect to the magnetic field) and one with different
components (non-isotropic). It is found that for the isotropic case the
steady-state turbulence spectra are  nearly isotropic in the inertial
range and can be fitted by a single power-law function with
a spectral index of $-3/2$, similar to the Iroshnikov-Kraichnan
phenomenology, while for the non-isotropic case the spectra vary greatly with
the direction of propagation. The energy fluxes in both cases are much
higher in the perpendicular direction than in the parallel direction due
 to the angular dependence (or inhomogeneity) of the components. In
 addition, beyond the MHD regime the kinetic effects make the spectrum
 softer at higher wavenumbers. In the dissipation range the turbulence
 spectrum cuts off at the wavenumber, where the damping rate becomes
 comparable to the cascade rate, and the cutoff
wavenumber changes with the wave propagation direction. The angle
averaged turbulence spectrum of the isotropic model resembles a broken
 power-law, which
cuts off at the maximum of the cutoff wavenumbers or the $^4$He cyclotron
frequency. Taking into account the Doppler effects, the model naturally
reproduces the broken power-law turbulence spectra observed in
the solar wind and predicts that a higher break frequency
always comes along with a softer dissipation range spectrum that
may be caused by the increase of the turbulence intensity, the
reciprocal of the plasma $\beta_{\rm p}$, and/or the angle between
the solar wind velocity and the mean magnetic field. These
predictions can be tested by detailed comparisons with more accurate
observations.


\end{abstract}





\section{Introduction}

Turbulence is ubiquitous in the universe and plays important roles in our
understanding
of many natural phenomena \citep{K41, I63, K65}. It
occurs
in highly non-equilibrium systems, where the microscopic viscous and/or
resistive
dissipation processes cannot effectively convert the free energy into the
internal energy of the fluid. Such systems usually have very high
Reynolds numbers and/or cover huge spatial and temporal scales, so that
the free energy is stored in the large scale
motions and/or magnetic fields. For plasmas, the free energy initially may also
be stored in non-equilibrium distributions of charged particles. Turbulence is
generated through a variety of instabilities related either to the large scale
magneto-hydrodynamic (MHD) processes or the microscopic collective plasma
effects
and/or plasma physics processes \citep{BF03}. These aspects have been
extensively investigated with the MHD and/or plasma physics theories. In
astrophysics,
most turbulence is carried by magnetized plasmas. It is responsible for
distributing
energies among different components of the plasmas, which may result in
distinct
emission characteristics or other observable features. Observations of these
radiations
can be used to study the corresponding astrophysical sources [e.g., \citet{L04,
L06} on $^3$He rich impulsive solar energetic particle events].

Plasma waves or turbulence once generated (usually on large scales comparable to
the dynamical scales given the large amount of energy inferred from
astrophysical observations) undergo two types of interactions: wave-wave
and wave-particle. The former dominates at large scales, where the
Reynolds (ordinary or magnetic or both)
number is large, and results in an almost dissipationless cascade primarily to
smaller scales and higher values of the wavenumber $k$. The interactions of
waves  with the more numerous low energy particles
gradually become more important as the cascade reaches short enough scales (and
the Reynolds number approaches unity and/or the kinetic effects set in)
and damp the waves. This results in plasma
heating and/or particle acceleration. The wave-particle interactions
also determine the spatial diffusion and energizing of high energy particles in
collisionless turbulent plasmas and are essential elements of the cosmic
ray theory \citep{GS69, YL02}. The spectral range from the turbulence
generation scale $k_0$ to the damping
scale $k_{\rm max}$ is called the {\it inertial range}, where the angle
averaged energy flux is independent of $k$.
Beyond this range, there are strong
couplings between the charged background particles and turbulent motions. For
collisional plasmas, where the Coulomb collision mean free path is much shorter
than other relevant spatial scales,
the energy from the turbulence dissipation is thermalized quickly and
the result is plasma heating.
Much astrophysical turbulence, however, is carried by
collisionless plasmas, where the Coulomb mean free path is
long. Although it is generally accepted that the particle distributions
in these plasmas are determined by couplings of
the (charged) particles with the turbulent electro-magnetic fluctuations, the
details
of these interactions are not well understood. The wave damping, which usually
occurs
on the smallest spatial scales, has been an essential part of plasma physics
theories for  collisionless but presumably {\it thermal} plasmas [see
e.g., \citet{B65, A85, GB04}]. There is little
work on
damping by nonthermal particle distributions.  Recently, \citet{P06}
derived the transit-time damping (TTD) rate of fast-mode waves by an arbitrary
energy
distribution of electrons and protons under typical solar-flare conditions.
Large scale
waves in this case can be damped by high energy particles through resonant
wave-particle couplings. The studies of the turbulent energy dissipation in
general and
the particle acceleration from a low energy background to high energies in
particular
require a more complete (beyond the MHD regime) treatment of the wave cascade
and
damping.

The cascade process has been an essential element in all kinds of turbulence
studies.
The
highly nonlinear nature of turbulence makes this a very challenging aspect of
the
problem. Nevertheless, the energy transfer of isotropic incompressible
hydrodynamical
turbulence, the simplest form of turbulence, has been described reasonably well
with
the Kolmogorov phenomenology that assumes a scale-independent self-similar
cascade
process, leading to the famous Kolmogorov power-law spectrum with a (one
dimensional
isotropic) spectral index of $-5/3$ in the inertial range
\citep{K41}.
Most astrophysical plasmas carry magnetic fields so that large scale MHD waves
and
small scale kinetic plasma fluctuations may be excited. The wave propagation
effects
on the cascade of MHD turbulence were first discussed by \citet{I63} and
\citet{K65}. Although MHD effects are expected to introduce anisotropy to
the
system that can affect the couplings between the turbulence and background
particles
significantly, these preliminary investigations assumed that the turbulence is
isotropic and reduced the cascade to a one-dimensional (1D) problem, yielding a
spectral index of $-3/2$ \citep{ZM90}. This is called the
Iroshnikov-Kraichnan (IK) phenomenology. Since both the Kolmogorov and IK
cascade
timescales ($\propto k^{-2/3}$ and $k^{-1/2}$, respectively) decline with
increasing  $k$ (or decreasing spatial scales) more slowly than
the periods of MHD waves ($\propto k^{-1}$),
the turbulence may be better described as spectra of waves
at higher values of $k$'s (smaller spatial scales), which play a
critical role in the
energizing of low energy background particles. However, at such high values of
$k$'s
one may be stepping beyond the MHD regime and must use more complex dispersion
relations to take into account the kinetic effects and the anisotropy of
the turbulence properly \citep{L00, M06}.

To study the anisotropic effects induced by the presence of large scale
magnetic
fields, MHD turbulence has been treated as an ensemble of linear wave modes
[e.g., \citet{ C03}]. Its nonlinear nature is revealed in
the
wave-wave couplings. Although this quasi-linear treatment may not be valid for
strong
turbulence, it is certainly a good approximation when the magnetic field
fluctuations
are much smaller than the large scale field and the wave periods are much
shorter than
the eddy turnover timescales \citep{Galtier06}. Significant
insight of properties of Alfv\'{e}n  and
magnetosonic turbulence has been obtained recently through this approach
\citep{SG94, GS95, G00, C05}.
However, the component of Alfv\'{e}n turbulence excitations nearly perpendicular
to a
large scale magnetic field is dominated by nonlinear effects \citep{MT81}. This
indicates inherent limitations of treating MHD turbulence as a
spectrum of
waves \citep{Galtier06}. As shown by \citet{MM95}, linear MHD wave modes do not
give a
complete description of turbulence excitations and their couplings,
and treating MHD
turbulence as an ensemble of linear wave modes may miss some critical nonlinear
effects \citep{NB97}. This is  more evident in hydrodynamic
simulations of
decaying turbulence, where the compressible wave component accounts for about
$10\%$ of
the turbulence kinetic energy \citep{P98, P99}. MHD simulations give
similar
results \citep{M96, V03}. These studies suggest
that
treating turbulence as a spectrum of linear wave modes may not describe the
dominant
energy component properly, which also reflects the highly non-linear nature of a
fully
developed turbulence, especially on large scales, where the wave periods
may be longer than the turbulence cascade timescale
\citep{BF03}.

Based on the dominance of nonlinear or wave-propagation effects, \citet{O06}
recently separated the Alfv\'{e}n turbulence into two interacting parts:
quasi-two-dimensional (2D) and wave-like fluctuations. The quasi-2D component
characterizes the nearly perpendicular excitations and may be described with a
quasi-2D
Kolmogorov phenomenology. The wave-like fluctuations may be described with an
IK
phenomenology with propagation direction dependent wave periods. The cascade of
turbulence then also depends on the assumed couplings of these two components.
It is
interesting to note that the two components are separated by the critical
balance
between linear wave periods and nonlinear eddy turnover timescales, and the
incompressible strong Alfv\'{e}n turbulence described by \citet{GS95}
appears
to be appropriate for the quasi-2D component that is dominated by nonlinear
effects.
Due to the suppression of cascade by wave propagation effects, Alfv\'{e}n
turbulence
with wave periods proportional to the parallel component of wavevectors
cascades
preferentially in the direction perpendicular to the mean magnetic field. The
wave-like
fluctuations need longer time to develop, which may explain the exponential
cutoff of
the power spectrum in the direction parallel to the mean large scale magnetic
field
observed in MHD simulations \citep{C02}.

Advances in computational power and numerical algorithms over the past few
decades have
made numerical simulations one of the important tools for quantitative
investigations
of magnetized turbulence \citep{S83, M98, C02, C03, CL03}. However, turbulence
usually covers a huge dynamical range from the
macroscopic
scales of turbulence generation to the microscopic dissipation scales. Current
simulations have a dynamical range of a few hundreds to a few thousands and have
not
been able to give a complete description of energy flows in magnetized
turbulence,  limiting their astrophysical applications. Moreover, most of
these
studies are limited to the MHD regime, where the background particles are
strongly
coupled with each other and can be treated together as a single fluid. The
electron magnetohydrodynamics (EMHD) treats electrons and ions as two fluids. It is only
applicable in a narrow frequency range between the electron and ion
gyro-frequencies,
where the whistler dispersion relation is valid \citep{B99,
PL04}. Recently Galtier (2006) showed that the transition from the MHD to the
EMHD may be studied with the incompressible Hall MHD for weak
turbulence. However, a quantitative weak turbulence theory
is still to be developed, and it is not obvious how
a complete theory may cover both the weak and strong
turbulence regimes.
To address the heating of background
particles by turbulence, one usually assumes
that most of the background particles reach thermal distributions and
arbitrarily
extrapolates the turbulence spectrum into the dissipation range, where
the MHD, Hall MHD, and EMHD
formalisms may be invalid [e.g., \citet{L99, T02, ZL04, WY06}]. The
particle-in-cell and kinetic simulations can usually
study processes near the gyro-scales of the background particles \citep{G08,
H08}.

Observations of solar corona, solar flares, solar winds, and space plasmas, on
the other hand, demand a detailed study of the turbulence evolution over a large
dynamical
range. There are no direct observations of turbulence in solar corona and solar
flares, but the observed high Reynolds numbers and models for
particle acceleration demand existence of turbulent plasma
waves \citep{BM77, MR95, FH08, KB08}.
{\it In situ} observations of space
plasmas reveal evidence of wave-particle interactions \citep{P06a, C07},
turbulence generation and dissipation, and the dominance of
the turbulence (in the  Earth's plasma sheet) by eddies instead of plasma
waves \citep{BF03}. In solar
winds, direct measurements of the turbulence spectrum over more than 6 orders of
magnitude in frequency challenge all existing theoretical models \citep{D83}.
Indeed, strong anisotropies are observed in almost all the relevant spatial
scales \citep{Ham08, OH07, D05, M90}. While the turbulence anisotropy at large
scales, which is likely related to the turbulence generation mechanism,
appears to be correlated with the solar wind speed \citep{D05},
at smaller scales the quasi-2D component always dominates,
which should be attributed to the dynamics
of the turbulence cascade \citep{Ham08}. At even higher frequencies, there is a
spectral break, which has been attributed to the onset of dissipation or kinetic
effects by the background particles \citep{L01, S01, Galtier06, H08a}.
In the dissipation range and kinetic effect dominant regime, background
particles with different charge to mass ratios interact with the
electromagnetic fluctuations quite differently \citep{PL04, L04,
L06, L98}.
This results in complicated wave dispersion relations,
namely the dependence of the wave frequency on the wavevector \citep{A85},
and
certainly affects the turbulence cascade. The couplings among waves and
particles are
also complicated \citep{S62, XO04, SG07}. Thus for practical
applications one requires a more readily solvable approach and algorithm, which
can utilize the  basic physics learned from detailed simulations
in an approximate but numerically effective manner.

The diffusion approximation for the power spectrum in the wavevector space has
been a
very efficient tool to study the turbulence cascade and dissipation
over a
large dynamical range \citep{ZK78}. Although the diffusion equation may
not be derived from the fluid dynamical equations, especially for weak
turbulence \citep{NB97, G00}, it has several achievements in
quantitative studies of the MHD turbulence.
The 1D models not only address the transition of
turbulence from
the large scale Kolmogorov phenomenology to the small scale IK cascade
\citep{ZM90}, but also are used to study the acceleration and heating of
background
particles by magnetized turbulence \citep{MR95, MLM96} and the damping of
waves
at small scales \citep{L01}. These studies have deepened our understanding
of
energy release processes during solar flares significantly.  However, they
encounter
difficulties in reproducing the observed broken power-law spectrum of solar wind
magnetic
fluctuations \citep{L01, S01}. Due to the nonlinear nature
of
these 1D diffusion models, the turbulence spectrum cuts off sharply at the wavenumber,
where the cascade timescale becomes comparable to the wave damping time.


Studies of weak Alfv\'{e}n turbulence with the quasi-linear treatment of
electromagnetic fluctuations have shown that the cascade is
anisotropic \citep{Galtier06}. It is
also
well-known that damping rates of different plasma wave modes by thermal
background
particles are very sensitive to the wave propagation direction [see  e.g.,
\citet{G61, GB04, P06}]. These anisotropies are
critical to study the energy dissipation through magnetized turbulence in a
collisionless
thermal plasma. Thus, 2D diffusion models are needed to advance our understanding of
magnetized turbulence and address the difficulties encountered with the 1D
models.
\citet{CV03} have shown recently that the heating of
the
background particles by Alfv\'{e}n turbulence are very sensitive to the 2D and
kinetic
effects. To recover the critical balance proposed by \citet{GS95}
and
partially take into account the kinetic effects, they constructed a complicated
advection-diffusion equation for the power spectrum with three dimensionless
coefficients. A similar quasi-2D model was proposed recently by \citet{H07,
H08a} to
explain the broken power-law character of the  power spectrum of the turbulent
magnetic field fluctuations in the solar wind.

In this paper, we study the general characteristics of the nonlinear 2D
diffusion model
with the kinetic and damping effects of Alfv\'{e}n-cyclotron fluctuations
treated
self-consistently.
In principle, all plasma mode branches, such as Alfv\'{e}n,
fast, and
slow
modes, and their extension beyond the MHD regime at small scales, should be
included to
have a complete description of the cascade and damping of magnetized turbulence.
One
then needs to solve a set of coupled diffusion equations for each wave mode
branch
\citep{A85}. This is a quite challenging task because the results will depend on
both
the
interactions within each wave mode branch and couplings among different
branches.
\citet{CL03} showed that couplings among different mode branches are usually
weak,
and \citet{LM06} showed that these couplings decrease toward small spatial
scales.
\citet{C05}, on the other hand, found that the couplings between Alfv\'{e}n
and
fast modes are significant in the direction parallel to the large scale magnetic
field,
where the frequencies of the Alfv\'{e}n and fast modes are comparable. The
couplings
among different branches therefore can be separated from other processes near
the
dissipation range, where the kinetic and damping effects dominate.

We use the exact dispersion relation for a cold plasma (to go beyond the MHD
regime),
which is a good approximation for the more general dispersion relation of a
collisionless thermal plasma. The diffusion coefficients are constructed as a
function
of the wave frequency, phase or group velocities so that the kinetic effects
are
treated self-consistently. In general, the linear Vlasov equation is
used to derive the thermal damping rates of the waves. Here we use the
so-called WHAMP code \citep{R82} for calculation of the damping
rate.\footnote{Since the damping rate increases sharply with the
increase of the wavenumber, the quasi-linear treatment is expected to
give a good approximation of the damping even for the nonlinear effect
dominated quasi-2D component.} 
The diffusion model
actually does not distinguish the quasi-2D and wave-like fluctuations
\citep{O06}. The balance between the eddy turnover and wave propagation is
revealed in
the
diffusion tensor.

We apply our formalism (for cascade and damping) to the
Alfv\'{e}n-cyclotron branch, which has been studied
extensively due to its simplicity and its prevalence in magnetized turbulence.
It is
chosen here to facilitate better comparisons of our models with previous
studies. In \S\
\ref{general}, we discuss how the turbulence cascade can be studied using the
diffusion
approximation, which reduces the turbulence evolution to a nonlinear 2D
diffusion
problem. We also investigate different forms for the diffusion tensor and include wave damping rate. For the
simplest cases, where the diffusion tensor is only a function of  time,
wavevector, and local turbulence power in the wavevector space, the nonlinear
diffusion equation can be solved numerically to obtain
the power spectrum from the MHD region to the $^4$He gyro-frequency, where the
Alfv\'{e}n dispersion surface cuts off. To better understand the cascade models,
we first
present results for the Alfv\'{e}n turbulence \S\ \ref{cascade}, where
the dispersion relation for Alfv\'{e}n waves is used. The exact dispersion
relation and the cascade beyond the MHD regime are discussed in \S\ \ref{disp}.
The damping effects are
investigated in \S\ \ref{result}. In \S\ \ref{obs} we apply models to solar
wind magnetic fluctuations and show that a locally isotropic diffusion model can
naturally explain the observations. We discuss the implication of this theory on
the study of
plasma heating and particle acceleration by magnetized turbulence and future
work, and
draw conclusions in \S\ \ref{dis}.

\section{Kinetic Equation}
\label{general}

Following the approach of \citet{ZM90}, we assume local interactions in
the wavevector ${\bf k}$ space so that the evolution
of the turbulence power spectrum ${\cal W}({\bf k}, t)$ only depends on its
properties near ${\bf k}$ at the time $t$:
\begin{eqnarray}
   {\partial {\cal W}({\bf k}, t) \over \partial t}
& = &\dot{Q}({\bf k}, t)
+ {\partial\over\partial k_i}\left[D_{ij}({\bf k}, t){\partial\over \partial
k_j}{\cal W}({\bf k}, t)\right]
- \Gamma({\bf k, t}){\cal W}({\bf k}, t)
- {{\cal W}({\bf k}, t) \over T_{\rm esc}({\bf k},t)}\,.
\label{WHOMOG}
\end{eqnarray}
Here $i, j$ indicate the three bases of the wave vector ${\bf k}$,
and the terms on the
right-hand-side represent the turbulence generation, cascade (through
diffusion), damping, and leakage processes, respectively. The rates of turbulence
generation ($\dot{Q}$), damping ($\Gamma$), and leakage ($T_{\rm esc}^{-1}$)
are mostly determined by the properties of the background plasma, whose
variation timescales, comparable to the large scale dynamical time, are usually
much longer than the turbulence evolution
timescales  so that the time-dependence of these rates can be
ignored. The diffusion tensor $D_{ij}({\bf k}, t)$ can be constructed with the
nonlinear timescale $\tau_{\rm NL}({\bf k}, t)$ associated with the eddy
turnover,
the wave crossing time $\tau_{\rm W}({\bf k})$, and the triple correlation time
$\tau_3=(\tau_{\rm NL}^{-1}+\tau_{\rm W}^{-1})^{-1}$. The turbulence cascade
rate then is defined as $\tau_{\rm cas}^{-1}\simeq\tau_{\rm NL}^{-2}\tau_3$, and
$D_{ij}\sim k^2 \tau_{\rm cas}^{-1}$.

More generally, one may also include an advection term to the above equation so
that the energy flux in the wavevector space
\begin{equation}
f_i({\bf k}, t) = -D_{ij}({\bf k}, t){\partial\over \partial k_j}{\cal W}({\bf
k}, t) + u_i ({\bf k}, t) {\cal W}({\bf k}, t)\,,
\label{flux0}
\end{equation}
where $u_i$ is a velocity field in the wavevector space \citep{CV03}. This
velocity field can be decomposed into a solenoidal
${\bf u}_{\rm s}$ and a compressible component ${\bf u}_{\rm c}$. The former is
divergence-free and can be attributed to the anti-symmetric part of the
diffusion tensor $D_{ij}({\bf k}, t)$. The latter is driven by an external
potential in the ${\bf k}$ space $\Phi({\bf k}, t)$, i.e., $u_{{\rm c}i} =
\partial \Phi/\partial k_i$. The highly non-linear phenomenon of intermittence
can also affect the energy diffusion of turbulence. However, intermittence
likely results from the non-local nature of interactions in the wavevector
space \citep{B02, C02}. One may partially take into account
the effects of this process
on the turbulence energy transfer with a nonlinear damping rate
[see  e.g., \citet{BL08}]. In
general, the diffusion tensor is a functional of ${\cal W}$. For local
interactions in the ${\bf k}$ space, it is reduced to a function of ${\bf k}$.
In what
follows, we will focus on the relatively simple case with ${\bf  u}_{\rm c}=0$,
and $D_{ij}$ symmetric and only functions of ${\bf k}$ and $t$, and ignore the
cascade and dissipation associated with intermittence.

Without excitation of waves, i.e., $\tau_{\rm W}=\infty$, the turbulence cascade
is well described with the Kolmogorov phenomenology, which assumes an
isotropic
incompressible fluid with local interactions of eddies in the wavenumber
 space so that $\tau_{\rm NL} = ({\cal W} k^5)^{-1/2}$ and $D_{ij}
= \delta_{ij}k^2/\tau_{\rm NL}$, where $\delta_{ij}$ is the Kronecker's delta
function. For strong turbulence with $\tau_{\rm NL}({\bf k})\ll\tau_{\rm W}({\bf
k})$, nonlinear processes dominate and one retains the Kolmogorov phenomenology.
In the weak turbulence limit, $\tau_{\rm NL}({\bf k})\gg\tau_{\rm W}({\bf k})$
and the turbulence may be described as a spectrum of linear wave
modes.
For fast mode waves in a cold or low beta plasma (i.e., when
the Alfv\'{e}n speed $v_{\rm A}\gg C_{\rm S}$, the sound speed), the
wave frequency $\omega({\bf k})\simeq v_{\rm A}k$, and
$\tau_{\rm W} =
\omega({\bf k})^{-1}\ll \tau_{\rm NL} = ({\cal W} k^5)^{-1/2}$, $D_{ij} =
\delta_{ij}k\tau^{-2}_{\rm NL}/v_{\rm A}$, and one recovers the IK
phenomenology.\footnote{Note that  for sound waves in a fluid,
$\omega({\bf k})= C_{\rm S} k$, one obtains results similar to fast mode waves.}
For Alfv\'{e}n waves, $\omega({\bf k})= v_{\rm A} k_{||}$, where $k_{||}$ is
the
component of the wavevector parallel to the large scale magnetic field,
wave packets propagate along the mean
magnetic field. Since only oppositely directed wave packets can interact
\citep{G00}, $\tau_{\rm W}\propto k_{||}^{-1}$. Although the weak
turbulence approximation may be valid for
parallel propagating waves, nonlinear processes always dominate
in the perpendicular direction. Therefore the quasi-linear theory
for linear wave modes does not give a complete
description of the Alfv\'{e}nic
turbulence fluctuations \citep{O06}.

We are interested in extending the turbulence cascade study beyond the MHD
regime by using the general dispersion relation for the Alfv\'{e}n-cyclotron
fluctuations and compressible plasma waves. The quasi-linear approaches for
MHD wave interactions then become extremely complex if not completely
unsolvable. On the other hand, the diffusion approximation can
provide simple solutions with a prescribed diffusion tensor and may help us to
uncover processes beyond the MHD regime. In the following, we will ignore the
leakage term in equation (\ref{WHOMOG}) and treat the source term as a
$\delta$-function at certain large scale. Thus, for the investigation of  the
evolution of the  turbulence power
spectrum we only need proper choices of the diffusion tensor $D_{ij}$ and the
damping
rate $\Gamma$.

\subsection{Diffusion Tensor}

We  adopt the recipe given by Zhou \& Matthaeus (1990) for the 1D cascade
of Alfv\'en waves caused by the wave-wave
resonances and eddy turnovers and generalize it to waves propagating in
arbitrary
directions with a propagating direction dependent diffusion tensor.
The eddy turnover time can be estimated with $\tau_{\rm NL}=(v_k k)^{-1}$, where
the eddy velocity \footnote{Here we have ignored the
potential anisotropy of the turbulence. In a more accurate treatment,
one also needs to distinguish contributions to the turbulence power spectrum
from the velocity and magnetic fields. The eddy velocity should only depend on
the turbulence kinetic energy \citep{Galtier06}.}
\begin{equation}
v_{\rm k}\simeq ({\cal W} k^3)^{1/2}\,\,\,\,\,\, {\rm so\ \
that}\,\,\,\,\,\,\tau_{\rm NL}= ({\cal W} k^5)^{-1/2}\,.
\label{eddytime}
\end{equation}
For the wave crossing time,
one must consider the anisotropy of the Alfv\'{e}n-cyclotron dispersion.
Wave packets with a size of $\sim k^{-1}_{||}$ cross each other at the
Alfv\'{e}n speed. We then have
$\tau_{\rm W}\simeq(v_{\rm A} k_{||})^{-1}$, and the wave propagation effect
reduces the triple correlation time from $\tau_{\rm NL}$ to
$\tau_3$ and gives a cascade rate $\tau_{\rm cas}^{-1}={\tau_{NL}^{-2}/(
\tau_{NL}^{-1}+\tau_{\rm W}^{-1}}$).\footnote{Note that,
following Kraichnan's argument (1965) for an
isotropic Alfv\'{e}n wave turbulence, Zhou \& Matthaeus (1990) obtained
$\tau_{\rm W}=(v_{\rm A} k)^{-1}$ for their 1D diffusion
model, which is appropriate for the isotropic sound and fast mode
waves.}
Then the
simplest (locally isotropic) diffusion tensor may be constructed as:
\begin{equation}
D_{ij}\equiv \delta_{ij} {C \over 4\pi} k^2 \tau_{\rm cas}^{-1}
=\delta_{ij}{C \over 4\pi}{{\cal W} k^7 \over ({\cal W} k^3)^{1/2} k + v_{\rm
A}k \cos\theta}=\delta_{ij}{C \over 4\pi}{{\cal W} k^6 \over v_ {\rm k }+ v_{\rm
A} \cos\theta}
\label{DijMHD}
\end{equation}
where $C$ is a dimensionless scaling constant corresponding to the Kolmogorov
constant for hydrodynamic turbulence, and $\theta$ is the angle between ${\bf
k}$ and the mean magnetic
field: $k_{||}=k\cos\theta$. In what follows we set $C=1$.\footnote{$C=
(3/11)(4\pi/C_0)^{3/2}\simeq 5.89$, where $C_0\simeq 1.62$ is  the Kolmogorov
constant \citep{YZ97}. If the eddy speed $v_k=({4\pi\cal W}k^3)^{1/2}$, $C\simeq
1.66$.} This diffusion tensor
is
locally isotropic in the wavevector space implying that
the energy of a wave packet at ${\bf k}$ can spread into its neighborhood in the
${\bf k}$ space without any preferred directions.
The wave
propagation effect just reduces the diffusion coefficient. This reduction is
higher for waves with higher phase speeds.

In more general cases, the diffusion tensor may not be locally isotropic in the
wave
vector space due to
resonance interactions.\footnote{Non-resonance interactions presumably lead to
an
isotropy.} One may construct the parallel and perpendicular components of the
diffusion tensor differently.
In the perpendicular
direction,
the turbulence cascade is determined by the eddy turnover time only. We have
$D_{\perp,\perp}=k^2\tau^{-1}_{\rm cas,\perp}=k^2 \tau^{-1}_{\rm NL}$.
The wave propagation effect reduces the cascade rate in
the parallel direction $\tau^{-1}_{\rm cas,||}=\tau^{-2}_{\rm NL}/(\tau_{\rm
NL}^{-1}+\tau_{\rm W}^{-1})$, and the
corresponding diffusion coefficient can be written as
$D_{||,||}=k^2\tau^{-1}_{\rm cas,||}$. Thus,
we obtain an anisotropic diffusion tensor
\begin{equation}
D_{ij} = {1 \over 4\pi}{ k^2\over \tau_{\rm NL}} \left[
\begin{array}{cc}
(1+ \tau_{\rm NL}/\tau_{\rm W})^{-1} & 0 \\
0 & 1\\
\end{array}
\right].
\label{Dij_ani.eq}
\end{equation}
There are also two choices for the wave interaction timescale. We can choose
\begin{equation}
\tau_{\rm W}^{-1}= {\bf k}.{\bf v}_{\rm gr}, \,\,\,\,\,\,
{\rm or} \,\,\,\,\,\, \omega({\bf k})\,,
\label{wavetimes}
\end{equation}
where ${\bf v}_{\rm gr}$ is the wave group velocity.
In the MHD regime they both give  $\tau_{\rm W}=kv_{\rm A}\cos\theta$ used in
equation
(\ref{DijMHD}). However, as mentioned above we are often interested in
acceleration of low energy particles, which requires cascade to high values of
$k$ regime.
In this case we should use the exact dispersion relation, going beyond the MHD
regime
($k\ge\Omega_p/v_{\rm A}$), where the above timescales are different. In what
follows
we will present results for both tensors. A brief review of the exact dispersion
relation and its decomposition into different modes is presented in Appendix A.
Figure  \ref{disp.ps} gives the $\omega-k$ relation for different modes at
various angles of
propagation with respect to the large scale magnetic field.

So far we have only used the dispersion relation of waves to construct the
diffusion
tensor. Physically the polarization state (including the compressibility) of the
waves may also affect the wave diffusion. An equation for the magnetic
helicity spectrum is also required \citep{Galtier06}.
These may become more significant when we consider the couplings between
different wave branches.
They may also explain the weak couplings between incompressible
Alfv\'{e}n wave modes and the compressible fast and slow mode
waves \citep{CL03}, and the relatively strong couplings between
parallel propagating Alfv\'{e}n and fast mode waves \citep{C05}.
Instead of
exploring more possibilities on the forms of the diffusion tensor, we defer to
observations of turbulence spectrum
in solar winds to point us the right direction. The detailed discussions
on this is given in  \S\ \ref{obs} below.

\subsection{Damping Rate}
\label{damp}

MHD turbulence in collisionless plasmas is subject to damping caused by both
resonance and non-resonance processes. The latter may be associated with current
sheets and is not well understood \citep{L00, M06}, and
the former likely dominates the damping of plasma waves \citep{CL06}.
It is well-known that charged particles can extract energy from plasma waves
through cyclotron resonances, transit-time damping (TTD), and Landau damping
(LD) processes. Cyclotron resonances by low energy background particles can only
lead to damping of a  high frequency wave with the wave frequency comparable to
the particle gyro-frequency.  Magnetic field fluctuations in the direction of
the mean magnetic field are required for the TTD processes to operate. The TTD
conserves the magnetic moment of the particles. Efficient pitch angle scattering
of the particles is also needed for this process to energize particles
effectively \citep{M97}. The LD is related to electric field fluctuations
along the mean magnetic field and therefore can be very efficient \citep{S62}.
Alfv\'{e}n waves in the MHD regime
don't have electric and magnetic field components along the mean magnetic field
and therefore are only subject to cyclotron damping by relatively high energy
particles,
whose gyro-frequency is relatively low.
At small scales and/or high frequencies, where the kinetic
effects become important, there are electric and magnetic field
fluctuations along the mean magnetic field, and the TTD and LD processes can
damp the Alfv\'{e}n wave branch.

The damping of waves energizes the plasma either by heating it or by production
of
nonthermal tails. The properties of the turbulence and background plasma
determine the
energy partition into these two channels. Therefore, in general, both the hot
plasma
and non-thermal particles contribute to the damping rate. Unfortunately there is
very
little work on determination of the damping rate by nonthermal particles. The
only
result we are aware of is that by \citet{P06}  [see also \citet{Y08}]
on the TTD of fast mode waves in low beta  plasmas;  $\beta_{\rm p}<1$. This
was done by using the fact that the rate of particle acceleration and the
damping rate
are related via the energy conservation \citep{MR95, MLM96}. It was shown
that for typical solar flare
conditions, except for very large flares, the damping is usually dominated by
thermal
particles. We believe that the same is also true in the solar wind, where we
apply our
formalism in this paper, as indicated by the relatively
low energy content of non-thermal
particles. Thus, in what follows, we include damping by thermal particles only,
in
which case the damping rate as a function of wavevector ${\bf k}$ is given by
the imaginary
part of the frequency obtained from the dispersion relation for a warm ($T\not=0$) plasma.

For thermal background particles with isotropic pitch angle distributions,
depending on the background magnetic field, particle abundances and
temperatures,
the different damping processes of the Alfv\'{e}n wave branch mentioned above dominate in
different wave propagating
directions \citep{GB04}. At low values of $k_{||}<\Omega_\alpha/v_{\rm A}$,
where $\Omega_\alpha$ is the non-relativistic gyro-frequency of the
$\alpha$-particle, these
waves are called the kinetic Alfv\'{e}n waves ({KAW) near $k_\perp \rho_p \sim
1$, where $\rho_p$ is the Larmor radius of the thermal background protons. They
play important roles in the heating of background particles \citep{L99,
CV05}.
The quasi-parallel propagating Alfv\'{e}n waves have very weak electric and
magnetic field fluctuations along the large scale magnetic field and therefore
are not
subject to the TTD and LD. These waves play important roles in the selective
acceleration of background ions with different gyro-frequencies to high energies
through cyclotron resonances \citep{R94, M02, L04, L06}.

\citet{S89}
gives a simplified expression for the damping rate by electron cyclotron
resonances of parallel propagating waves under the weak
damping approximation.
This result can be easily generalized
to calculate the cyclotron damping rate by all background particle species:
\begin{displaymath}
\frac{\omega_i}{\omega_r}=-\frac
{\displaystyle\sum_{s}^{} \frac{\sqrt{\pi}\omega^2_{p,s}}{\omega_r k v_{t,s}}
\exp\biggl[-\left(\frac{\omega-\epsilon_s\Omega_s}{k v_{t,s}}\right)^2 \biggr]}
{\displaystyle\frac{2k^2}{\omega^2}+
\displaystyle\sum_{s}^{}\frac{2\omega-\epsilon_s\Omega_s}{
\omega(\omega-\epsilon_s\Omega_s)^2}\omega_{p,s}\,}.
\end{displaymath}
Here subscripts $i$ and $r$ refer to the imaginary and real parts,  $s$ stands for different particle species and $\omega_p$, $\Omega$, $v_t$,
and
$\epsilon$ are the corresponding particle plasma frequency, non-relativistic
gyro-frequency, thermal speed, and sign of the charge, respectively.
For a given power spectrum of quasi-parallel propagating waves, this equation
can be used to study the relative acceleration of different background
particle species.

We use the Waves in Homogeneous Anisotropic Multitemperature Plasma (WHAMP)
code \cite{R82} to obtain the thermal damping rate numerically. When
$\Gamma=\omega_i \sim \omega_r$, the weak damping approximation is invalid and
the code fails to converge.
We use a power-law extrapolation with respect to the wavevector to approximate
the damping rate in the strong dissipation regime. This does not affect our
result because the turbulence spectra usually cut off at the wavevectors where
the damping rate is still much lower than the wave frequency.  For further
details see Appendix B and Figures \ref{kc_contour.ps} and \ref{taus_cut.ps}
below.

\section{Cascade of Alfv\'{e}n Turbulence}
\label{cascade}

In this section we consider the cascade in the {\it inertial range} ignoring damping and
leakage.
To compare with previous studies, we first consider Alfv\'{e}n turbulence in the
MHD regime
with $\tau_{\rm W}^{-1}=\omega({\bf k}) = k_{||} v_{\rm A}$.
In the strong turbulence limit one has $v_{\rm k} \gg v_{\rm A}$ and  $\tau_{\rm NL}\ll
\tau_{\rm W}$. As a result
the diffusion tensors of
equations (\ref{DijMHD}) and (\ref{Dij_ani.eq}) lead to the isotropic Kolmogorov
cascade. However, in the weak turbulence limit, i.e., $v_{\rm k} \ll v_{\rm A}$,
they
are different from the isotropic
IK phenomenology due to the dependence of the  Alfv\'{e}n
wave crossing time on the angle of propagation $\theta$. Since the Alfv\'{e}n
wave
crossing time $\tau_{\rm W}$ is
much longer in the perpendicular
direction than in the parallel direction, the wave cascade rate is high in the
perpendicular direction and
decreases dramatically with the increase of $k_{||}$.
We solve the 2D time-dependent wave kinetic equation (\ref{WHOMOG}) with
the alternative direction implicit scheme on a uniform log-log grid and with
\begin{eqnarray}
\dot{Q}_{\cal W}({\bf k}, t) & = & F_0\delta({\bf k}-{\bf
 k}_0)\Theta(t)\,,\nonumber \\
k_{||,0}&=& k_{\perp,0}=k_0=1.6\times 10^{-8}\Omega_p/v_{\rm
A}\,, F_0=9.55\times 10^{-10}v_{\rm A}^2\Omega_p,
\label{source}
\end{eqnarray}
 where $\Theta$ is the Heaviside step function and $\delta$ is the 3D Dirac-$\delta$-function.
A reflective boundary condition is used at the large scale boundary $k_{||,0},
k_{\perp,0}$.  In this section, we ignore
the damping and leakage and allow the  turbulence to escape
freely at the high end boundary of $k$.%
\footnote{When damping is included (see \S\ 5 below), it dominates the cascade  at high values of $k$ (or small scales). We then choose an upper boundary for $k$  large enough such that all
injected wave
energy is damped within the simulation domain.}
Because no damping is included, the time unit is set by the proton
gyro-frequency $\Omega_{p}$.
The turbulence power
spectral contours obtained with this method are shown in Figure \ref{Acon.ps}
for the two
diffusion tensors [eqs. (\ref{DijMHD}) and  (\ref{Dij_ani.eq})] at two
different times. Figure \ref{MHD_ani.ps} shows the corresponding
steady-state spectral contours 
(left panels) and power spectra at several propagation angles (right panels).

\begin{figure}[htb]
\begin{center}
\includegraphics[width=0.24\textwidth]{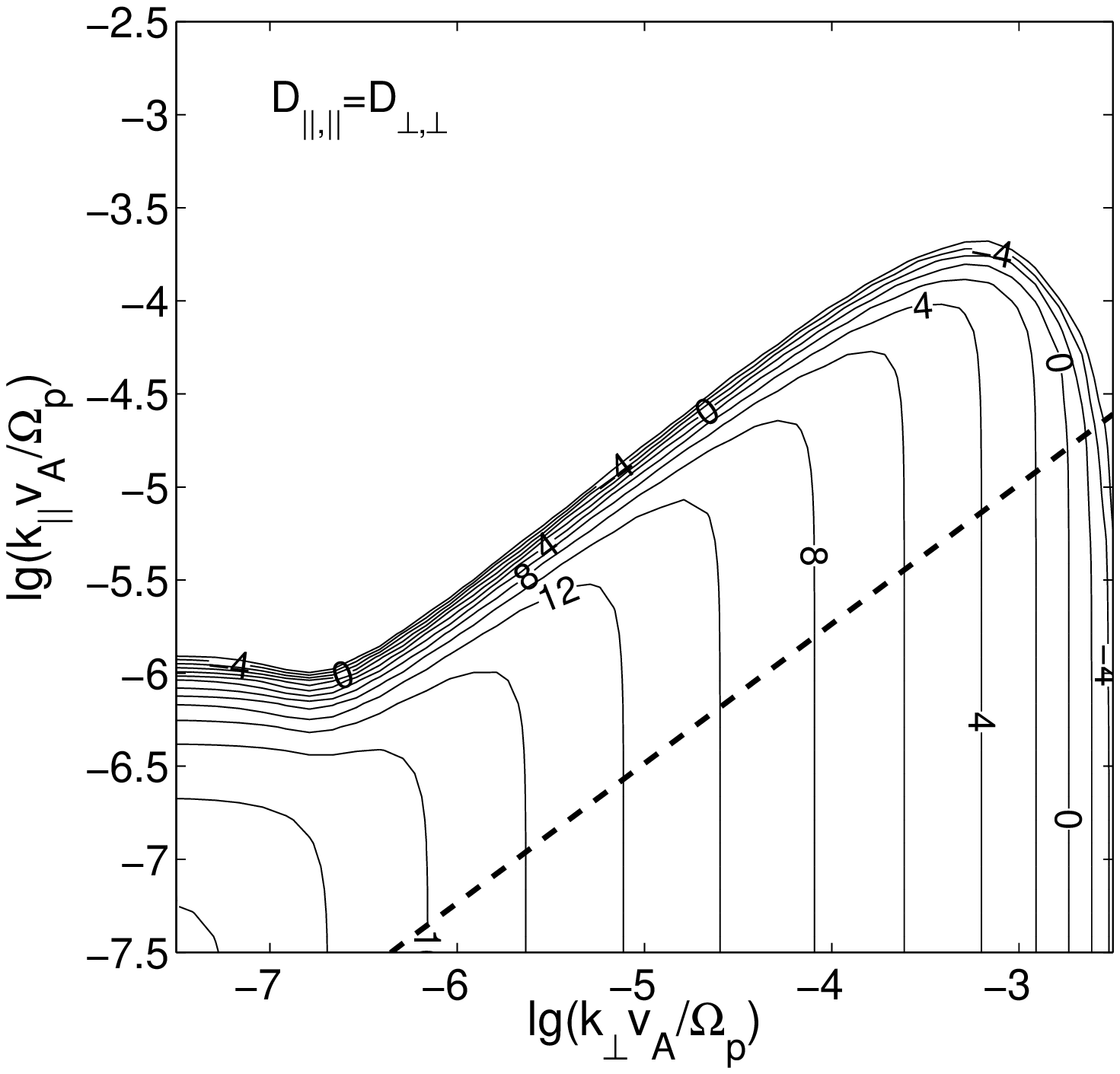}
\includegraphics[width=0.24\textwidth]{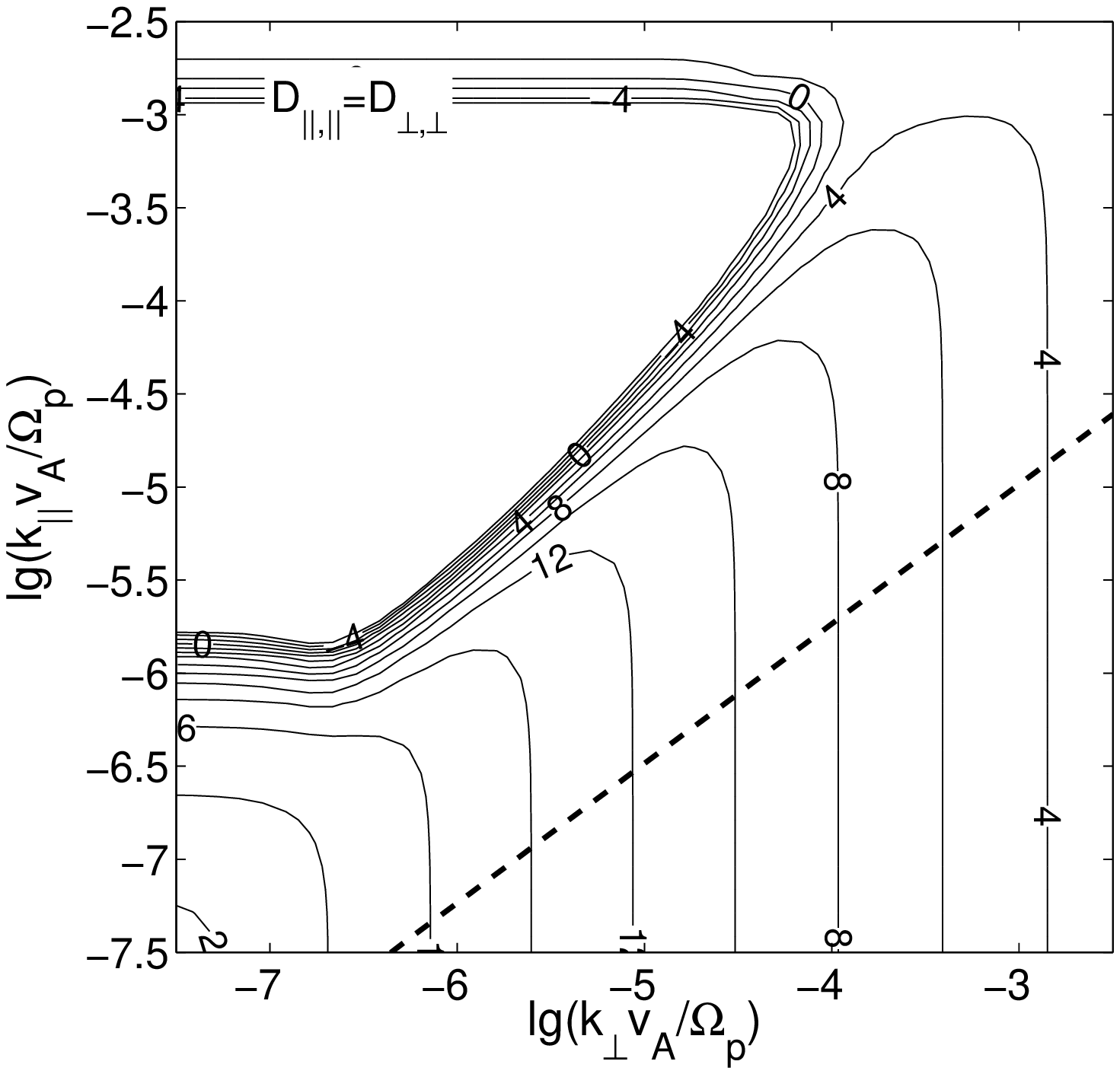}
\end{center}
\begin{center}
\includegraphics[width=0.24\textwidth]{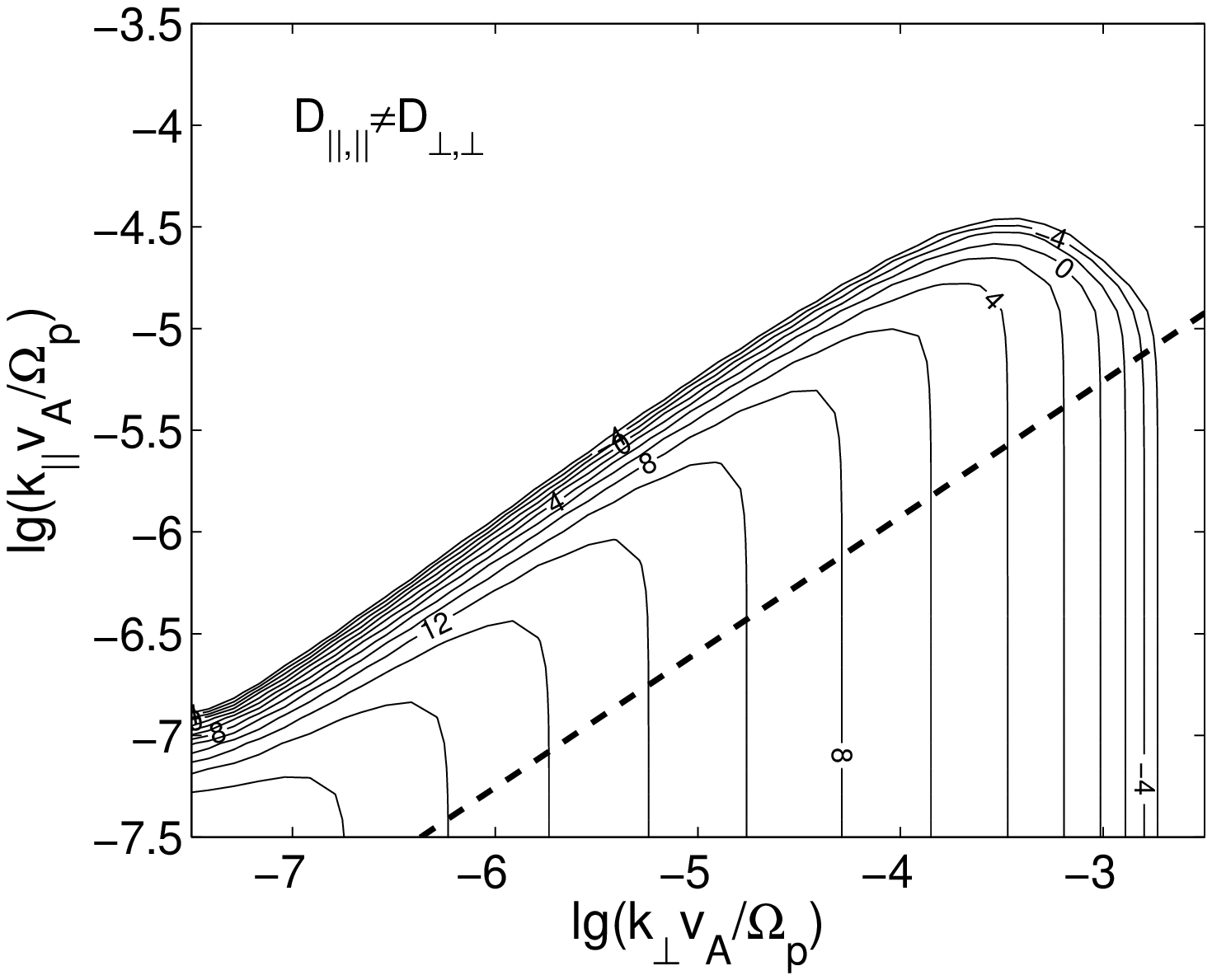}
\includegraphics[width=0.24\textwidth]{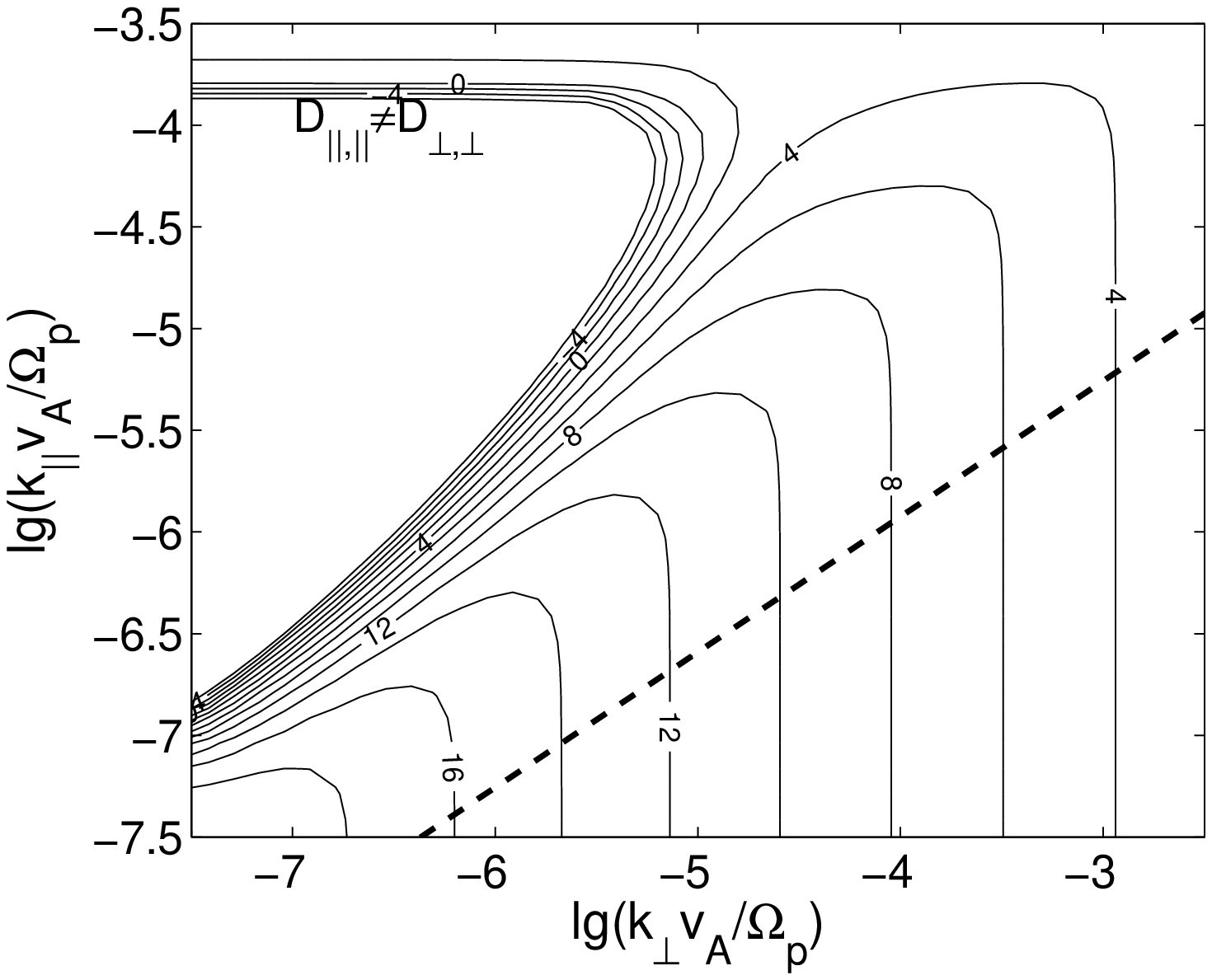}
\end{center}
\caption{ 
 Contours of the power spectrum ${\cal W}(k_{||},k_{\perp}, t)$ of the MHD range
Alfv\'{e}n
turbulence without damping at $t=3.46\times10^9\ \Omega_p^{-1}$
(top left),
$3.70\times10^9\ \Omega_p^{-1}$ (top right),  $1.16\times10^9\
\Omega_p^{-1}$ (bottom left), and $1.26 \times10^9\ \Omega_p^{-1}$ (bottom
right). The energy is injected uniformly at the surface of
$k_{||,0} =k_{\perp,0}= 1.6\times10^{-8}\Omega_p/v_{\rm A}$  with
an
injection rate $F_0=9.55\times 10^{-10}v_{\rm A}^2\Omega_p$. The logarithm to
the base 10
of the contour levels in units of $v_{\rm A}^5\Omega_p^{-3}$
are indicated in the figures.
The upper and lower panels show the results of the isotropic
($D_{||,||}=D_{\perp,\perp}$) and
anisotropic ($D_{||,||}\neq D_{\perp,\perp}$) diffusion models,
respectively. The dashed lines indicate the critical balance relation
discussed in the text. 
The power spectra in the left panels are similar to those uncovered by MHD
simulations \citep{C02}. The reverse cascade in the
perpendicular direction at high values of $k_{||}$ is evident in the right
panels.
}
\label{Acon.ps}
\end{figure}

As expected, the turbulence cascade in the parallel direction is initially
suppressed. The Alfv\'{e}n wave period at the injection scale is given by
$T_A=2\pi/k_{||,0}v_A= 3.9\times 10^8\Omega_p^{-1}$. It takes less than
$10\ T_A$ for the turbulence to reach the small scale boundary of the
simulation domain. Steady-states are reached in about $10\ T_A$. This
result is in conflict with simulations by Cho et al. (2002), where
the turbulence power spectrum at a given $k_\perp$ is found cutting off
exponentially with the increase of $k_{||}$ in the steady-state (Cho et
al. 2002). It is not obvious to us what prevents the generation of
waves in these simulations. Theoretically, if the turbulence cascades
toward small scales following the critical balance proposed by Goldreich
\& Sridhar 
(1995), as confirmed by these simulations, it should be able to reverse
cascade along lines (surfaces) of constant perpendicular wavenumbers
through three wave-interactions. Our time-dependent numerical
results in Figure \ref{Acon.ps} mimic such a reverse cascade process.
Over a period of less than $T_A$ (from the left panels to the right ones), high frequency nearly parallel
propagating waves are generated by small scale fluctuations of the
(quasi-2D) perpendicular component. This phenomenon is caused by the
increase of the diffusion coefficient with the increase of the
wavenumber and the suppression of 
the diffusion toward small parallel scales. 
At large scales, the wave energy can also diffuse slowly toward high
values of $k_{||}$. 

\begin{figure}[htb]
\begin{center}
\includegraphics[width=0.24\textwidth]{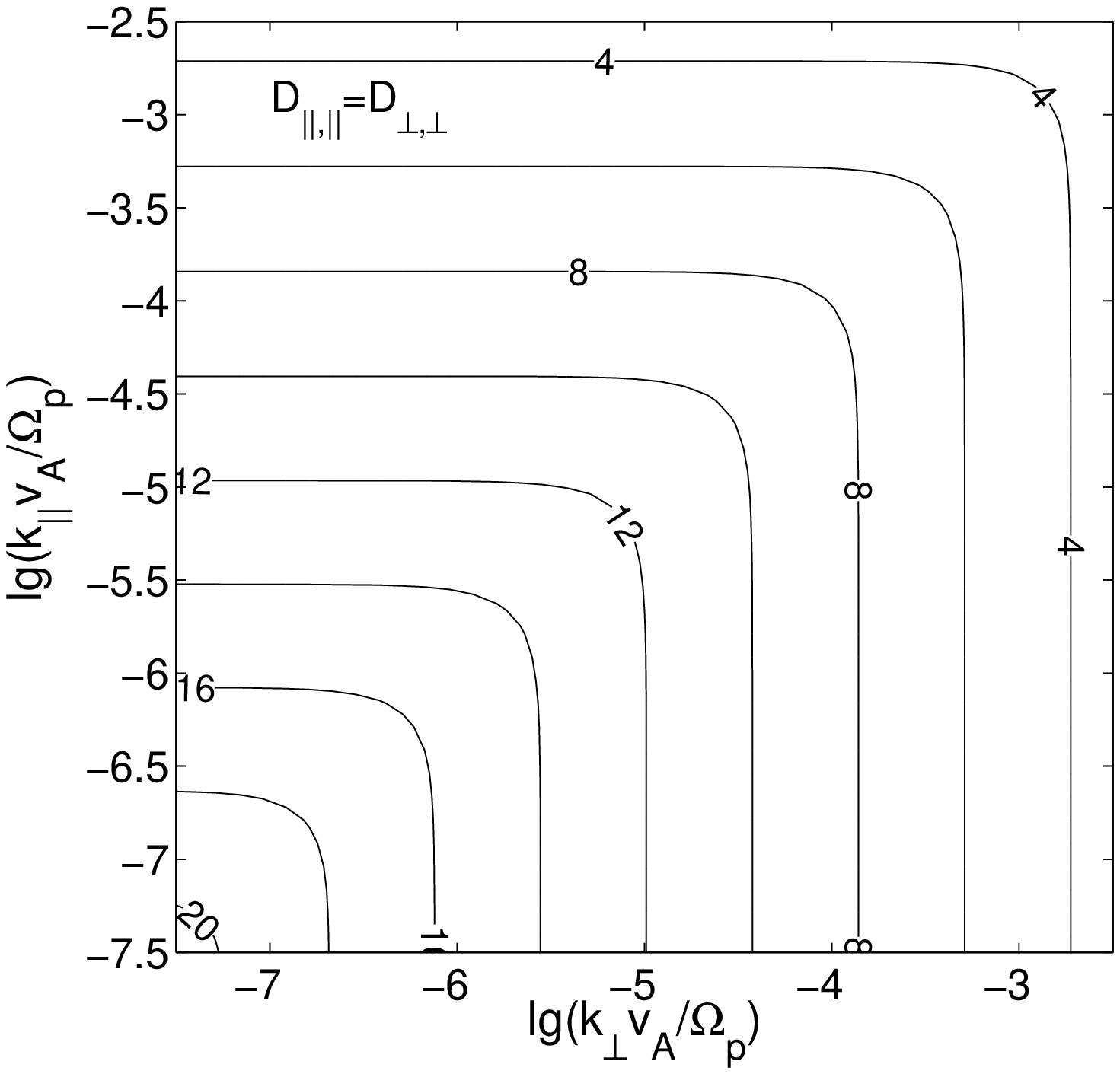}
\includegraphics[width=0.24\textwidth]{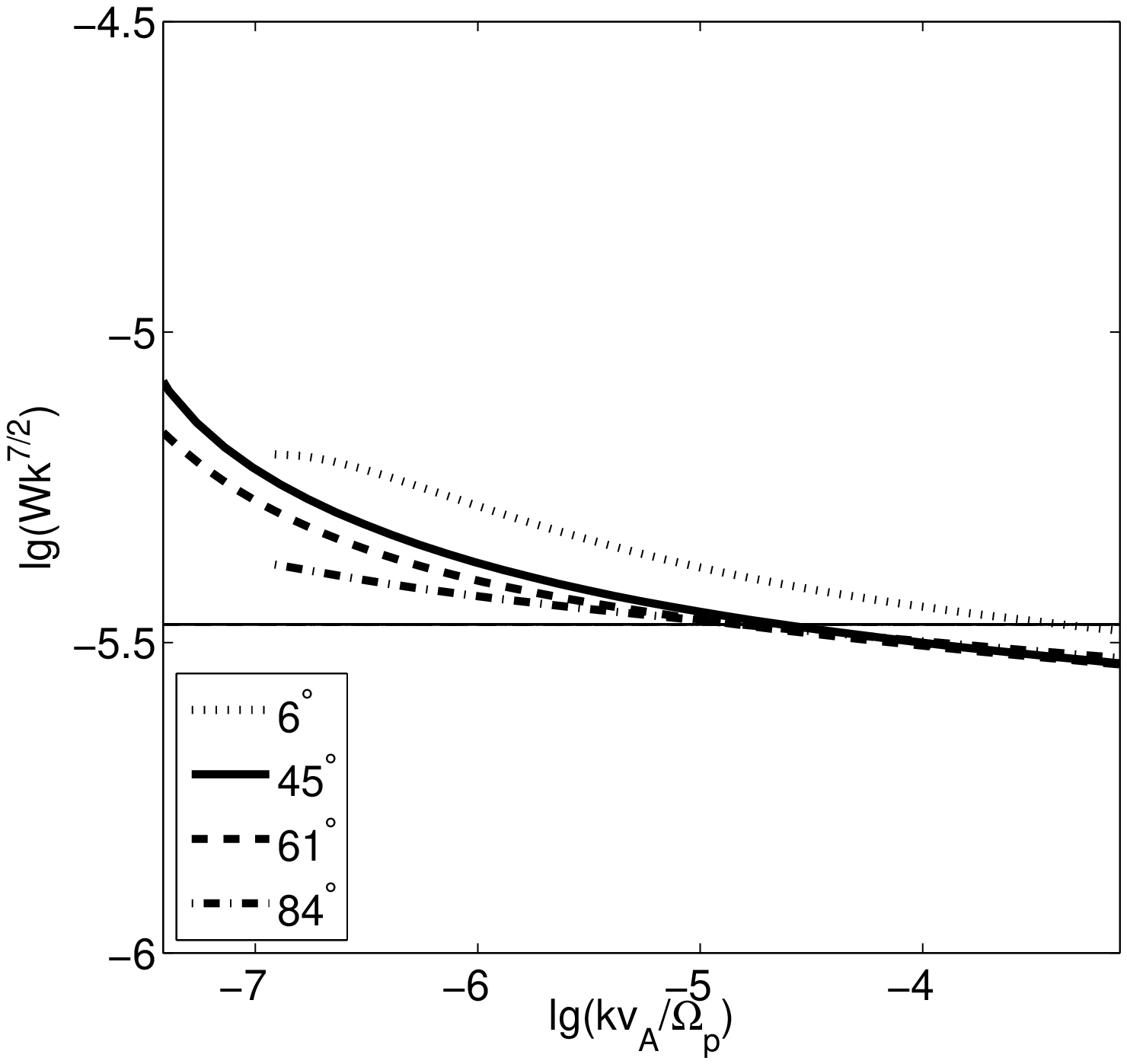}
\end{center}
\begin{center}
\includegraphics[width=0.24\textwidth]{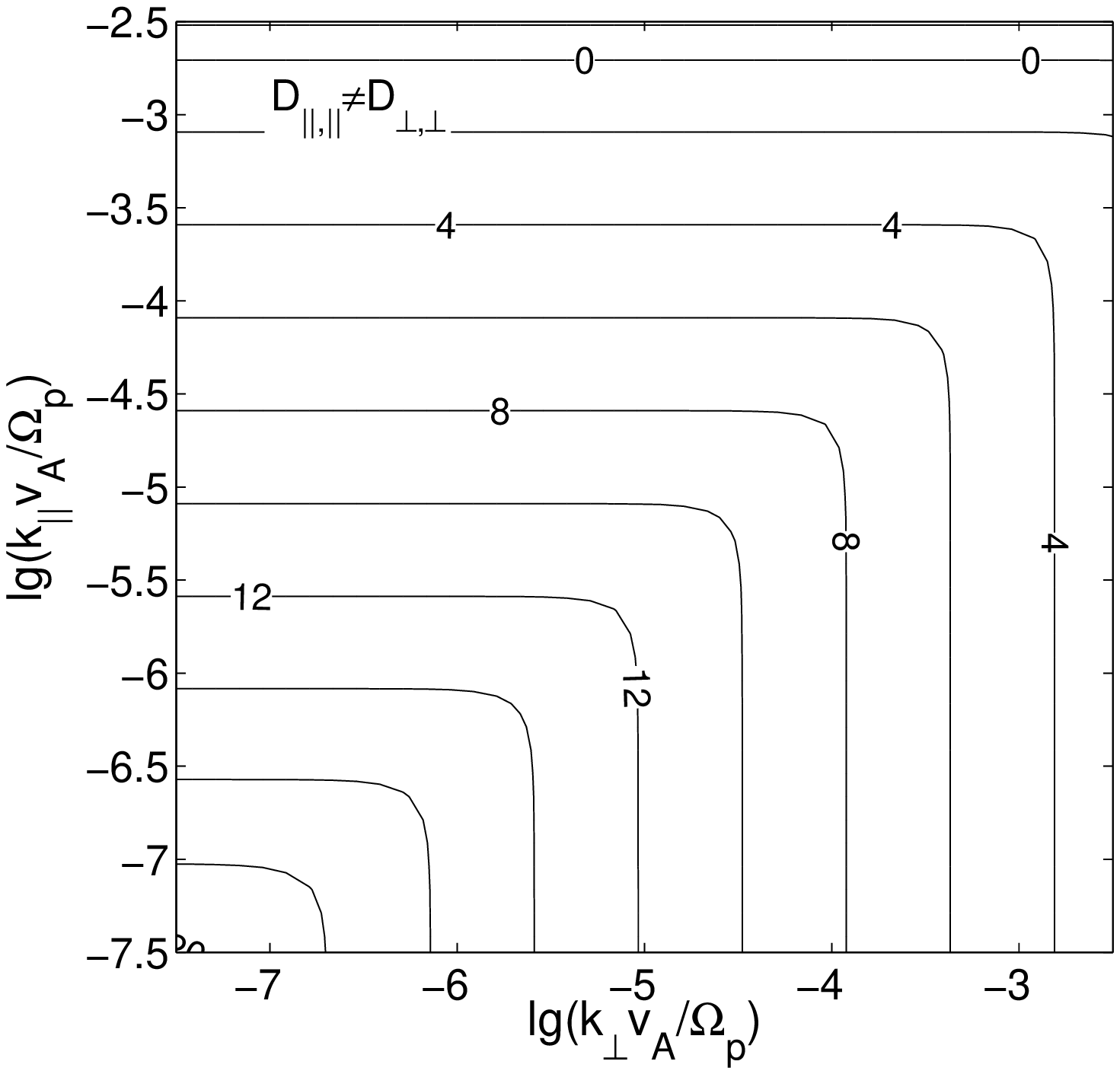}
\includegraphics[width=0.24\textwidth]{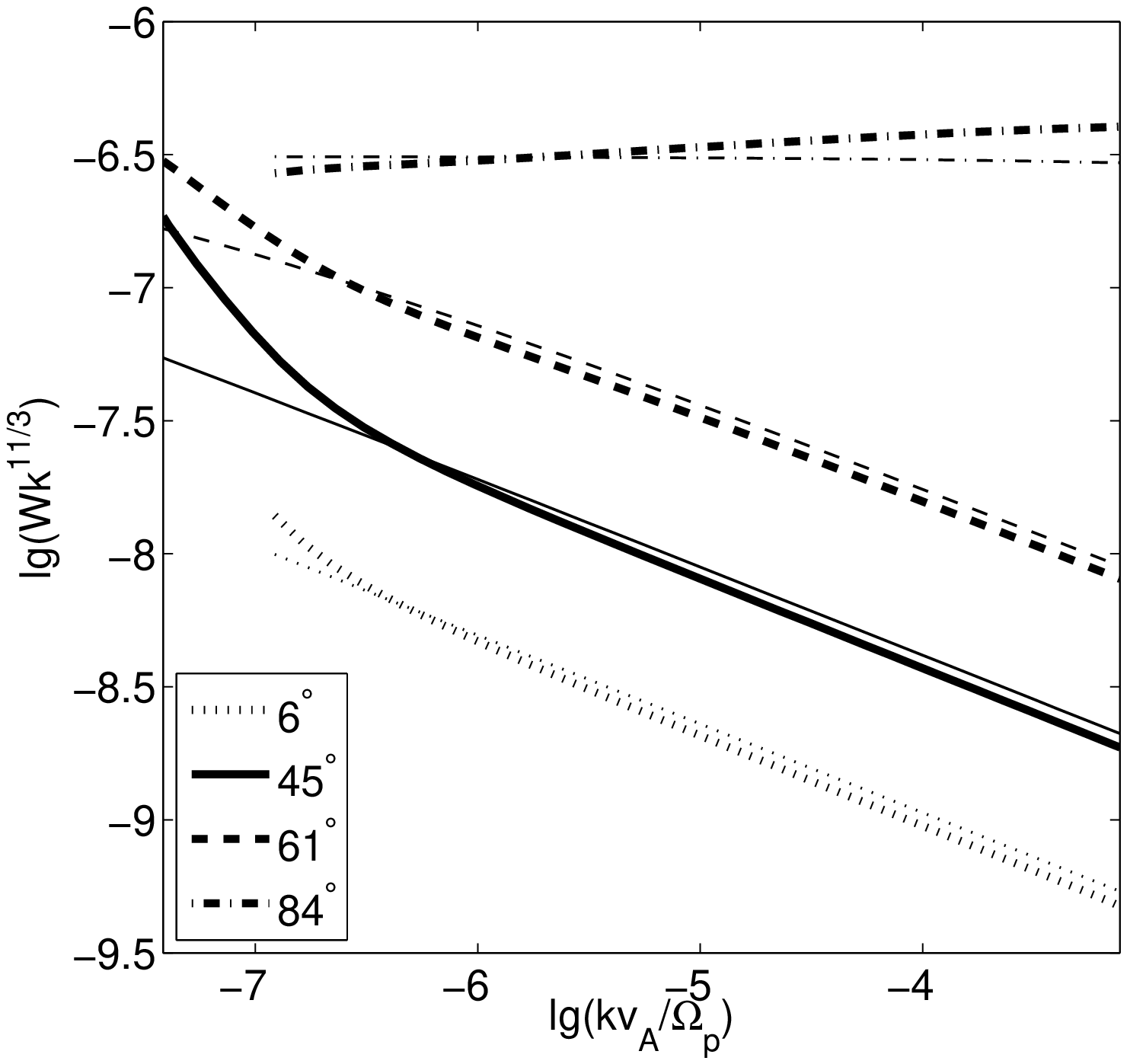}
\end{center}
\caption{
{\it Left:} Same as Figure \ref{Acon.ps} but for the steady-state power
spectral
contours. {\it Right:} The scaled steady-state power spectra in several
propagation
directions indicated in the legends. The  Alfv\'{e}n Mach number $M_{\rm
A}=0.30$  for
the top panels and $M_{\rm A}=0.23$ for the bottom panels.
The thick lines in the right panels
are simulation results.
The isotropic diffusion coefficient produces a nearly isotropic spectrum, which
is very
similar to the isotropic
Kraichnan spectrum (thin horizontal lines), and the anisotropic diffusion
coefficient produces spectra
steeper
than the Kolmogorov spectrum in directions other than the perpendicular
direction. The thin straight lines (power-laws) are the approximate spectra
based on equation (\ref{specfit}). The
spectral features at low values of $k$ may depend on the geometry of the
surface, where energy is injected into the system.
}
\label{MHD_ani.ps}
\end{figure}

\subsection{Results from Isotropic Diffusion Tensor}

Let us first consider
the {\it locally} isotropic diffusion tensor (top panels). Note that the
magnetic Mach number $M_{\rm A}=v({k_0})/v_{\rm
A}=0.30$, which means that our numerical solution is in the wave effect dominant
domain. The steady-state power
spectrum is nearly isotropic and is very similar to the IK spectrum between
$10^{-6}\Omega_p/v_{\rm A}$ and $10^{-3}\Omega_p/v_{\rm A}$, ${\cal W}={\cal
W}_0 (k/k_0)^{-7/2+\alpha}$, where
$\alpha(k)\ll1$ is a slowly varying function of wavenumber $k$ ($\alpha(k\rightarrow \infty)\rightarrow 0$).
Substitution of  this  spectrum and $D_{ij}$ given in equation (\ref{DijMHD}) into equation (\ref{flux0}) gives, the radial component of the
slightly anisotropic energy flux density  as
\begin{equation}
f(k) = (1/4\pi) (7/2-\alpha-k\alpha^\prime\ln (k/k_0)){\cal W}^2k^5[v_{\rm k} +
v_{\rm A} \mu]^{-1}\,,
\label{fluxd}
\end{equation}
where $\mu=\cos\theta$ and the prime indicates  derivative with respect to $k$. The total
energy flux through spheres centred at the origin of the ${\bf k}$ space is then
\begin{eqnarray}
F_0 &=&  (7/2-\alpha-k\alpha^\prime\ln (k/k_0)){\cal
W}^2k^7\int_{0}^1{\rm d} \mu[(v_{\rm k} + v_{\rm A}  \mu]^{-1}
\nonumber \\
&=&(7/2-\alpha-k\alpha^\prime\ln (k/k_0)){\cal W}^2 k^7v^{-1}_{\rm A} \ln[1 +
v_{\rm A}/v_{\rm k}],
\label{flux}
\end{eqnarray}
which should be independent
of $k$ in the steady-state case.\footnote{Note that the IK spectrum (${\cal W}\propto k^{-7/2}$) does
approximately satisfy this constraint for $v_{\rm A}\gg v$, and
for $v\gg v_{\rm A}$,
$$
F_0 \simeq (7/2-\alpha-k\alpha^\prime\ln (k/k_0)){\cal W}^{3/2}k^{11/2},
$$ one recovers the Kolmogorov spectrum.}

For a nearly isotropic solution, the constancy of $F_0$ can be used to derive
the weak
dependence of $\alpha$ on $k$:
\begin{equation}
7/2-\alpha(k) -k\alpha^\prime\ln (k/k_0)= F_0v_{\rm A}/[{\cal W}^2 k^7 \ln(1 +
v_{\rm A}/v_{\rm k})]\,
\end{equation}
and the radial energy flux density becomes
\begin{equation}
f={F_0\over 4\pi k^2(v_{\rm k}/v_{\rm A} + \cos\theta) \ln(1 +
v_{\rm A}/v_{\rm k})},
\end{equation}
which deviates from the $k^{-2}$ dependence expected for exactly isotropic
solutions for $M_{\rm A}<1$, especially near the perpendicular directions. So
there is no exactly isotropic solution for this diffusion tensor.\footnote{For
locally isotropic diffusion tensors with their dependence on the wavenumber and
the wave propagation direction separable, one can always find an isotropic
solution for the steady-state power spectrum in the inertial range \citep{C05}.
In this
case, the diffusion equation can be simplified as
\begin{equation}
4\pi k^2 D({\bf k}, k, {\cal W}) {{\rm d} {\cal W}(k) \over {\rm d} k} =
F_0({\bf k}/k)\,,
\end{equation}
where $F_0$ gives the steady-state radial energy flux and is independent of $k$
in the inertial range. Let $F_0({\bf k}/k)$ have the same propagation direction
dependence as $D$, the corresponding formal solution is then given by
\begin{equation}
{\cal W}(k) = \int^{k} {\rm d} k F_0({\bf k}/k)k^{-2}[4\pi D({\bf k}/k, k,
{\cal W})]^{-1}\,.
\end{equation}
} The steady-state solution we get here is just approximately isotropic, and
there is no feature associated with the critical balance proposed by
\citet{GS95}. It is  similar to the steady-state fast mode solution given by
\citet{C05} for coupled weak Alfv\'{e}n-fast mode turbulence.
As expected, for $v_{\rm
k}\gg v_{\rm
A}$, $f =
F_0/4\pi k^2$, the turbulence is isotropic.

On the other hand, in the wave effect dominant regime if we assume an isotropic
power-law spectrum with $\alpha=0$, the Alfv\'{e}n wave crossing process starts
to affect the wave cascade at $k_{||} \propto k^{3/4}$, which is similar to the
critical balance relation described by Goldreich \& Shridhar (1995) and explains
the $k_{||} \propto k^{3/4}$ scaling of the cascading wavevector front in the
time-dependent power spectrum (Fig. \ref{Acon.ps}). This cascading wavevector
front extends until it reaches the numerical boundary or the physical limit
set by damping or non-MHD effects at large $k$. Because our eddy turnover
timescale has different $k$ and ${\cal W}$ dependence's than that given by
Goldreich \& Shridhar (1995), we obtain a different critical balance relation.
For ${\cal W}\propto k_\perp^{-11/3}$ with $v_{\rm k}^2 = {\cal W} k_\perp^3$
and
$\tau_{\rm NL} = (v_{\rm k}k_\perp)^{-1}$, one can recover the Goldreich \&
Shridhar
critical balance relation $k_{||}\propto k_\perp^{2/3}$.

The critical propagation direction, which separates the linear and nonlinear
effect dominant regimes,
satisfies $\cos\theta_c=v_{\rm k}/ v_{\rm A}\propto (k/k_0)^{-1/4}$. Although
the
cascading energy flux is highly
concentrated in the perpendicular directions with small values of $k_{||}$, the
diffusion process eventually  carries wave
energy to large values of $k_{||}$.
The turbulence fills all possible wavevector space in the steady-state.  The
fraction of energy, which becomes linear Alfv\'{e}n waves, is given by the wave
efficiency
\begin{equation}
\eta_{\rm W}=\int_{\cos\theta_c}^1f{\rm d}\mu/\int_{0}^1f{\rm
 d}\mu = 1- \ln2/\ln[1 + v_{\rm A}/v_{\rm k}] \,,
\label{ew}
\end{equation}
which approaches 1 as
$k\rightarrow\infty$. So the wave like component dominates
at small scales as expected from the scalings of the cascade and wave crossing
timescales with $k$ mentioned above.

When $v_{\rm k}\ge v_{\rm A}$, $\eta_{\rm W} \le 0$ and the above formula for
the wave
efficiency is invalid.  The turbulence then  is dominated by nonlinear effects.
 One therefore can define a transition wave
number $k_{\rm tr}$ so that $v(k_{\rm tr}) =[{\cal W}(k_{\rm tr}) k_{\rm
tr}^3]^{1/2}= v_{\rm A}$. Then
\begin{equation}
{\cal W}(k, F_0, v_{\rm A}) \simeq \left({3F_0\over 11}\right)^{2/3}
k_{\rm tr}^{-11/3}
(k/k_{\rm tr})^{-7/2+\alpha}
\simeq v_{\rm A}^{11} \left({11\over 3 F_0}\right)^{3}(k/k_{\rm
tr})^{-7/2+\alpha}\,,
\label{Wapp}
\end{equation}
where
\begin{equation}
\alpha(k) + k\alpha^\prime\ln (k/k_0)\simeq 7/2-(11/3)
(k_{\rm tr}/k)^{2\alpha}/\ln[1+v_{\rm A}/v_{\rm k}]
\end{equation}
and $k_{\rm tr}\simeq 3F_0/11v_{\rm A}^3$. We note that $\alpha\rightarrow 0$
for $k\gg k_{\rm tr}$
or $v_{\rm k}\ll v_{\rm A}$ (IK spectrum), and $\alpha\rightarrow -1/6$ for
$k\ll k_{\rm tr}$ or $v_{\rm k}\gg
v_{\rm A}$ (Kolmogorov spectrum). The transition scale is proportional to the
cube of the Alfv\'{e}n
velocity and inversely proportional to the energy injection rate. It gives the
coherent length of the magnetic field (Cho \& Vishniac 2000) and may determine
the maximum energy that charged particles can reach through resonant
interactions with
the turbulent magnetic field (Liu et al. 2008).
The top right panel
of Figure \ref{MHD_ani.ps} compares the numerical results with the IK spectrum,
${\cal W}(k, F_0, v_{\rm A}) \simeq
 0.22 v_{\rm A}^{11} \left({11/3 F_0}\right)^{3}(k/k_{tr})^{-7/2}$ (thin solid
line).
The deviations of the spectra in different wave propagation directions from the
IK spectrum at  $k<10^{-6}\Omega_p/ v_{\rm A}$ shown in Figure \ref{MHD_ani.ps}
are likely caused by the cylinder geometry of the surface with
$k_{||0}=k_{\perp0}=k_0= 1.6\times10^{-8}\Omega_p/v_{\rm A}$, where turbulence
energy is injected into the system. For the parameters of this numerical
calculation, $k_{\rm tr} = 2.60\times 10^{-10} \Omega_p/v_{\rm A}$, and equation
(\ref{Wapp}) with $\alpha=0$ predicts an $M_{\rm A} = (k_{\rm
tr}/k_0)^{1/4}\simeq
0.36$ at $k_0=1.6\times 10^{-8}\Omega_p/v_{\rm A}$, which is in agreement with
the numerical results.

\subsection{Results from Anisotropic Diffusion Tensor}

For the {\it  locally} anisotropic diffusion tensor (\ref{Dij_ani.eq})  the
diffusion
coefficient in the perpendicular direction is higher than that of the isotropic
diffusion model. Thus, for a given energy injection rate,  in the
steady-state case, the turbulence contains slightly less energy than that of the
isotropic model. We find  that for $M_{\rm A} = 0.23$,  the steady-state power
spectral contours (Fig.
\ref{MHD_ani.ps}, bottom panels) can be fitted with
\begin{equation}
{\cal W}({\bf k}) \simeq 0.62 \left({3F_0\over 11}\right)^{2/3}k_{\rm
tr}^{-11/3}\left[\left({k_{||}\over k_{||, tr}}\right)^{\alpha_{||}} +
\left({k_{\perp}\over k_{\perp, tr}}\right)^{\alpha_{\perp}}\right]^{-\alpha}
\label{specfit}
\end{equation}
with $\alpha = 1$, $\alpha_{\perp} = 11/3$, $\alpha_{||} = 4$, $k_{||, tr} =
0.5k_{\perp, tr}$, and $k_{tr}=(k_{||, tr}^2+k_{\perp, tr}^2)^{1/2}\simeq
3F_0/11v_{\rm
A}^3$.
The bottom right panel of Figure \ref{MHD_ani.ps} compares this analytical fit
(thin lines)
 to the numerical results (thick lines).  The significant differences between
the
fit and numerical results at low values of $k$ may also be related to the
assumed cylindrical
geometry of the injection surface. As evident for $\theta\sim \pi /2$ we have
a spectrum  very similar to the  Kolmogorov spectrum, but away from the
transverse
direction the spectrum
at high $k$'s becomes $\propto k^{-4}$ very quickly.
In the limit of $k\gg
k_{\rm tr}$, the angle averaged spectrum $W(k) \equiv \int
{\cal
W}({\bf k})k^2 {\rm d} \Omega \propto
k^{1-\alpha\alpha_\perp+\alpha_\perp/\alpha_{||}}=k^{-7/4}$.
These results are similar to Chandran's steady-state solutions, but there are
significant quantitative differences. The corresponding critical balance $v_{\rm
A} k_{||}= v_{\rm k} k$ leads to
\begin{equation}
k_{||}=0.79 k_{\rm tr}^{1/3} k_\perp^{-11/6}
k^{5/2}[1+2^{11/3}(k_{||}/k_\perp)^4 (k_\perp/k_{||, tr})^{1/3}]^{-1/2}\simeq
0.79
k_{\rm tr}^{1/3} k_\perp^{2/3}.
\end{equation}
This  is almost identical to the
critical balance
relation proposed by \citet{GS95}.\footnote{The critical balance
only exists in the regime of $v_{\rm k}\ll v_{\rm A}$, which implies $k_{||}\ll
k\simeq
k_\perp$.}  The
spectrum of the nonlinear effect dominant 2D component $W_{2D}(k_\perp) \equiv
2\pi k_\perp\int_{v_{\rm A} k_{||} < v k} {\cal W}({\bf k}) {\rm d}
k_{||} \propto
k_\perp^{-2},$ which is different from the result of \citet{GS95}.
As for the isotropic case, this is due to the difference in the scaling
of the eddy turnover timescale with ${\bf k}$ and ${\cal W}$. In our model
$v_{\rm k}^2\propto {\cal W}k^3$, while $v_{\rm k}^2\propto {\cal W}
k_{||}k_{\perp}^2$ in
Goldreich \& Sridhar model. The fact that the spectrum of the 2D component is
softer than the overall turbulence spectrum  is also consistent with the
assessment that the wave component dominates at small spatial scales. The energy
flux  of the 2D component $F_{2D}(k_\perp) \equiv 2\pi k_\perp\int_{v_{\rm A}
k_{||} < v k}D_{\perp,\perp}[\partial {\cal W}({\bf k})/\partial k_\perp] {\rm
d} k_{||}\propto k_\perp^{-1/3}$, which decreases with $k_\perp$, and as
the turbulence cascades towards small 
spatial scales, there is continuous energy conversion from
the 2D perpendicular component to the wave component.

On the other hand, based on the anisotropy of the steady-state spectrum, one may
define a perpendicular component of the turbulence with $k_\perp/k_{\perp,tr}
>k_{||}/k_{||,tr}$, the corresponding spectrum $W_\perp(k_\perp) \equiv 2\pi
k_\perp \int_{k_\perp k_{||, tr} > k_{||}k_{\perp, tr}}{\cal W}({\bf k}){\rm d}
k_{||}\propto k_\perp^{-5/3}.$ Due to the anisotropy of the diffusion tensor,
this component develops quickly as can be seen from the time-dependent results
(Fig. \ref{Acon.ps}). The perpendicular energy flux $F_\perp (k_\perp)  \equiv
2\pi k_\perp \int_{k_\perp k_{||, tr} > k_{||}k_{\perp,
tr}}D_{\perp,\perp}[\partial {\cal W}({\bf k})/\partial k_\perp] {\rm d}
k_{||}$ $\propto k_\perp^{-1/3}$, which also carries most of the energy flux from
large scales to small scales. The corresponding parallel component with
$k_\perp/k_{\perp,tr} <k_{||}/k_{||,tr}$ is dominated by linear waves and has a
much steeper spectrum $W_{||}(k_{||})\equiv  \int_{k_\perp k_{||, tr} <
k_{||}k_{\perp, tr}}2\pi k_\perp{\cal W}({\bf k}){\rm d} k_{\perp}\propto
k_{||}^{-2}$. 
The parallel energy flux $F_{||} (k_{||})
\equiv  \int_{k_\perp k_{||, tr} < k_{||}k_{\perp, tr}}4\pi k_\perp
D_{||,||}[\partial {\cal W}({\bf k})/\partial k_{||}] {\rm d} k_{\perp}\propto
k_{||}^{-1}$. Since both the perpendicular and parallel energy fluxes decrease
with the increase of $k$, the total energy is not conserved, which reflects the
inaccuracy of the analytical fit to the numerical results (Fig.
\ref{MHD_ani.ps}). The energy is better conserved with a slightly higher value
of $\alpha_\perp$. Instead of pursuing even better analytical fits to the
steady-state spectrum, which most likely will be complicated, we focus on
numerical
results in what follows.

The most notable difference between these results and recent theoretical
and numerical studies of Aflv\'{e}n turbulence is the presence of a prominent
parallel wave component in our model.  Such a component is found to be
absent or dynamical unimportant in many studies (Galtier 2006; Cho et
al. 2002; Goldreich \& Sridhar 1995). On the other hand, these parallel
wave modes may play a dominant role in the selective acceleration of
ions from the background plasma (Mason et al. 2002; Liu et al. 2004,
2006) and scattering of high energy particles. They have been the
subject of extensive studies (Yan \& Lazarian 
2002; Farmer \& Goldreich 2004; Lazarian \& Beresnyak 2006). Since the 
diffusion process tends to fill the wave vector space with energy, the
presence of a strong parallel wave component is not a surprise in our
model. 
They are
also not subject to the turbulence damping studied by Farmer \&
Goldreich (2004) and Lazarian \& Beresnyak (2006). This damping is
caused by the interactions of the waves with the quasi-2D strong
turbulence component in the perpendicular direction, which makes the
waves cascade toward small perpendicular scales. Our parallel wave component
is produced by the cascade (or reverse cascade) itself. So the cascade
process generates these waves instead of damping them. With
the strong turbulence theory proposed by Goldreich \& Sridhar (1995),
high values of $k_{||}$ can be reached through the critically balanced
cascade $k_{||}\propto k_\perp^{2/3}$. The turbulence power spectrum is
given by $W\propto k_\perp^{-10/3} g(k_{||}/k_\perp^{2/3})$. If the
turbulence can cascade along lines of constant values of $k_\perp$ as
required by the three-wave resonance conditions, one would expect that
at a given value of $k_{||}$, the spectrum is flat with the increase of
$k_\perp$ until the critical balance condition is satisfied.
Therefore $W\propto k_\perp^{-10/3}\propto k_{||}^{-5}$. 
This result is similar to the result of our anisotropic
cascade model. However, they are quite different from the power spectrum
derived from MHD simulations by Cho et al. (2002), where it is found
that the spectra cut off exponentially toward small parallel scales. It
remains to be seen what prevents the presence of the parallel wave
component 
in these simulations. 

\section{Turbulence Cascade beyond the MHD Regime}
\label{disp}

So far we have assumed the simple MHD dispersion relation $\omega=kv_{\rm A}\cos
\theta$,
which is valid on large scales. For a more exact treatment of both cascade and
damping
we need to go beyond the MHD regime and use a more complete description of
the
dispersion relation, which is well known.
Linear waves in collisionless magnetized plasmas have been studied extensively
with the linear Vlasov theory (Andr\'{e} 1985). Here, for the purpose of
comparing with the observed  turbulence in the  solar wind, we focus on the
dominant
Alfv\'{e}n-cyclotron fluctuations.
Observations suggest that the Alfv\'{e}n turbulence can cascade to spatial
scales below the MHD regime and the kinetic effects are important \citep{D83,
L98, L99, B05, OH07}. For fully ionized plasmas
with the solar abundance,
the dispersion relation deviates from the simple
MHD relation  significantly near the
$^4$He nucleus (or $\alpha$-particle) cyclotron frequency
$\Omega_\alpha = 0.5\Omega_p=0.076 (2\pi) (B/10 {\rm nT})$Hz
(the Alfv\'{e}n-cyclotron fluctuations exist only below this frequency).

\begin{figure}[htb]
\begin{center}
\includegraphics[width=0.5\textwidth]{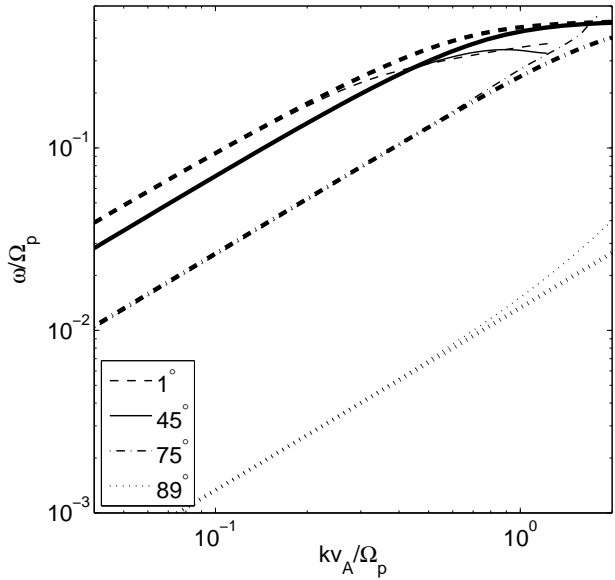}
\end{center}
\caption{
Dispersion relations (the real part of the wave frequency) of
Alfv\'{e}n-cyclotron
fluctuations in several wave propagation directions. The thick lines give
the cold plasma relations and the thin lines are
obtained
numerically with the WHAMP code. We assume  $\Omega_p = 0.096(2\pi)$ Hz ($B =
6.3$nT), the electron and ions temperatures $k_{\rm B} T_e = k_{\rm B}T_p
=k_{\rm B} T_\alpha= 3.35$ eV, and densities $n_e = 12$ cm$^{-3}$ ($\beta_{\rm
p}=0.41$),
$n_p = 10$ cm$^{-3}$, $n_\alpha =1$ cm$^{-3}$, which are typical for the solar
wind (Leamon et al. 1998, 1999).
Note that   the Lauren series are used in the code to approximate the
plasma
dispersion function Z, that is  valid when the damping rate, i.e., the imaginary
part of the wave frequency $\omega_i=\Gamma$, is much less than the real wave
frequency $\omega_r$. Thus,
the solution at high frequencies may not be valid. The missing segments of the
dispersion relation in the figure are related to this caveat.
}
\label{disps.ps}
\end{figure}

Our diffusion tensor is constructed from the wave dispersion relation
$\omega_r({\bf k})$, where the subscript ``$r$'' refers to the real part of the
frequency.  (As discussed in \S2.2 and  below, the imaginary part determines the damping
rate.). For an accurate determination of $\omega_r$ we use the  WHAMP code to
obtain the ``hot-plasma''  Alfv\'{e}n wave
dispersion surface (R\"{o}nnmark 1982). The thin lines in Figure
\ref{disps.ps} show this relation for different angles of propagation under
typical solar wind
conditions (temperature $T\sim 4\times 10^4$ K). These match very closely the
``cold plasma'' ($T=0$) dispersion relation, shown  by the thick lines (see
Appendix A), up to $k\sim 0.6\Omega_p/v_{\rm  A}$ with less than 30\% deviation
at highest frequencies.\footnote{Note that the hot plasma dispersion curves
terminate at some finite values
of $k$. This is because currently there are
 no simple schemes, which can lead to reliable dispersion relation (and
damping rate) in the high frequency range, where damping dominates the wave
effect. At such high frequencies thermal damping
dominates and, as we will see below, the turbulence spectrum cuts off sharply.}
At a higher (lower) values of temperaure the deviation will start at a lower (higher) value of $k$.
In what follows we use the cold plasma dispersion relation without loss of much
accuracy but with a great gain in computational simplification because these
dispersion relations can be described analytically. This greatly simplifies the
construction of the diffusion tensor at all the wavevector grid points.

For Alfv\'{e}n waves,
the turbulence cascades preferentially in the perpendicular direction. This
is still true when we consider the kinetic effects on the plasma wave dispersion
relations. However, as mentioned above in the MHD regime the  Alfv\'{e}n waves
have a well defined wave crossing time. The kinetic effects introduce
ambiguities on the definition of $\tau_{\rm W}$. One therefore can identify
$\tau^{-1}_{\rm
W}$ with either $\omega_r({\bf k})$ or ${\bf v}_{\rm gr} {\bf \cdot k}$.
Because the Alfv\'{e}n wave packet becomes stationary near $\Omega_\alpha$
(where $k\rightarrow \infty$), one may  favor the latter choice that implies
negligible wave propagation effects.
It is not obvious which one of these two choices is physically more reasonable.
We first try both forms still without the damping (or leakage) term.
The corresponding steady-state spectra are shown in Figure \ref{ND.ps}. The
spectra are nearly identical at large scales (the spectra for $\tau_{\rm W}^{-1}
= \omega$ are shifted upward by a factor of 5 for the illustrative
purpose) and, as expected,  both contain
a spectral break at $k_{||} v_{\rm A} \sim \Omega_{\alpha}$, where the wave
dispersion surface starts to deviate from the MHD relation.
These breaks, though appealing, only have a spectral index change of less than
one. They may explain a few observations of solar wind turbulence \citep{B05}.  However, neither of these breaks can account for the more commonly observed steep spectra at high frequencies \citep{D83, L98, CV03, OH07}. In the
kinetic effect dominant regime, the wave propagation effects vanish for
$\tau^{-1}_{\rm W} = {\bf v}_{\rm gr}{\bf
\cdot k}$, the spectra approach, once again,  the Kolmogorov spectrum at high
values of $k$.

We will show below (\S\ \ref{result}) these two
cases cannot be distinguished
with most observations of the solar wind turbulence since the strong thermal damping
cuts off the spectra at the wavevectors before the spectral deviations due to
the kinetic effects shown above set in. [For other conditions, e.g., with lower
temperatures, the
break due to kinetic effects may appear before the damping cuts off the
spectra, see e.g., \citet{B05}.]  The kinetic
effects, however, play crucial
roles in the thermal damping processes.
In what follows, we leave the subtle differences between these two possible
choices of the wave coupling times to further study, and adopt the first
choice, i.e., $\tau_{\rm W}=\omega^{-1}$ obtained from the
exact cold plasma dispersion relation for both the isotropic and anisotropic
diffusion tensors [eqs.
(\ref{DijMHD}) and (\ref{Dij_ani.eq})].

\begin{figure}[htb]
\epsscale{0.6}
\centerline{\plotone{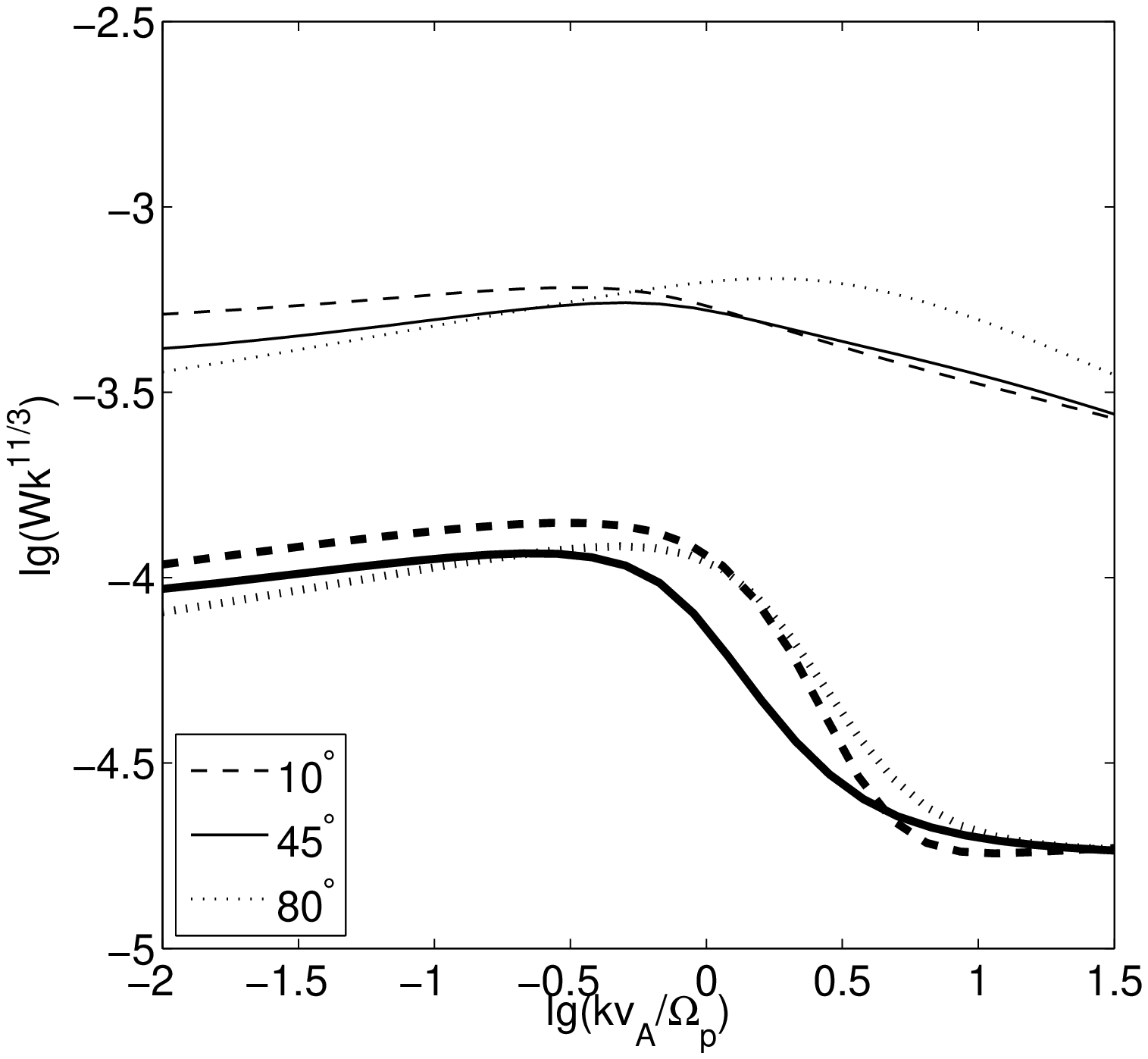}}
\caption{The steady-state wave power spectra in several wave propagation
directions. The energy is injected at $k_{||0} = k_{\perp0} = 2.01 \times
10^{-4}
\Omega_p/v_{\rm A}$ with ${\cal W}({\bf k_0})=1.44\times 10^9 v_{\rm
A}^5\Omega_p^{-3}$. The corresponding Mach number $M_{\rm A} = 0.18$.
The upper and lower lines are for $\tau_{\rm W}^{-1} = \omega_r({\bf k})$ and
 $ {\bf k}\cdot
{\bf v}_{\rm gr}$, respectively.
To avoid confusion the former has been
shifted upward by a factor of 5.
The spectral breaks at $k_{||} v_{\rm A} \simeq 0.5 \Omega_p$ in both cases are
due to
the kinetic effects on the wave dispersion relation.
}
\label{ND.ps}
\end{figure}

\section{Turbulence Spectra in the Dissipation Range}
\label{result}

It is well-known that the damping of waves in a magnetized plasma by thermal
background particles is not isotropic \citep{GB04}. In combination with the
anisotropy of the
turbulence cascade, we would expect strong anisotropies in the turbulence
spectrum in the dissipation range. These anisotropies have significant effects
on the energizing of background particles by turbulent plasma waves and have
broad implications on the energy dissipation in collision-less astrophysical
plasmas.
Although these issues have been well-recognized, there are only preliminary and
approximate investigations of the relevant processes \citep{L99, CV03, H07,
H08a}. Compared with previous
studies, our model has less assumptions with most of the related physical
processes treated more self-consistently. We present the corresponding results
in this section.

\subsection{Start of the Dissipation}

As mentioned above the damping rate can be calculated from the WHAMP code for a
thermal plasma with $\Gamma({\bf k})=\omega_i({\bf k})$, the imaginary part of
the wave frequency. To our knowledge  there are
no equations or programs for damping by an arbitrary nonthermal particle
distributions
except the recent result from \citet{P06} for the TTD damping of fast
mode waves in low beta plasmas.  As shown in this paper the damping rate
increases with increasing $k$ more rapidly than the cascade rate so that it
becomes important beyond some critical wavevector ${\bf k}_c$ obtained from
$\Gamma({\bf k}_c)=\tau_{\rm cas}^{-1}({\bf k}_c)$.  And one expects sharp
cutoff of the turbulence
spectrum for $k>|{\bf k}_c|$ at a given wave propagation direction, which
corresponds to the dissipation range. In what
follows we first follow this procedure assuming a thermal plasma to estimate the
dissipation range turbulence spectrum and then show
exact numerical results from the solution of the wave equation including both
the diffusion and damping terms (with the leakage term still ignored).

For the locally isotropic diffusion model, the cascade rate is uniquely defined
as
$\tau^{-1}_{\rm cas}=\tau_{NL}^{-1}/(1+\tau_{NL}/\tau_{\rm W})$. The
left panel of  Figure \ref{kc_contour.ps} shows this cascade rate (from the
above numerical results with damping excluded) and the thermal damping rate (for
the plasma in Figure \ref{disps.ps}) in several wave propagation directions. We
see that with the increase of $k$ the damping rate increases much faster
than the cascade rate. The cutoff wave
number
$k_c$ is given by the intersection of  these two rates. The solid line in the
middle panel shows the dependence of $k_c$ on $\theta$. The dotted and dashed
lines
give  contours of constant cascade and damping rates, respectively. These
anisotropic
damping and cascade make $k_c$ change by more than one order of magnitude with
the
change of the wave propagation direction. This is distinct from the 1D
turbulence model, where the dissipation range covers a very narrow spatial
scales with the Reynolds number on the order of 1. The right panel shows the
dependence of $k_c$ on
$\beta_{\rm p}$ and the turbulence intensity. Higher values of $\beta_{\rm p}$
lead to stronger damping and lower values of $k_c$. With the increase of the
turbulence intensity, the dissipation range shifts to smaller spatial scales.
For the model with the anisotropic diffusion tensor, the cascade rate in
the parallel direction is different from that in the perpendicular direction.
The model
expectations are also more involved. However, numerical evaluations of both
models are  straightforward.

\begin{figure}[htb]
\begin{center}
\includegraphics[width=0.28\textwidth]{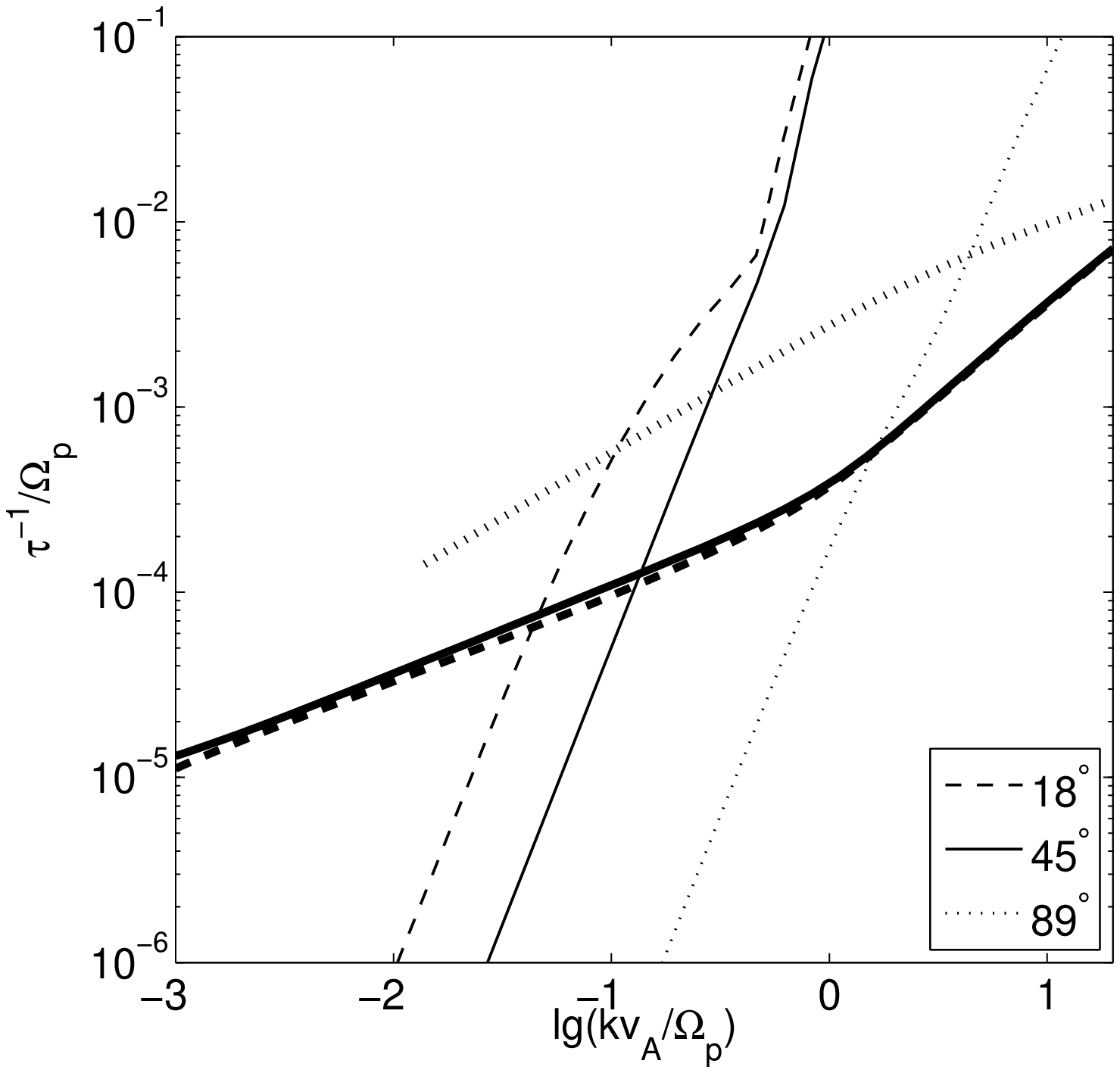}
\includegraphics[width=0.28\textwidth]{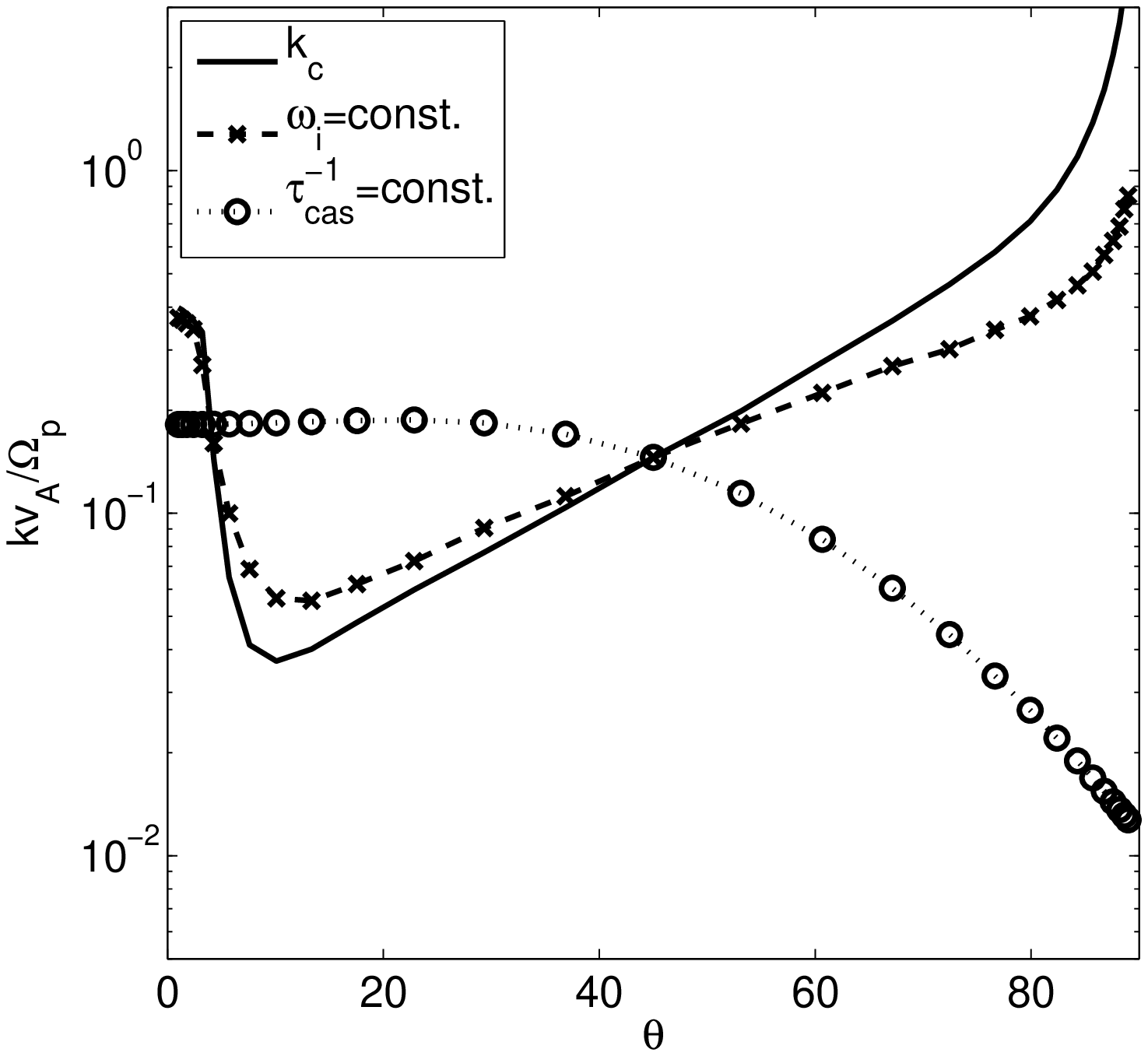}
\includegraphics[width=0.28\textwidth]{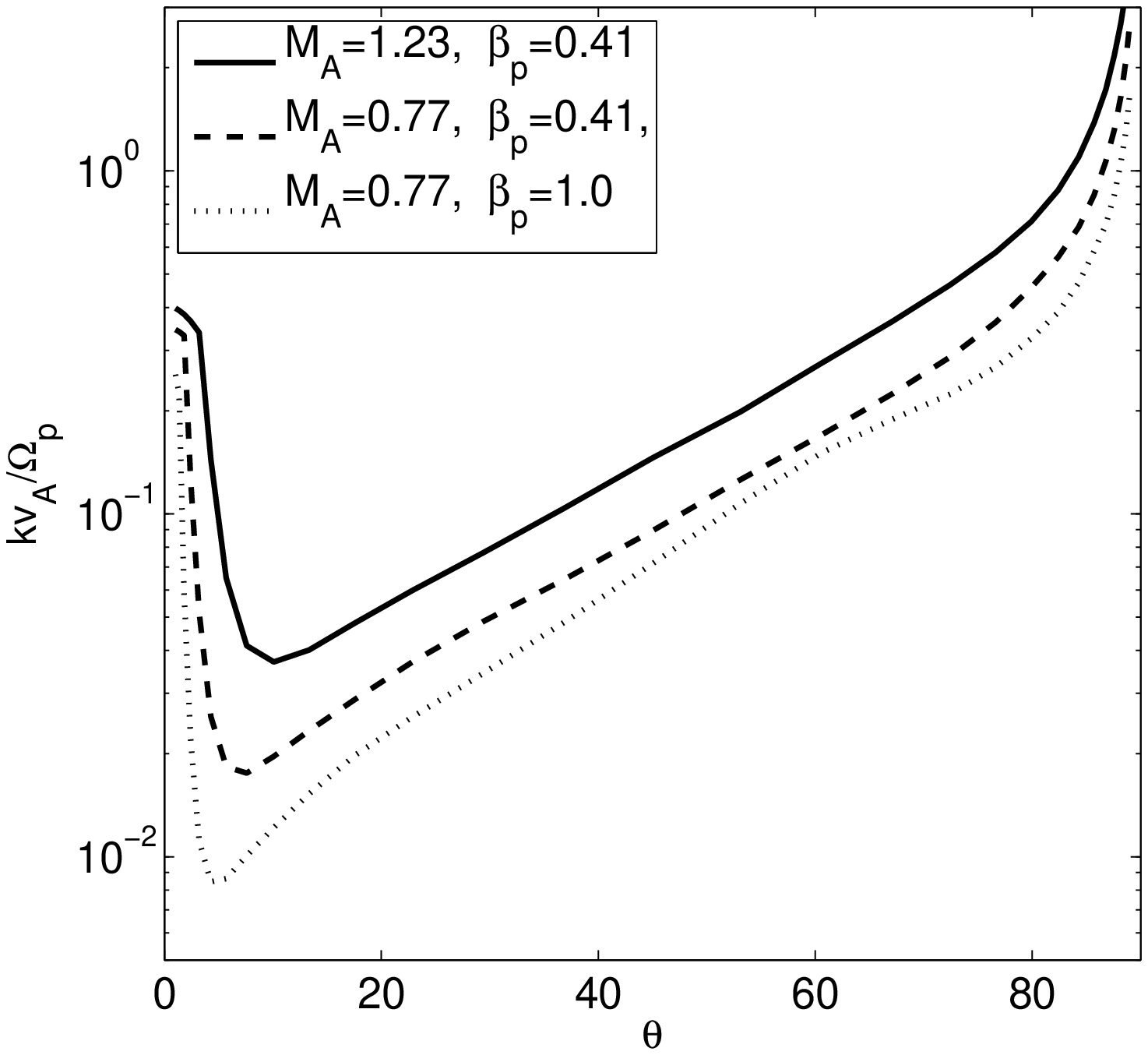}
\end{center}
\caption{
{\it Left}:
The wave cascade rates obtained without damping (thick) and damping rates
($\omega_i$, thin) as
functions of $k$ for three wave propagation directions. The properties of
the plasma are the same as in Figure \ref{disps.ps}. The energy is injected at
$k_{||0}=k_{\perp0}=2.01\times 10^{-4}\Omega_p/v_{\rm A}$ with $F_0=1.89\times
10^{-3}v_{\rm A}^2\Omega_p$, which implies ${\cal W}({\bf k_0}) = 6.62 \times
10^{10} v_{\rm
A}^5\Omega_p^{-3}$. The corresponding $M_{\rm A}=1.23$.
{\it Middle}:
The dotted and dashed lines show a sample contour of the above cascade
and damping rates in the wavevector space ($k, \theta$), respectively, for $\tau^{-1}=10^{-4}\Omega_p$.
The critical wavenumber $k_c$ defined as the location where these two rates
are equal is indicated by the solid line. At a given $\theta$, $\omega_i$
increases monotonically with $k$. The turbulence spectrum should cut off at
$k_c$,
which varies by more than one order of magnitude with the change of  the wave
propagation direction.
The nearly parallel propagating waves are damped through cyclotron resonances at
$\omega\simeq\Omega_\alpha$. Obliquely propagating waves are subject to the TTD
and LD. The nonlinear effects dominate in the nearly perpendicular directions.
{\it Right}: The dependence of $k_c$ on $\beta_{\rm p}$ and $M_{\rm A}$. Solid,
dashed, and dotted lines show the cases with ($M_{\rm A}=1.23$, $\beta_{\rm
p}=0.41$), ($M_{\rm A}=0.77$, $\beta_{\rm p}=0.41$), and
($M_{\rm A}=0.77$, $\beta_{\rm p}=1.0$), respectively. With the increase of the
turbulence intensity, the cascade rate increases leading to higher values of
$k_c$. The damping rate increases with $\beta_{\rm p}$ so that a higher
$\beta_{\rm p}$ leads to lower values of $k_c$.
}
\label{kc_contour.ps}
\end{figure}

\subsection{More Exact Spectra}

We now present some numerical results on the spectrum and anisotropy of the
turbulence in the dissipation range where we include the thermal damping term using the WHAMP code as  described in \S\
\ref{damp}. The left and middle
panels of Figure
\ref{taus_cut.ps} show the steady-state turbulence power spectrum for the
isotropic
diffusion model. For the plasma parameters chosen here the break due to damping
comes before the spectral break associated with the kinetic effects shown in the
previous section.
We therefore do not expect spectral features caused by the kinetic effects on the
turbulence cascade \citep{S01}. The spectra cut off sharply when the damping
dominates, which is consistent
with the results of the previous 1D models (Miller et al. 1995, 1996; Li et al.
2001;
Stawicki et al. 2001). This is mainly due to the non-linear nature of the
diffusion
equation. Once the damping rate is high enough to make the turbulence spectrum
deviates
significantly from the inertial range spectrum, the cascade, whose rate
decreases with
the decrease of the local turbulence power in the wavevector space, is
suppressed,
driving the cascade and damping processes out of balance quickly with the
increase of
$k$ and resulting in a sharp spectral cutoff. Therefore, it is very difficult to
produce a broken-power law spectrum or even a gradual cutoff through the thermal
damping
with the 1D model except in some unusual scenarios, where the
damping
processes themselves are also nonlinear and the damping rate scales the same way
with
$k$ and ${\cal W}$ as the cascade rate \citep{M06}. We also don't see the
gradual spectral cutoff due to the gradual onset of damping with the increase of
$k$ as suggested by \citet{GB04}. Such more gradual cutoffs may appear with
different properties of the background plasma \citep{H08a}.

\begin{figure}[htb]
\begin{center}
\includegraphics[width=0.3\textwidth]{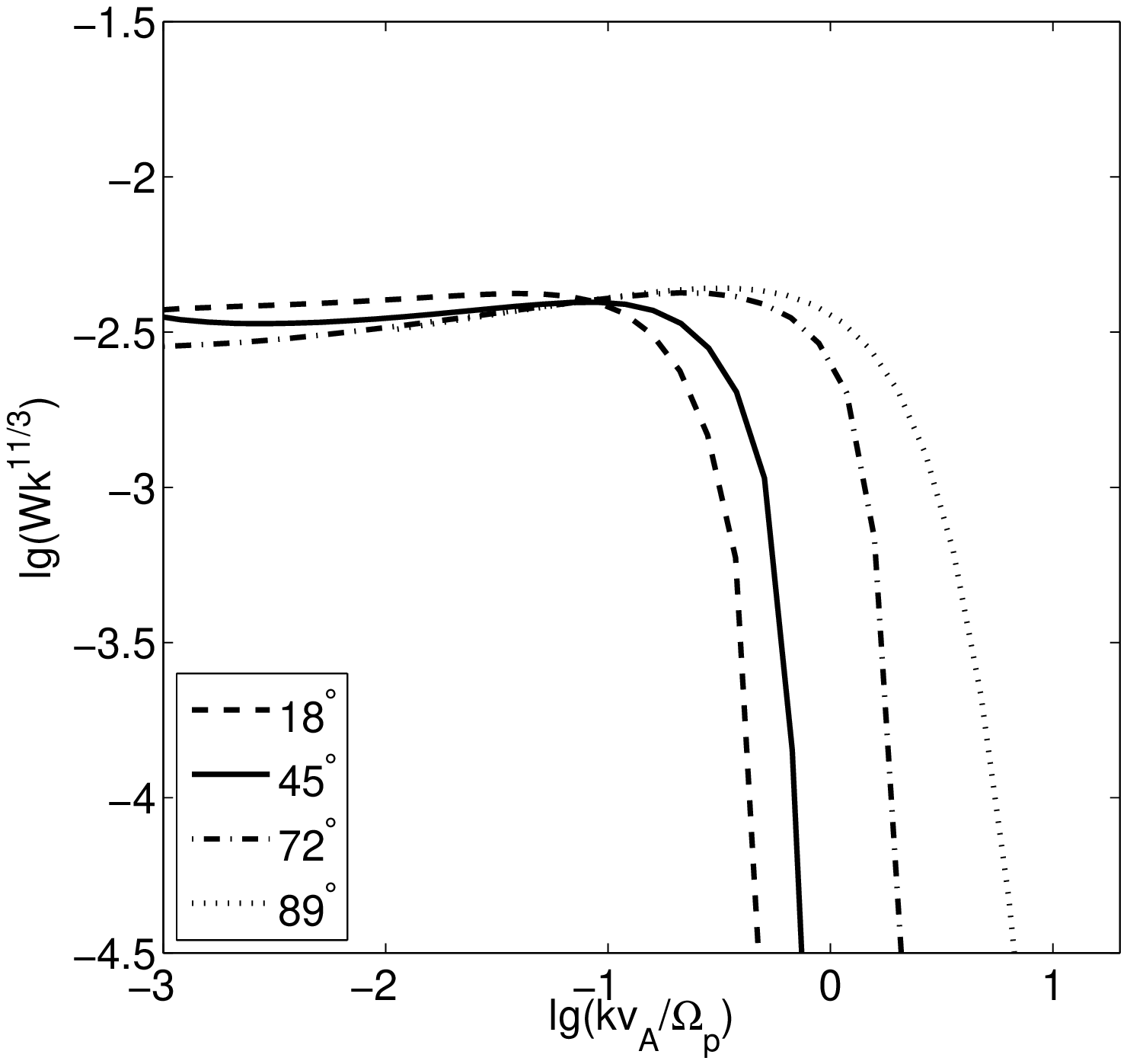}
\includegraphics[width=0.3\textwidth]{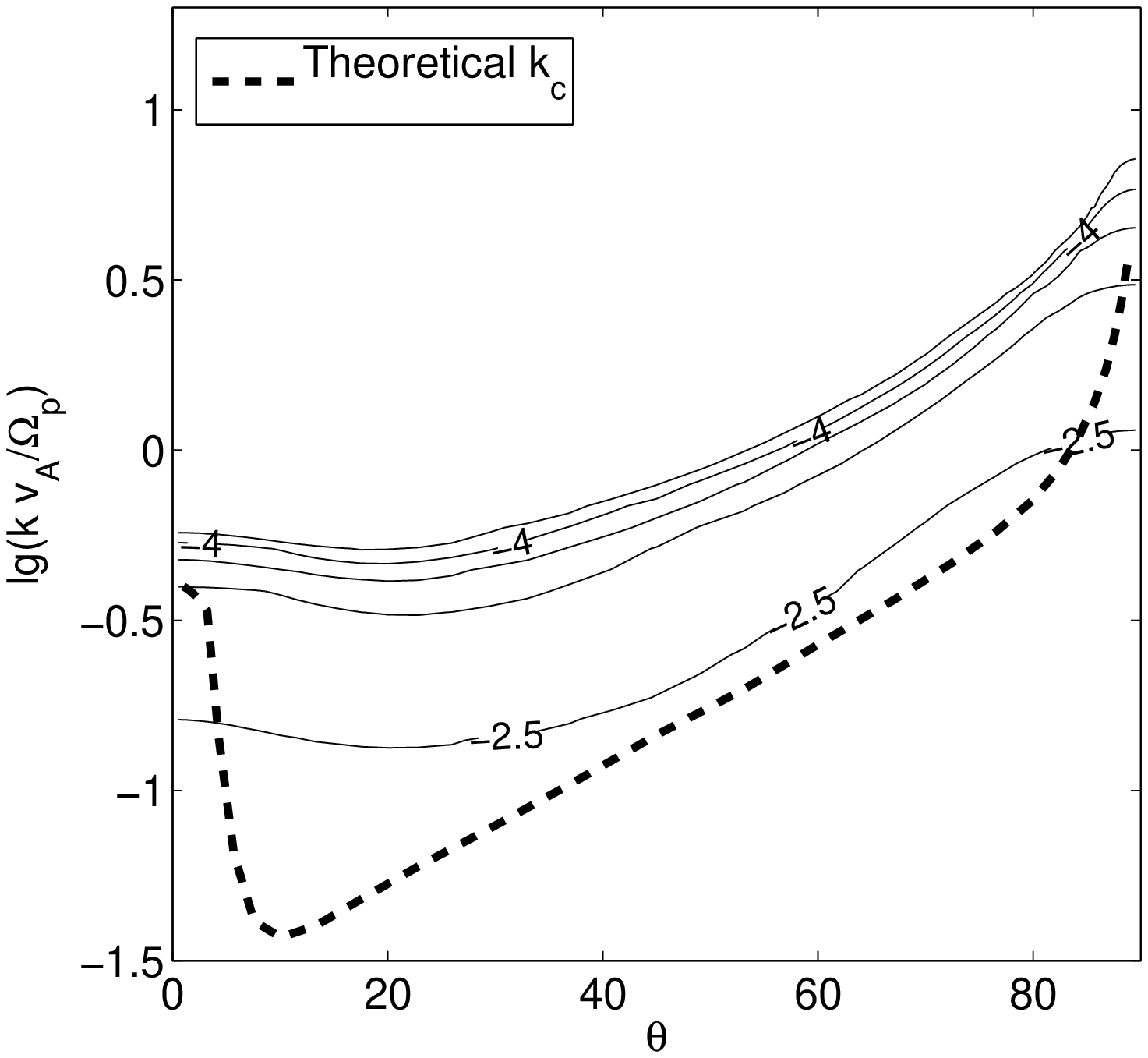}
\includegraphics[width=0.3\textwidth]{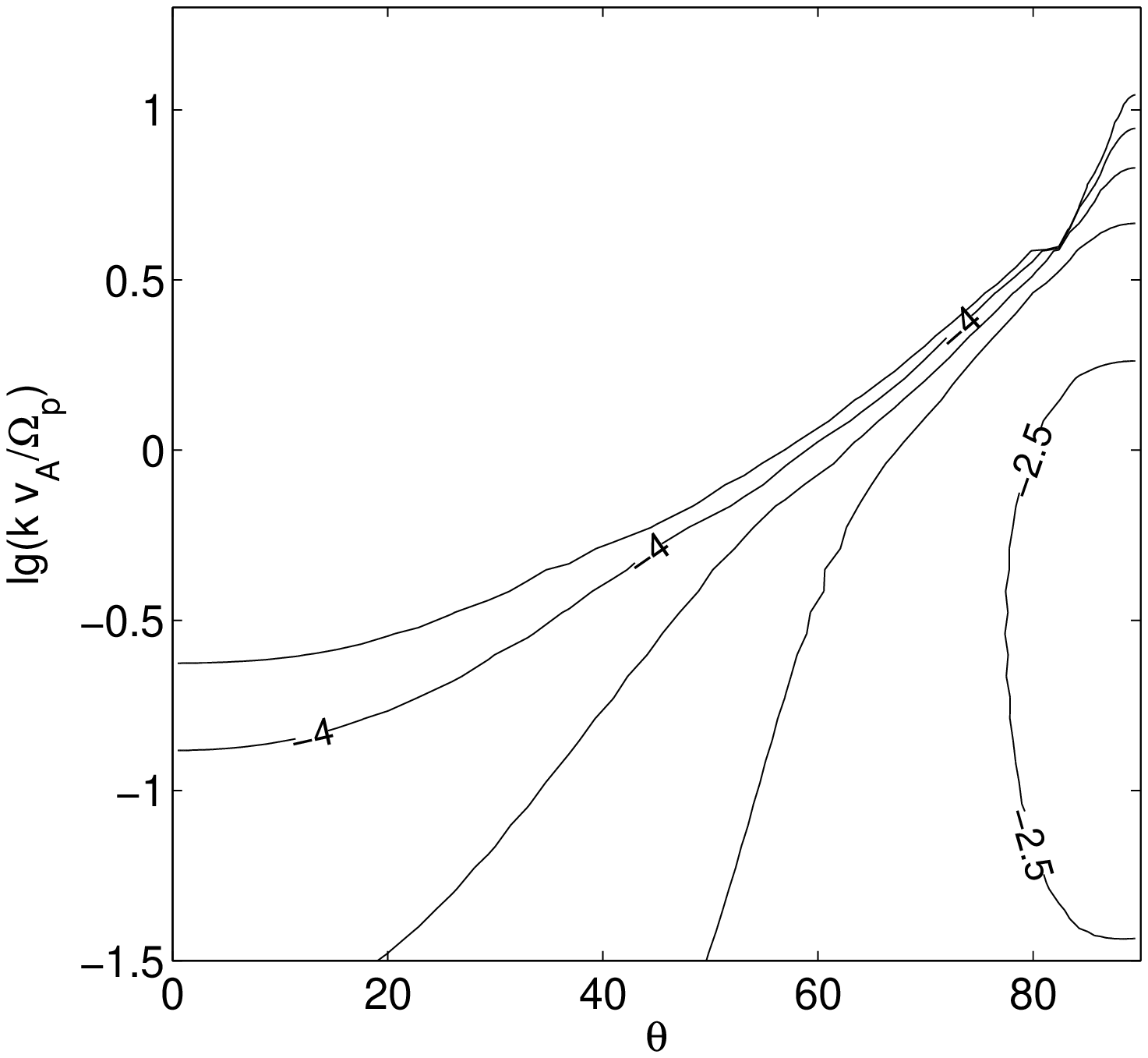}
\end{center}
\caption{
{\it Left}:
Same as Figure \ref{ND.ps} with the thermal damping included. The turbulence
injection
and properties of the background plasma are the same as that in the {\it left}
panel of Figure \ref{kc_contour.ps}. As expected, the power spectrum cuts
off sharply at high values of $k$, and the location of the cutoff wavenumber
$k_c$ is very sensitive to the wave propagation direction.
{\it Middle:}
The corresponding  contours of $\log_{10}(Wk^{11/3})$ (solid lines,
starting at -2.5 and decreasing in steps of
0.5)
of the {\it Left}
panel in $k-\theta$ plane. The spectrum breaks near $-2.5$. The dashed line indicates $k_c$ obtained from the
prescription given in Figure \ref{kc_contour.ps}. The  diffusion process
smoothes
out the valley near $\theta=10^\circ$ in $k_c(\theta)$.
{\it Right}:
Same as the {\it Middle} panel but for the anisotropic diffusion tensor. The
turbulence is
dominated by the perpendicular component, which is damped over a narrow
wavenumber
range.
}
\label{taus_cut.ps}
\end{figure}

As expected, the actual location of the cutoff wavevector depends on the wave
propagation
angle. As a consequence of weaker damping and stronger cascade in the
perpendicular directions, the perpendicular wave spectra cut off at scales
(wavenumbers) about one order of magnitude
smaller (larger) than that of the parallel waves. Using the equality of the
cascade and damping rates as the criterion for spectral break one would expect
strongest damping for
waves with $\theta\sim 10^\circ$; the minimum of  $k_c$ vs. $\theta$ plot
 (in middle panel of Fig. 5).
However, the valley here is smoothed out due
to the nature of the diffusion processes. Waves with higher and lower values of
$\theta$ can
cascade to smaller scales and induce waves with $\theta\sim 10^\circ$ through
the
diffusion processes. The isotropy of the diffusion tensor makes the
spectrum of waves subjected to heavy damping cut off at a value higher than
$k_c$.
Note also that the cascade rate in Figure \ref{kc_contour.ps} is
evaluated without the damping. The actual cutoff wavenumber should be lower
than $k_c$ in general.
The right panel shows the spectral contours for the anisotropic
diffusion tensor. The parallel component is damped  even more quickly
because the turbulence spectrum is
dominated by the perpendicular component whose damping sets in within a
relatively
narrow wavenumber range. The turbulence is therefore strongly anisotropic in
the
dissipation range.

\begin{figure}[htb]
\begin{center}
 \includegraphics[width=0.24\textwidth]{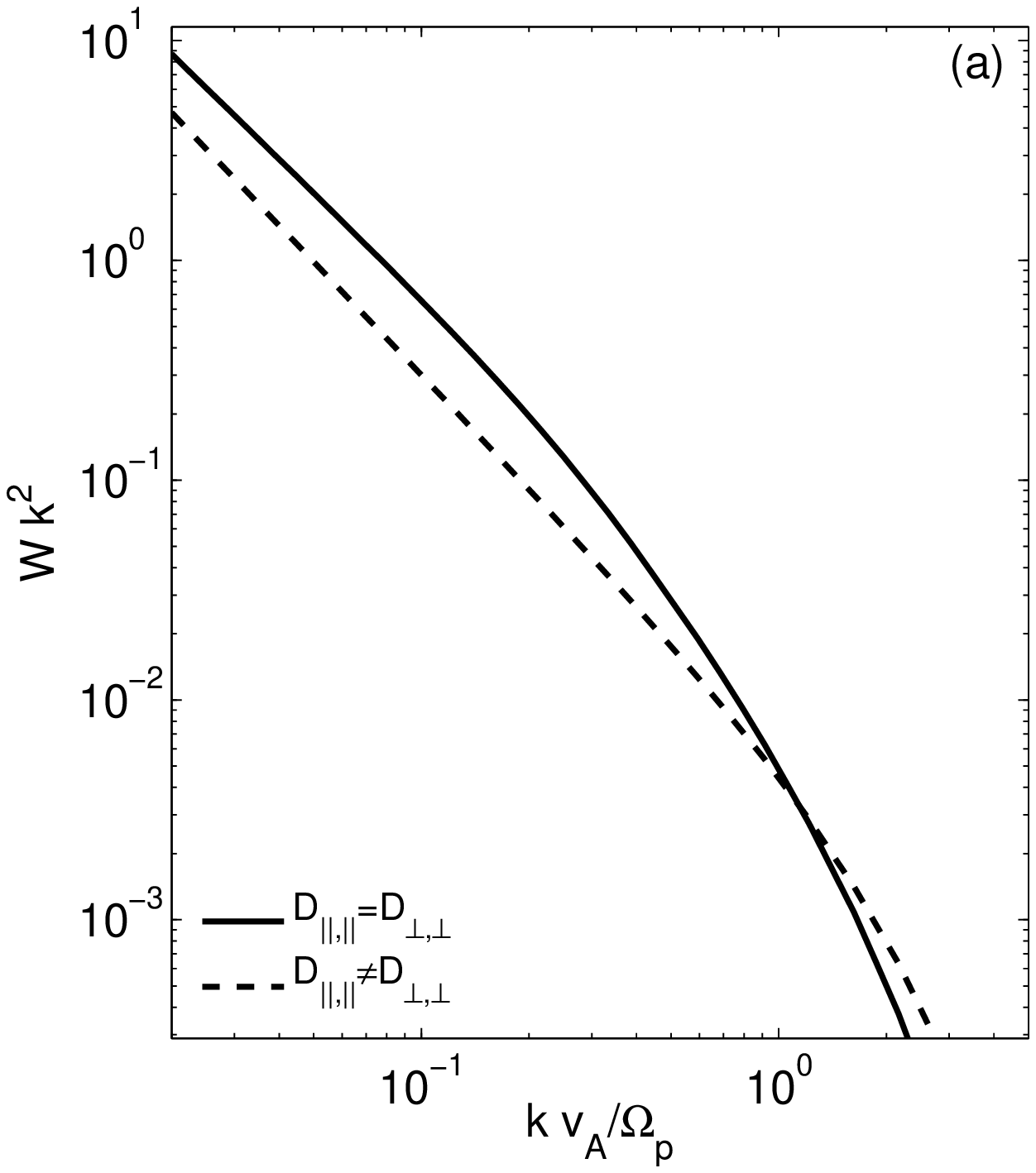}
 \includegraphics[width=0.24\textwidth]{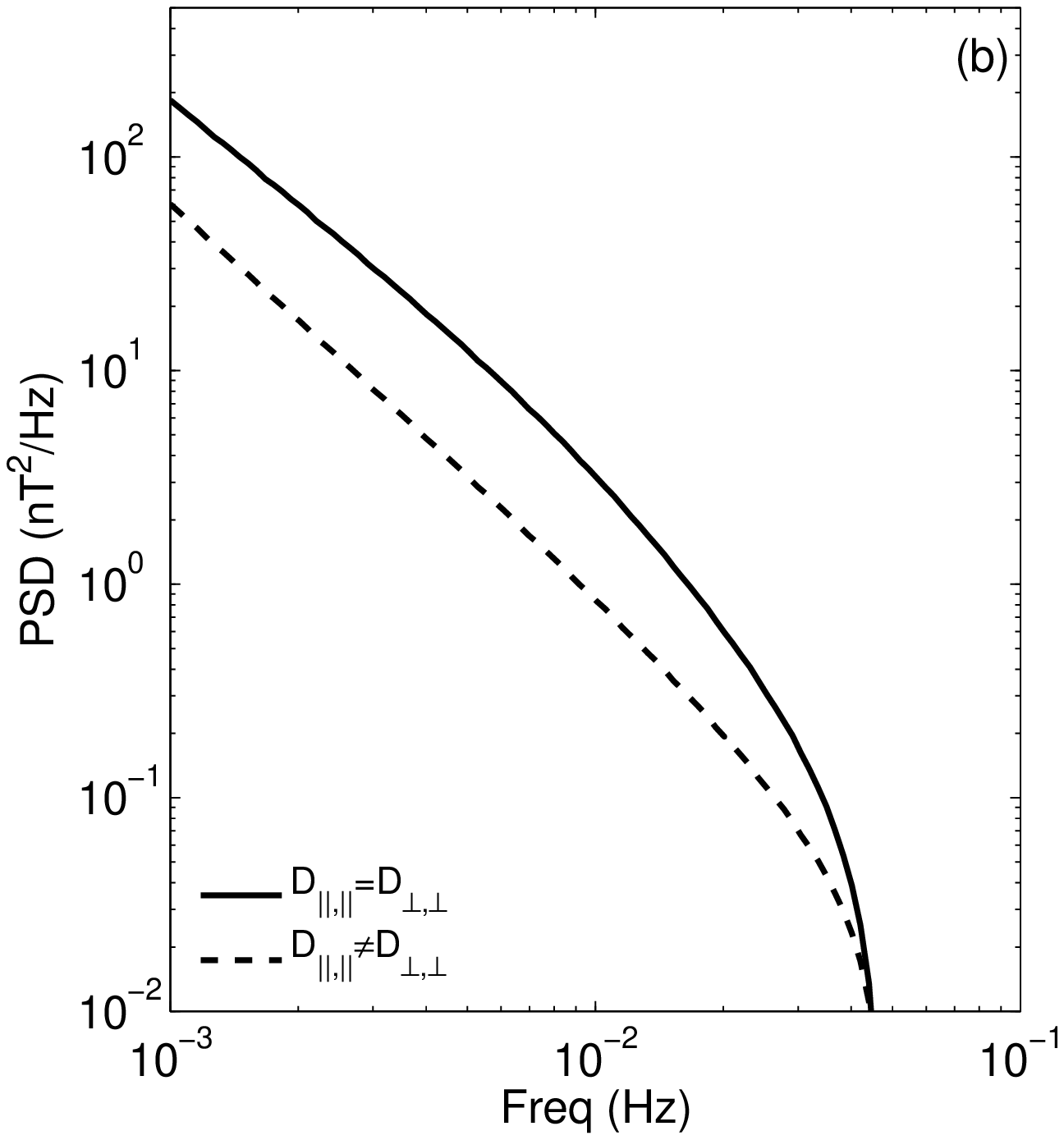}
 \includegraphics[width=0.24\textwidth]{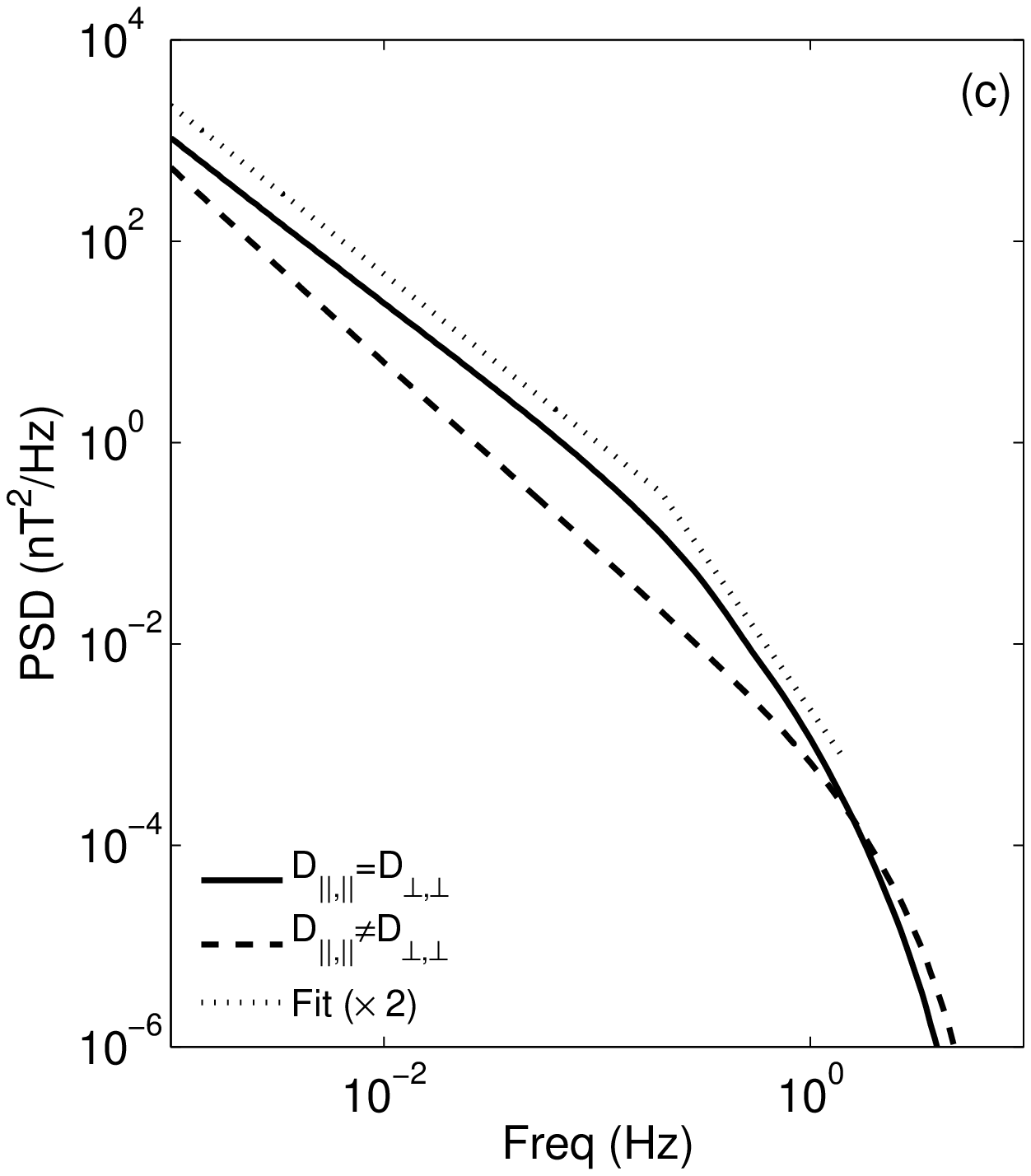}
\end{center}
\caption{
{\it Left}: Angle averaged turbulence power spectra as
functions of $k$ for the spectra in Figure \ref{taus_cut.ps}. The isotropic
diffusion tensor (solid) gives a gradual cutoff
with
the dissipation range covering about one order
of magnitude in the wavenumber as expected from the {\it Middle} panel of
Figure \ref{taus_cut.ps}.
The overall spectrum mimics a broken power law with a spectral break near
$k\simeq 0.5\Omega_p/v_{\rm A}$ (for the assumed plasma parameters).
The anisotropic tensor (dashed) gives a much sharper cutoff  at $k\sim1/\rho_p$
in agreement with the results of Howes et al. (2007), where $\rho_p$ is the
gyroradius of the thermal protons. Note that $v_{\rm
A}\approx v_{\rm th,p}$ for the assumed plasma parameters.
{\it Middle}: Same as the {\it Left} but plotted as a function of the
frequency.
Both spectra cut off at $\Omega_\alpha$ though the cutoff of the anisotropic
model is sharper. The isotropic model has a broader dissipation range with the
overall spectrum mimics a broken power-law with a sharp high frequency cutoff.
{\it Right}: Same as the {\it Middle} panel but with the Doppler shift due to
the relative motion between the solar wind and the spacecraft included. The
angle between the magnetic field and solar wind velocity $\theta_{BV}=38^{\rm
o}$ and the solar
wind speed
$V_{SW}=517$km/s. As in  the {\it Left} panel,
the spectrum of the isotropic diffusion tensor can be fitted with a broken power
law (dotted)
with the break frequency $\nu_{bf}=0.2$Hz and the lower and higher frequency
spectral indexes $\gamma_1=-1.7$ and $\gamma_2=-3.1$, respectively.
}
\label{W-k-f_spec.ps}
\end{figure}

\subsection{Total Spectra}

The angle averaged turbulence spectra are shown in the left
(as a function of the wavenumber) and middle (as a function of the frequency)
panels of Figure \ref{W-k-f_spec.ps}. The spectra of the isotropic diffusion
model mimic broken-power laws with softer dissipation range spectra.
The dissipation range extends nearly one
order of magnitude in both wavenumber and frequency, which may explain the
observed broken power law spectrum of the solar wind Alfv\'{e}n
turbulence. However, due to the low value of $\Omega_\alpha$, the break
frequency is
about one order of magnitude lower than the observed value. Although the
anisotropic diffusion model has a broader dissipation range as shown in the
right panel of Figure \ref{taus_cut.ps}, due to the dominance of the
perpendicular component the angle averaged spectrum can be
fitted with a single power law, which cuts off when the damping of the
perpendicular
component becomes significant.
The spectrum of the anisotropic diffusion model therefore appears to cut off
much sharply with a narrow dissipation range, which is similar to that of the 1D
diffusion models. The spectrum of the isotropic diffusion model is also slightly
harder and higher than that of the anisotropic model in the ``inertial'' range,
which is in agreement with our study of the Alfv\'{e}n turbulence in \S\
\ref{cascade}.

In summary, the above results demonstrate the possibility of a rich variety of
spectra for turbulence determined primarily by  the plasma parameters;  the
density
$n$ , temperature $T$, magnetic field $B$, source  size $L$, and the injected
energy
flux $F_0$ and wavenumber $k_0\sim L^{-1}$ \citep{S06, M08}. These parameters
can be derived from MHD simulations, which, in combination with the present
diffusion model, can be used to study the free energy dissipation processes in
magnetized plasmas quantitatively. The uncertain part of this
procedure is the choice of the diffusion tensor. As discussed above we have
constructed tensors that produce many features of turbulence spectra obtained
with simulations based on more basic physics. Such simulations are too
cumbersome or limited
to be useful for comparison with observations of astrophysical sources.  But
they are important in guiding us to derive the appropriate form of the diffusion
tensor that lies at the core of our approach. For a given diffusion tensor, the
above
discussion demonstrates that one can readily (that means no need of
supercomputing power) produce spectra for turbulence, for varied background
plasma conditions, which can then be used to determine the heating of the
background plasma and acceleration of particles and the resultant radiations.
The validity of the chosen diffusion tensor form can be established by comparing
the model predictions
with existing observations or some simulations \citep{H08, G08}.  In the next
section we compare the predictions of
our models with the observed spectra of magnetic fluctuations in the solar wind.

\section{Application to the Solar Wind Turbulence}
\label{obs}

To apply the model to observations of the solar wind turbulence, one must take
into account the Doppler shift due to the relative motion of the spacecraft with
respect to the solar wind (Leamon et al. 1999). The observed frequency spectrum
$P(\nu)$ is related to the turbulence spectrum as:
\begin{equation}
\label{transf}
P(\nu) = \int {\cal W}({\bf k}) \delta \left\{\frac{1}{2\pi}\left[{\bf k} \cdot
{\bf V}_{SW} +
\omega({\bf k}) \right] - \nu \right\}d{\bf k}
\end{equation}
where $\nu$ is the spacecraft-frame frequency and
$\delta$ is the Dirac-$\delta$ function. The right panel of Figure
\ref{W-k-f_spec.ps} shows the $P(\nu)$ for the angle between the magnetic field
and solar wind velocity $\theta_{BV}=38^{\rm o}$ and the solar wind speed
$V_{SW}=517$ km$/$s. The rest of the model parameters are the same as the middle
panel, which would be the observed spectrum for $V_{SW} = 0$ km/s. [The spectrum
with respect to the wavenumber $k$ in the left panel can be obtained by
replacing
the argument of the $\delta$ function with $(k-k^\prime)$, where $k^\prime$ is
the variable of the integration.]

All these spectra can be fitted with broken power-laws with steeper high
end declines. To understand this result, we may make the approximation: ${\cal
W}({\bf
k}) \propto k^{-11/3} \Theta[k_c(\theta)-k]$. Then the spectrum on the left
panel of Figure \ref{W-k-f_spec.ps} is
given by  $P(k) \propto k^{-5/3}\cos\theta_c(k)$, where $\theta_c(k)$ is the
cutoff propagation angle in the dissipation range and is given by the inverse of
the function $k_c(\theta)$ shown in Figure \ref{kc_contour.ps}, which increases
monotonically with $\theta$, except for  a small range near $\theta=0$.
Clearly, $P(k)$ in the dissipation range is determined by $k_c(\theta)$. Since
$k_c(\theta)$ covers more than one order of magnitude, the same is the
dissipation range of $P(k)$.  The spectrum in the middle panel ($V_{SW}=0$) is
obtained by
changing the variable from $k$ to frequency according to the dispersion relation
$\omega({\bf k})$. In the MHD regime $\omega({\bf k}) = v_{\rm A} k_{||}$, then
we have $P(\nu) \propto \nu^{-5/3} \cos^{8/3}\theta_c(k)$.  Beyond the MHD
regime $\omega({\bf k})$ flattens and approaches  $\Omega_\alpha$ as
$k\rightarrow
\infty$. As a result $P(\nu)$ cuts off at $\Omega_\alpha/2\pi$.  For the
spectrum in the right panel we must carry out the integration in equation
(\ref{transf}). For solar wind conditions in general, and for the observations
we
compare with below in particular, the Alfv\'{e}n velocity ($v_{\rm A} = 36.8$
km/s) is more than one order of magnitude lower than solar wind velocity
($V_{SW}=517$ km/s). One may then ignore the $\omega({\bf k})$ term in the
argument of the $\delta$-function in the above integration. Then for an
isotropic power-law turbulence spectrum, the shape of $P(\nu)$ will be identical
to $P(k)$ in the inertial range. In general, $P(\nu) \propto
\nu^{-5/3}\int_0^{\cos\theta_c(k)} \cos^{5/3}(\theta_{BV}-\theta){\rm
d}\cos\theta$, where $k=2\pi\nu/V_{SW}\cos(\theta_{BV}-\theta_c)$. The
dissipation range of $P(\nu)$ starts at $\nu_d \sim
V_{SW}k_c(0^\circ)\cos\theta_{BV}/2\pi \sim 0.1$ Hz for the parameters in Figure
\ref{W-k-f_spec.ps}, where we have assumed that $k_c(\theta)$ increases
monotonically with $\sin\theta$.


The
break frequency of the broken power law fit  will be at a higher frequency;
$\nu_{bf}\sim V_{SW}k_c(\theta_{BV})/2\pi$. By fitting $P(\nu)$
with a broken power law model down to the power spectral level of $2\times
10^{-4}$ nT$^2/$Hz, we obtain the dotted line in the right panel of Figure
\ref{W-k-f_spec.ps} with $\nu_{bf} = 0.2$ Hz, and  the lower and higher
frequency spectral indexes $\gamma_1=-1.7$ and $\gamma_2=-3.1$, respectively.
(For clarity the dotted line is shifted upward by a factor 2.) This broken power
law spectrum is very similar to those observed in the solar wind (Leamon et al.
1998, 1999).

\begin{figure}[htb]
\begin{center}
\includegraphics[width=0.8\textwidth]{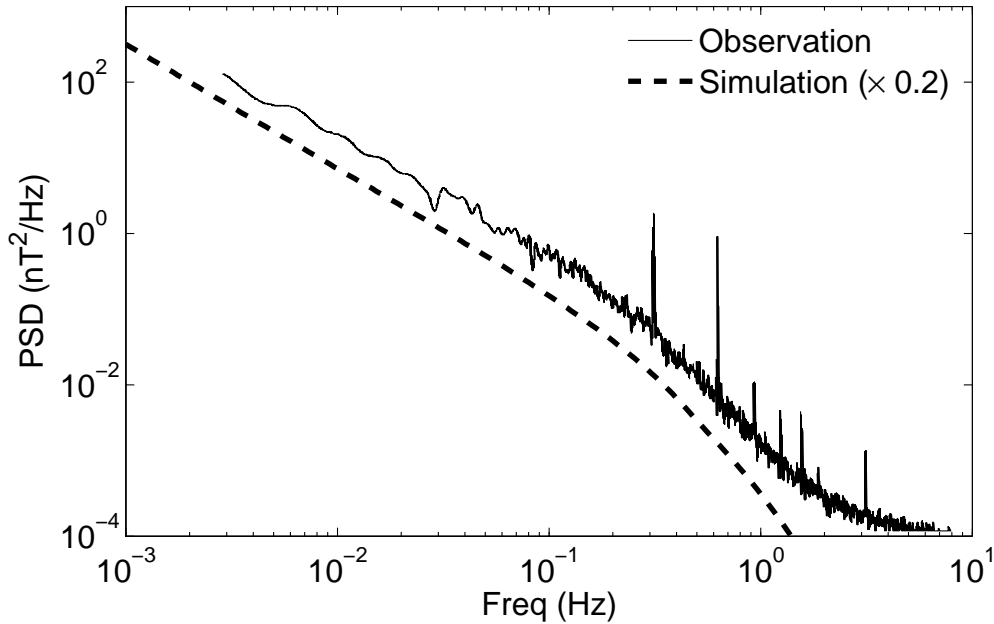}
\end{center}
\caption{
Observed interplanetary turbulence power spectrum (solid line; extracted from
Figure 1 of Leamon et. al. 1999) compared with our model spectrum (dashed line;
scaled downward by a factor of 5 for clarity) based on the locally isotropic
diffusion tensor and the
model parameters
$B=6.3 $nT, $\beta_{\rm p}=0.48$, $\theta_{BV}=38^{\rm o}$,
$V_{SW}=517$km/s given by the observations. The Alfv\'{e}n velocity is not well
determined. Our best fit model has an Alfv\'{e}n velocity of $34$ km/s. The
observed spectral
flattening at high frequencies is related to a spacecraft noise.
As evident this model with essentially only one free parameter provides an
excellent fit to the observed spectrum. We note that this model is
incompatible with the known properties
of Alfv\'{e}n turbulence, such as those studied by \citet{GS95}.
}
\label{break.ps}
\end{figure}

\subsection{Fit to Observed Spectra}

As an example, we fit  the observed solar wind turbulence spectrum at 2200UT,
January
11, 1997 at 1 AU from the Sun (Leamon et al. 1999), as shown by the solid line
in Figure \ref{break.ps}. The observation shows that $\theta_{BV} = 38^\circ$,
$V_{SW} = 517$ km$/$s, $B=63\mu$G [$\Omega_p =0.096(2\pi)$Hz], and $\beta_{\rm
p} = 0.48$. The Alfv\'{e}n velocity (or the plasma density or temperature) is
not given by the authors. By adjusting $v_{\rm A}$ and the energy flux  $F_0$
carried by the turbulence, we obtain the best fit to the observed spectrum {\it
for the locally isotropic diffusion tensor model} as shown by the dashed line in
the Figure \ref{break.ps}.
Leamon et. al. (1999) fit the
observation with a broken power law model
with indexes $\gamma_l=-1.67$, $\gamma_h=-2.91$, and a break frequency
$\nu_{b}=0.235$Hz.
Our simulated spectrum, when fitted by a broken power law model down to
$2\times 10^{-4}$nT$^2/$Hz, gives $\gamma_1=-1.67$, $\gamma_2=-2.97$, and
$\nu_{bf}=0.200$Hz.
$F_0$ is determined by the turbulence spectrum in the inertial range. With
$v_{\rm A}$ as the only free model
parameter, the model reproduces not only the observed break frequency
$\nu_{bf}$, but also the
power-law index in the dissipation range. This is the most appealing success of
the model. Our model fit gives $v_{\rm A}= 34$ km$/$s, $n_e=14$ cm$^{-3}$ for
$k_{\rm B}T_e=3.35$ eV, which can be compared with observations.\footnote{The
cascading constant $C$ in equation (\ref{DijMHD}) should be considered as
another parameter. There  is also an uncertainty in the definition of the eddy
speed. For this simulation, we set $C=1$. A complete measurement of
the properties of solar wind plasmas and the turbulence carried by them can be
used to determine $C$ and test the model \citep{YZ97}.}

We also find that the model based on the anisotropic diffusion tensor cannot fit
the observed spectrum very well. This is primarily due to the dominance of the
perpendicular component and the much sharper cutoff in the dissipation range as
shown by the dashed lines in Figure \ref{W-k-f_spec.ps}
for the angle averaged spectra.
We also consider the Kolmogorov phenomenology with the isotropic cascade rate
$\tau_{\rm cas}^{-1}\propto k^{2/3}$
in the inertial range. The damping rate is very sensitive to $k$ but has
relatively weak dependence on $\theta$ (Fig. \ref{kc_contour.ps}). So $k_c$,
which is identical to the contour of the damping rate, has
a weaker dependence on $\theta$ than the isotropic diffusion model studied
above. As a result, the angle averaged spectrum has a sharper
cutoff at high frequencies similar to the anisotropic diffusion model.

Considering the kinetic effects on the wave cascade and damping in the
perpendicular
direction along the line defined by the critical balance proposed by
\citet{GS95}, \citet{H07, H08a} argue that a 1D diffusion model can also produce
a
broken-power like spectrum with a cutoff at small spatial scales. In this model,
the kinetic Alfv\'ev wave (KAW) has a
steeper spectrum than the Alfv\'{e}n waves. However, to have significant change
in the spectral shape between the Alfv\'{e}n and KAW waves, the electron
temperature needs to be much higher than the proton temperature to make
the kinetic effects significant. This is quite different from our model, where a
broken-power like spectrum is produced by the anisotropic damping and the
inhomogeneous cascade in the wavevector space. The broken power-law spectrum
they produced is with respect to the perpendicular component of the wavevector.
To compare with solar wind observations, one needs to take into account the
Doppler effects.
The KAW starts at $k_{\perp} \rho_p \approx 1$. The corresponding
Doppler-shifted frequency
\begin{equation}
\nu \approx {\bf k}\cdot {\bf V}_{SW}/2\pi \ge k_{\perp}\rho_p
{V_{SW,\perp}\over v_{\rm A} \beta_{\rm p}^{1/2}}{\Omega_p\over 2\pi}\,.
\end{equation}
For typical solar wind conditions with $\beta_{\rm p}\le 1$ and $V_{SW}\sim 10
v_{\rm A}$, this frequency is more than 10 times higher than the proton
gyro-frequency, which disagrees with the observed fact that the break
frequency is usually slightly higher than $\Omega_p/2\pi$ (Leamon et al. 1998,
1999; Bale et al. 2005).
With the 1D model, Stawicki et al. (2001) argues that the broken power-law
spectrum of the solar wind turbulence may be attributed to the dispersive
effects alone as the turbulence cascades from the Aflv\'{e}n wave domain to the
whistler wave regime. However,
the diffusion coefficient must be prescribed properly to reproduce the observed
spectrum.

From the above discussion we conclude that the model based on the locally
isotropic diffusion tensor provides the most natural explanation of
the observed solar wind turbulence spectrum. All other models need fine-tuning
of the model parameters or relevant physical processes to give acceptable fits
to the observed turbulence spectrum.

\subsection{Spectral Features and Model Parameters}

The previous section shows the success of the model in accounting for a specific
observation interval. To understand the nature of the observed broken power
spectrum, statistical studies of a large sample of events are necessary. Some of
these studies \citep{L98, S06, M08, Ham08} have uncovered the nonlinear nature
of the dominant dynamics and have shown that several factors may affect the spectral
shape in the dissipation range. For a qualitative  comparison with these observations,  in this
section we explore the dependence of the spectral features on the model
parameters.

\begin{figure}[htb]
\begin{center}
  \includegraphics[width=0.45\textwidth]{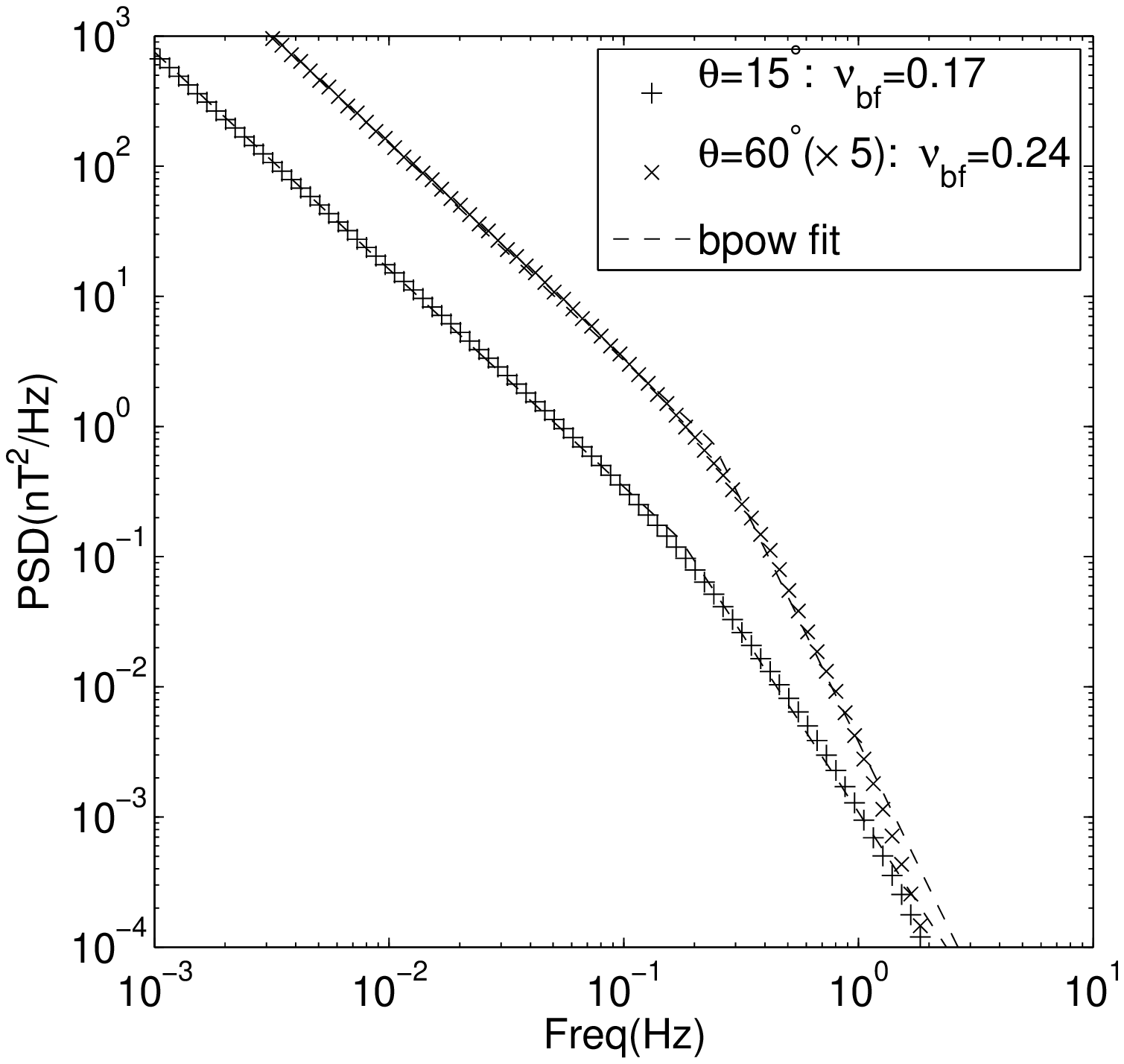}
 \includegraphics[width=0.45\textwidth]{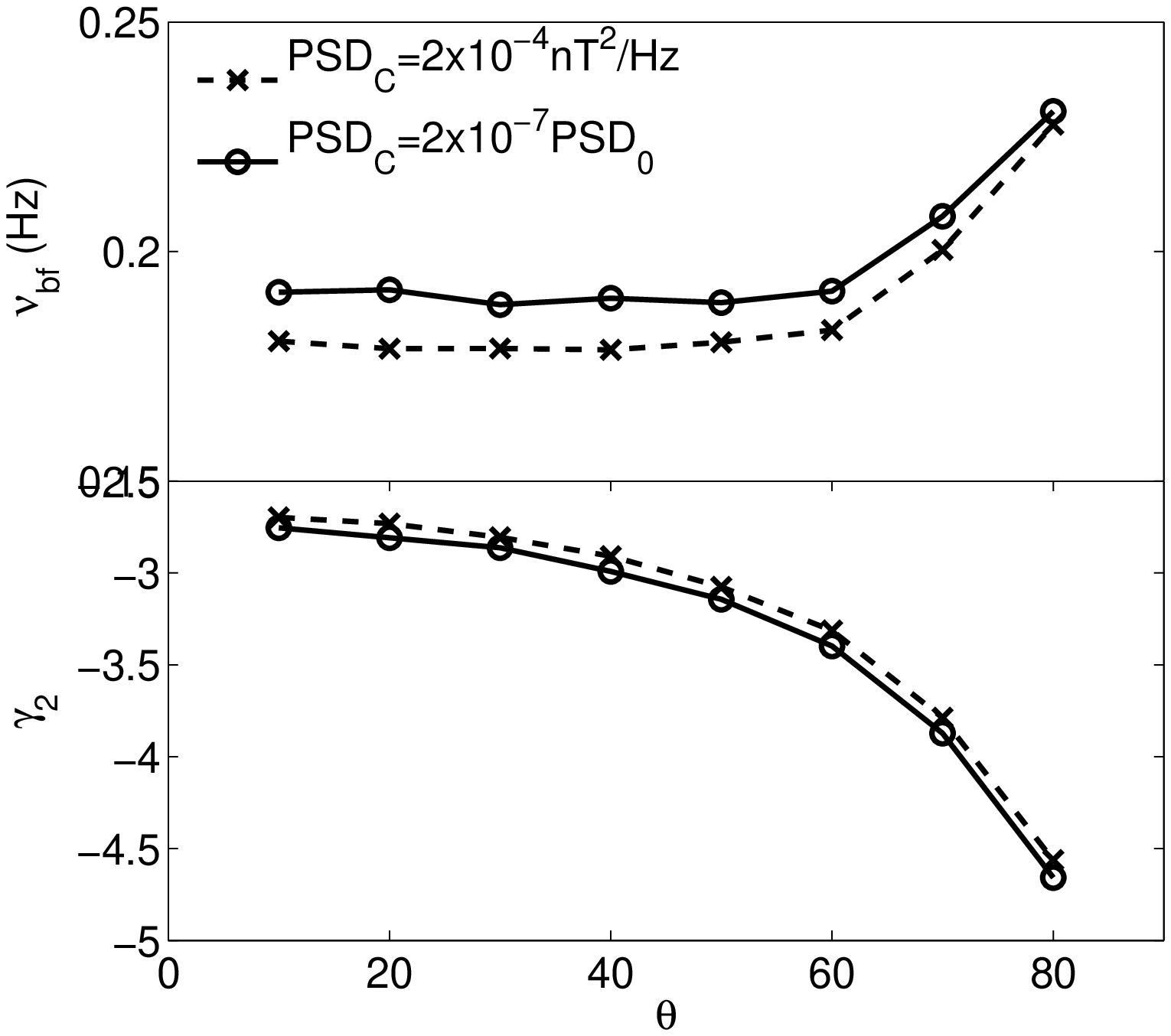}
 \end{center}
\caption{Dependence of the angle averaged spectrum on $\theta$
the angle between the magnetic field and solar wind velocity (the same as the
$\theta_{BV}$ in the text).
{\it Left}: Two sample spectra with $M_{\rm A}=1.23$,  $\beta_{\rm p}=0.41$, and
 two values of  $\theta_{BV}$
as indicated in the legend. The other parameters are the same as in the right
panel of Figure \ref{W-k-f_spec.ps}.
Broken power law fits to these spectra down to $2\times10^{-4}{\rm
nT^2Hz^{-1}}$ give a break frequency $\nu_{bf}=0.17$ Hz
for $\theta_{BV}=15^\circ$ and $\nu_{bf}=0.24$Hz for $\theta_{BV}=60^{\circ}$.
{\it Right}: Dependence of $\nu_{bf}$ and $\gamma_2$ on $\theta_{BV}$.
Two lower limits of the power spectrum are considered while fitting the spectrum
with broken power laws. The cross signs are
obtained by fitting the model
spectrum down to a turbulence power level of $2\times10^{-4}{\rm
nT^2Hz^{-1}}$, and
the circle signs correspond to a low limit of $2\times10^{-7}P(\nu=10^{-3}{\rm
Hz})$.
}
\label{ang_dep.ps}
\end{figure}

As indicated above, most observed spectra can be fitted with three parameters;
$\gamma_1$ that
is always nearly equal to -5/3, $\gamma_2$, and the break frequency $\nu_{bf}$.
We
have explored the dependence of the latter two on the observable  parameters
using the locally isotropic diffusion tensor scheme. Figures \ref{ang_dep.ps},
\ref{beta_dep.ps}, and \ref{MA_dep.ps} show the dependence of the spectrum on
$\theta_{\rm BV}$, plasma beta $\beta_{\rm p}$, and Mach number $M_{\rm A}$,
respectively. The other parameters of the plasma and the injected turbulence
are the same as that in the right panel of Figure \ref{W-k-f_spec.ps} unless
specified otherwise. To describe the spectrum quantitatively, we fit the
numerically calculated
spectra to  a broken power-law model down to a turbulence power level of
$2\times10^{-4}\ {\rm nT^2Hz^{-1}}$ and $2\times10^{-7}P(\nu=10^{-3}{\rm Hz})$.
The former is appropriate to compare with observations made by instruments with
a given sensitivity to the magnetic field turbulence power. The latter is
appropriate for observations with a given dynamical range in the turbulence
power spectrum. In general, $\nu_{bf}$ and $\gamma_2$ are very similar for these
two fits, and as stated above, $\gamma_1$  is always very close to
$-5/3$ for the
parameter space explored. The most distinct feature of these results is the
correlation between the two observables $\nu_{bf}$ and $\gamma_2$ (Fig.
\ref{cor.eps}). A higher
$\nu_{bf}$ always comes along a softer dissipation range spectrum.  The
correlation caused by variations in $\beta_{\rm p}$ and $M_{\rm A}$ is related
to the kinetic effects of the waves beyond the MHD regime. The damping rate is
very sensitive to $\beta_p$, as can be seen from the right panel of Figure
\ref{W-k-f_spec.ps}. The change in $M_{\rm A}$ leads to different dissipation
scales, where the kinetic effects are different. The resultant correlation
between $\nu_{bf}$ and $\gamma_2$ is mostly caused by the fact that the cascade
rate becomes more isotropic near $\Omega_\alpha$ (the left panel of
Fig. \ref{kc_contour.ps}).  With the decrease of the dissipation scales caused
by
either higher values of $M_{\rm A}$ or lower values of $\beta_{\rm p}$, the
dissipation range covers a narrower spatial extension, giving rise to a softer
dissipation range spectrum. The dependence on $\theta_{BV}$, on the other hand,
is mostly due to the Doppler shift. Since $\nu_{bf}\sim
V_{SW}k_c(\theta_{BV})/2\pi$, $\nu_{bf}$ increases with $\sin\theta_{BV}$, and
the
spectral range beyond $\nu_{bf}$ becomes narrower leading to lower values of
$\gamma_2$. These explain the nearly identical correlation between $\gamma_2$
and $\nu_{bf}$
caused by variations in $\beta_{\rm p}$ and $M_{\rm A}$ while a quite distinct
correlation due to changes in $\theta_{BV}$, as shown in Figure \ref{cor.eps}.

\begin{figure}[htb]
\begin{center}
 \includegraphics[width=0.45\textwidth]{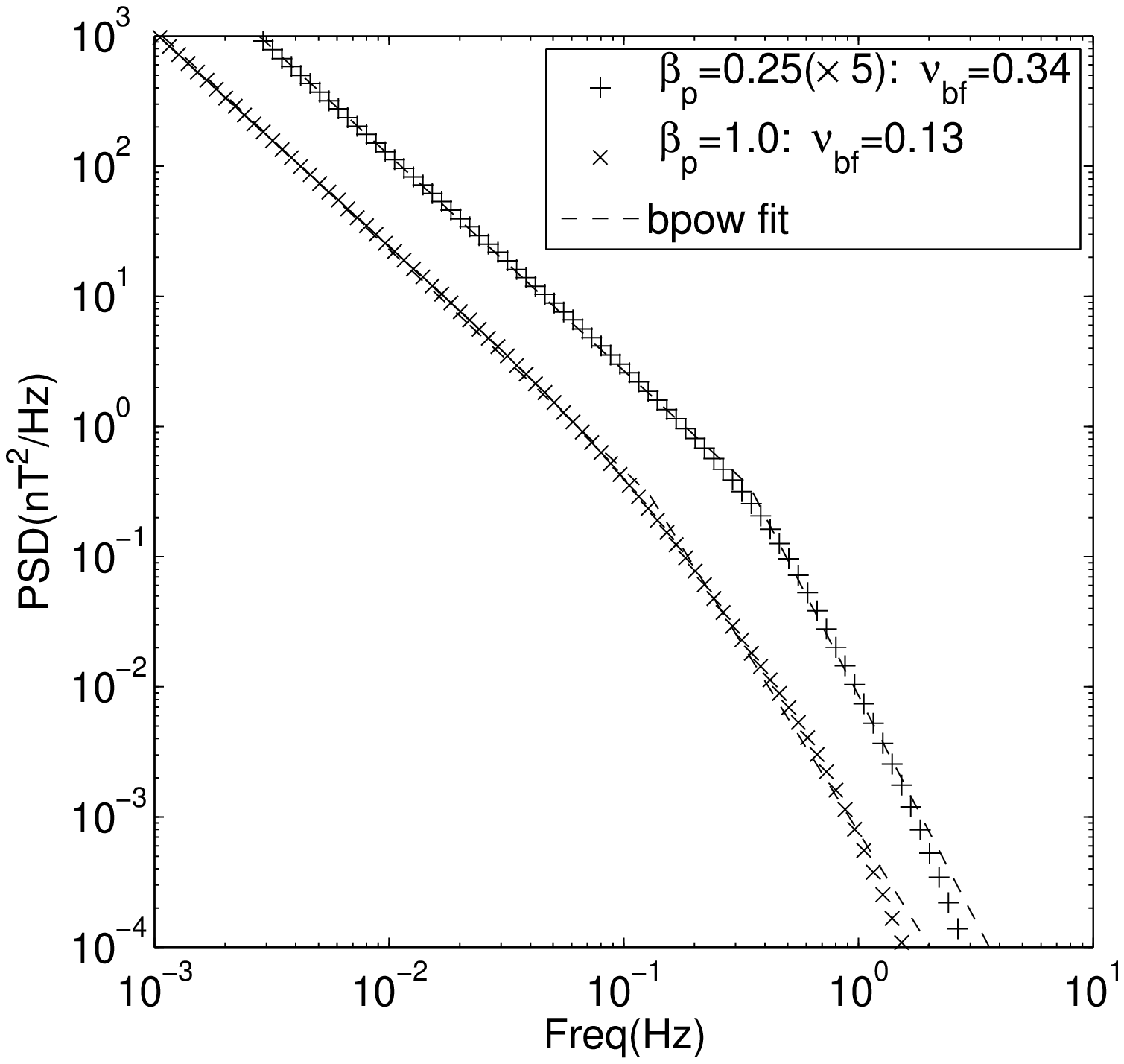}
 \includegraphics[width=0.45\textwidth]{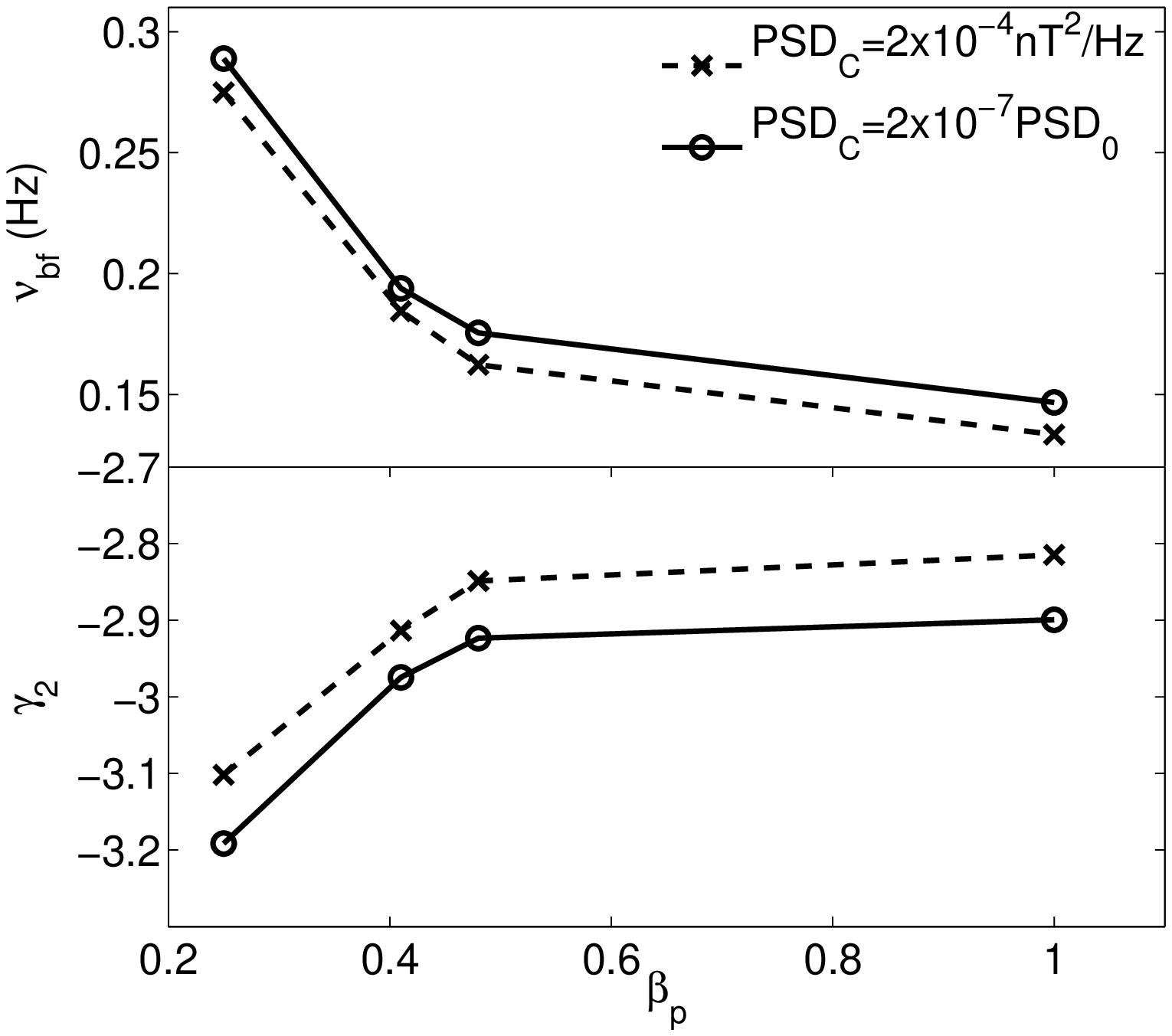}
\end{center}
\caption{
Same as Figure \ref{ang_dep.ps} but for the
dependence on $\beta_{\rm p}$. Here $M_{\rm A} = 1.23$ and $\theta_{BV} =
38^\circ$.
}
\label{beta_dep.ps}
\end{figure}


However, the dependence of the spectrum on $\theta_{BV}$ in our model is much
weaker than any quasi-2D
turbulence models with the energy cascading only in the perpendicular direction
\citep{H07, H08a}.
If most of the wave energy resides in the perpendicular direction, as suggested
by the
critically balanced cascade proposed by Goldreich \& Shridhar (1995),
the $k_c$ contour will be highly prolonged in the perpendicular direction and
$\theta_{BV}$
will strongly affect the break frequency. The observations by Leamon et al
(1998,
Fig. 6) and by \citet{OH07} appear to favor the isotropic diffusion model used
here \citep{Ham08}.
The dependence of the spectrum on the solar wind speed is relatively simple.
Obviously, a faster solar wind gives a  higher break frequency.
Clearly, more observations and/or systematic analyses of existing data are
required to test the validity of these models.

\begin{figure}[htb]
\begin{center}
 \includegraphics[width=0.45\textwidth]{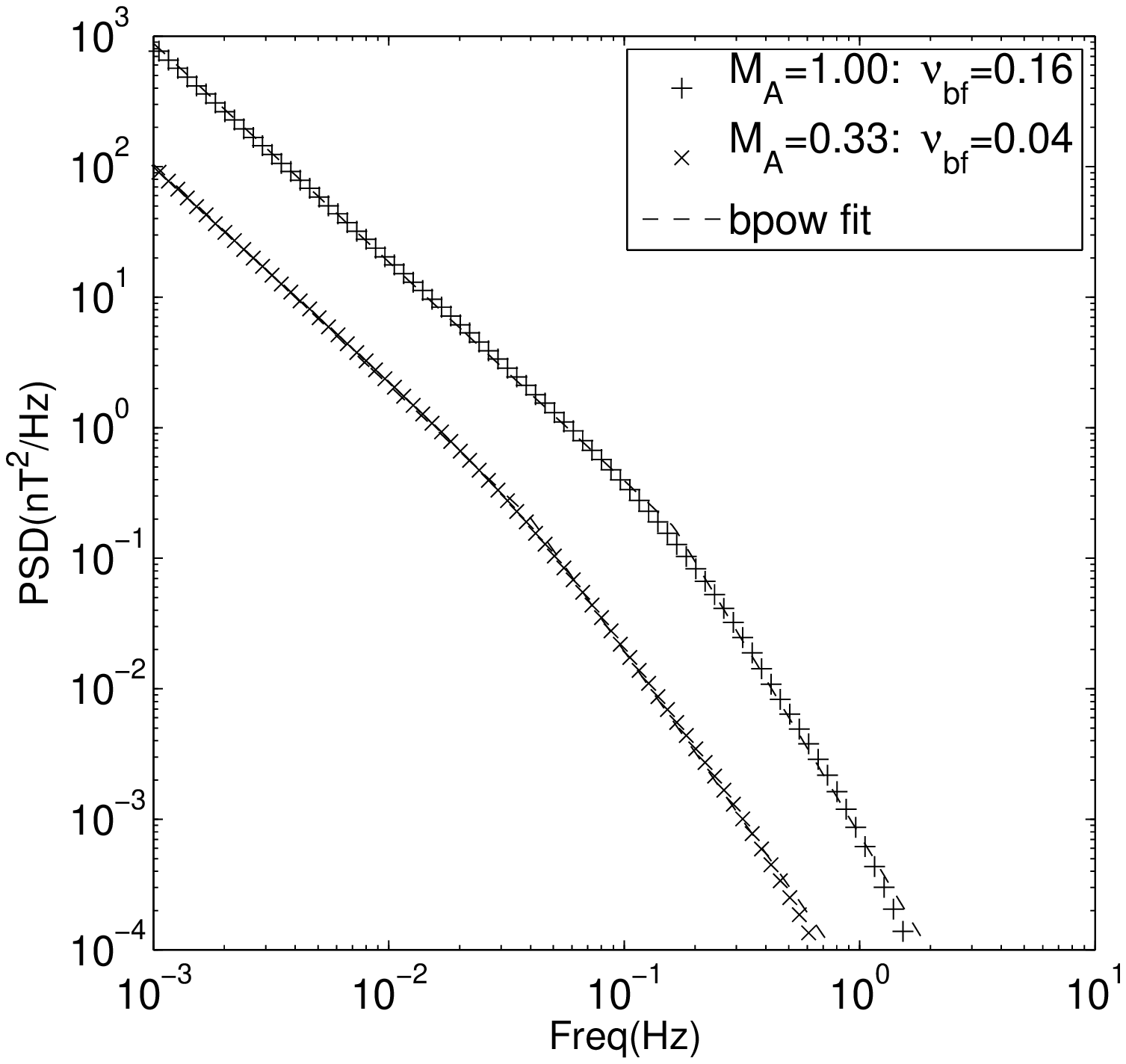}
 \includegraphics[width=0.45\textwidth]{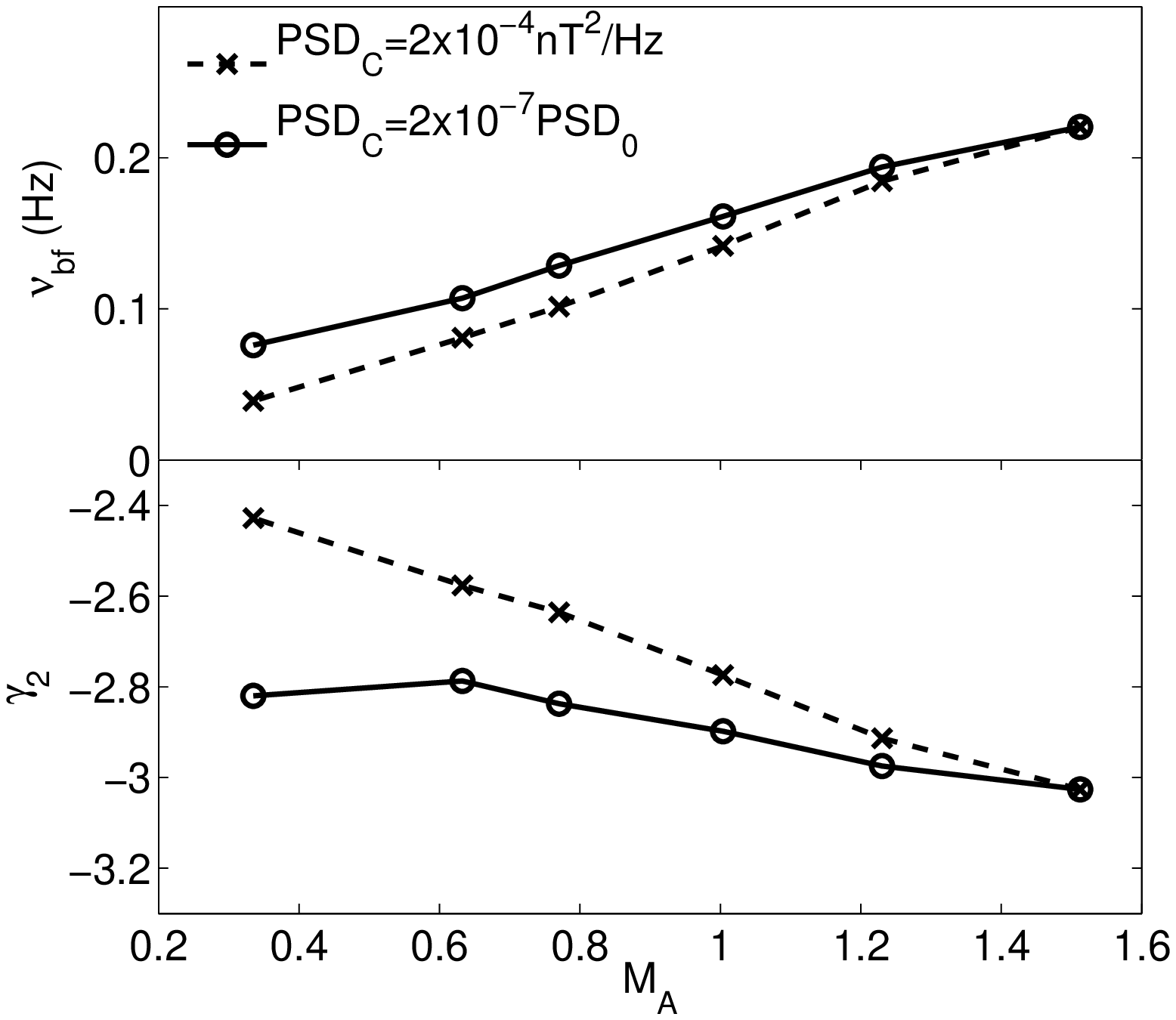}
\end{center}
\caption{
Same as Figure \ref{ang_dep.ps} but for the
dependence on $M_{\rm A}$. Here $\beta_{\rm p} = 0.41$ and $\theta_{BV} =
38^\circ$.
}
\label{MA_dep.ps}
\end{figure}


In fact there are already some preliminary results showing some weak
correlations (with lots of scatters) between the spectral features and other
parameters. For example, the Figure 3 of \citet{L98} shows a correlation between
the break frequency and the proton cyclotron frequency. For a given thermal
pressure, higher values of proton cyclotron frequency imply lower values of
$\beta_{\rm p}$, and as shown in Figure \ref{beta_dep.ps},  we expect higher
break frequencies as observed.
Their Figure 6 shows two events with distinct values of $\theta_{BV}$ but
similar values for the rest of the parameters. This observation suggests the
dissipation range spectrum is harder for higher values of $\theta_{BV}$ in
contradiction with our Figure \ref{ang_dep.ps}. However, their spectrum with a
higher value of $\theta_{BV}$ shows prominent noise at high frequencies so that
the spectral break is obscured. Their Figure 2 suggests that the dissipation
range spectrum is softer for higher values of proton temperature $T_p$, which
implies higher values of $\beta_p$ for a given Alfv\'{e}n speed. While we expect
the opposite, we notice that the higher values of $\beta_p$ likely result from
higher values of $M_{\rm A}$, as indicated by the convergence of their inertial
range spectral index with the increase of $T_p$,\footnote{The inertial range
spectrum can be fully developed in shorter periods for turbulence with higher
values of $M_{\rm A}$.} and simultaneous increases of $M_{\rm A}$ and $\beta_p$
can make the dissipation range spectrum change in either direction. Indeed,
\citet{S06} found that the dissipation range spectrum becomes softer with the
increase of the energy cascade rate in the inertial range. According to our
model, the energy cascade rate is determined by $M_{\rm A}$ and $k_0$. Thus this
observation has the same trend as the dependence of $\gamma_2$ on $M_{\rm A}$
shown in Figure \ref{MA_dep.ps}.

\begin{figure}[htb]
\begin{center}
 \includegraphics[width=0.5\textwidth]{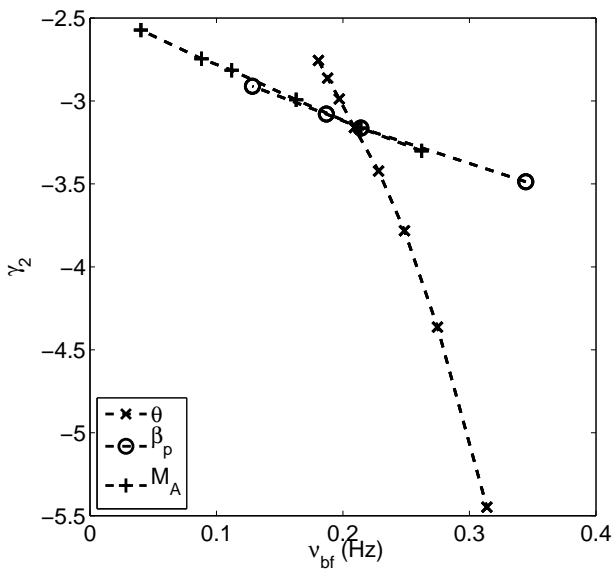}
\end{center}
\caption{
The correlation between $\nu_{bf}$ and $\gamma_2$ caused by variations in
$\theta$ (crosses), $\beta_{\rm p}$ (circles), and $M_{\rm A}$ (pluses). The
model parameters are the same as those in the right panel of Figure
\ref{W-k-f_spec.ps}.
}
\label{cor.eps}
\end{figure}

Leamon  et al. (1999, Fig. 2) showed a weak correlation
between  $k_{\rm diss}$ (the same as our $k_c$) and the inverse of the proton
gyro-radius $R_L$ (the same as our $\rho_p$). For damping dominated by cyclotron
resonances, $k_{\rm diss} C_S \simeq \Omega_p$, where $C_S$ is the sound speed.
Therefore $\nu_{bf}\propto k_{\rm diss} \propto  \Omega_p/v_A\sqrt{\beta_p}$,
which is similar to the trend  we find in  Figure \ref{beta_dep.ps}.
In general, higher values of $R_L$ imply higher values of $\beta_{\rm p}$ and
longer dissipation scales.
\citet{Smi01} studied a solar wind interval with very high Aflv\'{e}n speeds and
found that the cutoff frequency $\nu_{bf}$ decreases with the increase of
$v_{\rm A}$. They concluded from this that the dissipation is related to the ion
inertial length $v_{\rm A}/\Omega_p$, instead of $R_L$. We note that the
turbulence intensity was  very low during this interval, which  implies
low values of $\nu_{bf}$ according to our Figure \ref{MA_dep.ps}.
More recently, \citet{M08} correlated the break frequency $\nu_{bf}$ with
several  parameters and found some
weak correlations. For example, their Figures 4 and 7 also show a decline of
break frequency with the plasma beta;  $\nu_{bf}\propto \beta_p^{-x}$ with
$x\sim
0.24$, in agreement with our Figure \ref{beta_dep.ps}. On the other hand, their
Figure 6 can be interpreted as  $\nu_{bf}\propto
1/M_{\rm A}$, which has the opposite trend to what we find in Figure
\ref{MA_dep.ps}.

We only have qualitative comparisons of these observations with our model here
since the turbulence spectrum depends on several parameters,  $M_{\rm
A}$, $\theta_{BV}$, $V_{SW}$, $\beta_{\rm p}$, $k_0$ etc. Simple correlations
between $\gamma_2$, $\nu_{bf}$ and any one of these quantities are not expected
because variations caused by the other parameters may produce significant
scatters in the observed data. Clearly, one has to apply the model directly to
the observed data to assess its merit.
For some events with detailed observations over a large dynamical range, one may
use the large scale turbulence spectrum as input to test the diffusion
approximation we have promoted here. In some cases, the time dependence of the
turbulence
spectral evolution may also need to be considered, especially in cases with
strong anisotropy observed at large scales \citep{M90, D05, OH07}.

Our current model only consider the Alfv\'{e}n-Cyclotron branch. To explain the
observed spectrum above $10$ Hz \citep{D83}, one may have to include the fast
wave mode branch, which extends up to the electron cyclotron frequency (Appendix
A).
Another important observational feature is the high magnetic helicity  and the
enhancement of the parallel magnetic field fluctuations in the dissipation range
as compared with the inertial range \citep{L98, L99, Ham08}. Considering the
polarized characteristics of the waves, our diffusion model can readily produce
results that can be compared with these observations.

\section{Conclusion}
\label{dis}

The free energy dissipation in collisionless astrophysical plasmas plays crucial
roles in our understanding of all kinds of nonthermal phenomena. Given
the large amount of energy inferred from observations of these nonthermal
astrophysical sources, the free energy must come from large scale
structures, while the plasma heating and particle acceleration usually
start on microscopic scales. An energy cascade from large to small scales is therefore
necessary. Strong turbulence is expected to mediate this energy dissipation in many astrophysical
sources, and high energy particles, which are responsible for most of the
observed emission, compete with the low energy background particles to share
this energy. There are currently no theories, which can address the energy
partition between low and high energy particles under different astrophysical
conditions. And the major challenges reside in the not-well-known nature of
turbulence in magnetized collisionless plasmas. Although there is already much
research both numerical and theoretical on turbulence, none of them gives
quantitative predictions on the nature of the high energy particles produced in
this free energy dissipation process. In this paper we have proposed the
treatment of turbulence cascade using the diffusion approximation in the
2D wavevector space, with the aim of treating the free energy
dissipation mediated by
magnetized turbulence self-consistently. To simplify the problem, we assume that
the low energy particles reach a thermal equilibrium. One therefore can study
the turbulence cascade and damping by solving the kinetic equation for the
turbulence power spectrum numerically.

The diffusion tensor is the central part of the model.
Based on previous analytic and numerical results, we have tested two
forms for the diffusion tensor; one locally isotropic and one anisotropic. We
compare the cascade process of Alfv\'{e}n-cyclotron turbulence for these two
models. As expected the resultant spectra tend to be more anisotropic for the
anisotropic tensor but in both cases the energy flux is predominately in the
direction perpendicular to the large scale magnetic field, which is
qualitatively (but not quantitatively)
similar to the Goldreich \& Sridhar (1994, 1995) critical balance cascade. The
isotropic model for Alfv\'{e}n turbulence has a critical balance of
$k_{||}\propto k^{3/4}$ determined by the diffusion tensor chosen. The critical
balance of the anisotropic tensor is almost identical to the \citet{GS95}
relation.

As cascade proceeds to smaller scales ($k\rightarrow \Omega_p/v_{\rm A}$), the
dispersion relation begins to deviate from the simple form valid in the MHD
regime. We have included this effect (using two reasonable forms for $\tau_W$) and shown that the turbulence
spectrum steepens when we reach this regime. We also show that the inclusion of
thermal damping (based on the hot plasma dispersion relation)  induces
cutoffs
in the spectrum as the waves and turbulence dissipate and heat the plasma or
accelerate particles. These cutoffs are also very anisotropic and occur at
different wavenumbers for different angles of propagation. As a result of these
anisotropies, the angle averaged spectra resemble a broken power
law
both in the wavenumber or frequency domain. We emphasize that this broken power
law spectrum requires anisotropy in the cascade as well as damping.

We compare these spectra with those observed in the solar wind
by Leamon et. al. (1999) and find a good fit to the observations with the
isotropic
diffusion tensor model with essentially one free parameter, i.e., the gas
density, which is readily observable with the current instruments in space.
Based on this model,
we also make predictions on the variation of the break frequency and spectral
``index'' in the dissipation range with the angle between the magnetic field and
solar wind
velocity,  Alfv\'en Mach number, and  plasma beta. Some of these results appear
to be in agreement with the weak correlations found recently with the
statistical analyses of a large sample of events \citep{M08}.

Our model is distinct from existing 1D models \citep{S01, Galtier06,
H08a} for the observed broken power law spectrum of solar wind
turbulence. \citet{S01} essentially used the observed spectrum to infer
the required diffusion coefficient for the wave power spectrum. The weak
turbulence theory of \citet{Galtier06} retains only cascade in the
perpendicular direction. In \citet{H08a} model, a critical balance is
assumed, which essentially leads to a 1D problem. Due to
difficulties in producing a broken power law spectrum with the
dissipation processes, all these models invoke some kinetic effects of
the wave dispersion on the turbulence energy cascade.
Although for the particular event observed by \citet{B05}, it appears that
kinetic effects play dominant roles. The varieties of the observed high
frequency solar wind turbulence spectrum suggest that dissipation
processes should play an important role since the kinetic effect induced
spectral break does not vary significantly for different observation
intervals. In \citet{H08a} model, both
kinetic and dissipation processes are considered to produce a
varieties of solar wind turbulence spectrum.
\citet{CV03} were the first to study a truly 2D model. However, the model was
constructed with a much complicated approach to recover the critical
balance proposed by \citet{GS95} and it hasn't been applied to the
solar wind turbulence directly. With the diffusion approximation, our
model is simplified significantly. Although the diffusion approximation
may not be well justified, especially for weak turbulence, observations
can be used to guide the construct of the diffusion tensor and the wave
kinetic equations. Direct and more systematic applications of these
models to the observed solar wind spectra will be able to test them.

\acknowledgments

We thank Christopher L. Fryer, Hui Li, and Alex Lazarian for helpful
discussions and the referee for a critical review. The research at Stanford is partially supported
by NSF grant ATM-0312344, NASA grants NAG5-12111, NAG5 11918-1.
This work at the Los Alamos National Laboratory was carried out
under the auspices of the National Nuclear Security
Administration of the U.S.
Department of Energy  under Contract No.
DE-AC52-06NA25396,  and received partial support from the EU's Solaire
Research Training
Network at the University of Glasgow (MTRN-CT-2006-035484).

\appendix

\include{appendix}

\end{document}

%% file: appendix.tex
\begin{center}
    {\bf APPENDICES}
\end{center}

\section{Cold Plasma Dispersion Surface}
\label{cold.sec}

The stochastic particle acceleration theory is built on wave-particle resonant interactions, and the plasma turbulence 
cascade also highly depends on the wave-wave resonances. To investigate these resonant interactions, and  the acceleration and 
cascade processes, an exact form of the plasma dispersion relation is required. However, a complete treatment of the coupled 
oscillations of particles and fields is too complicated to provide even a stable numerical solution that can be integrated into
the numerical study of wave-particle or wave-wave resonant interactions. By introducing a cold plasma approximation, Stix (1962) 
solved the complete dispersion relation 
analytically. In this section we discuss the solution of the cold plasma dispersion relation and the procedure of isolating individual 
wave mode for the turbulence  cascade and particle acceleration study.

Although the solution of cold plasma dispersion relation is segmented and has poles at 
particle cyclotron frequencies \citep{S89}, each physical mode of dispersion surface is both 
continuous and smooth almost everywhere. Thereafter, one can isolate each 
continuous mode with segmented functions that gives $k(\omega,\theta)$ relationship by
one-to-one mapping. In this subsection, we discuss the dispersion relation and the range of
each wave mode specifically with the notation given by \citet{S89}.

Alfv\'{e}n-He Cyclotron branch,
\begin{equation}
k= \omega \sqrt{B-F \over 2A} \qquad \omega\in[0,\Omega_{\alpha})
\end{equation}
$R$ and $L$ term reach their first pole ($\Omega_{\alpha}$) and $\displaystyle\lim_{\omega\to\Omega_{\alpha}^-}k=\infty$.
This cutoff frequency is the end of the Alfv\'{e}n branch. 

Fast-Proton Cyclotron branch,
\begin{displaymath}
k=\left\{ \begin{array}{ll} \displaystyle \omega \sqrt{B+F \over 2A} & \omega\in[0,\Omega_{\alpha}^-]\\
\displaystyle \omega \sqrt{B-F \over 2A} & \omega\in[\Omega_{\alpha}^+,\Omega_{p})\\ \end{array} \right.
\end{displaymath}
Similarly, the fast branch cuts off at $\Omega_p$ with $k_{||}\to\infty$.
Although $\Omega_{\alpha}$ is a pole for $R$ and $L$, $\displaystyle\lim_{\omega\to\Omega_{\alpha}^-}{B + F \over 2A}=
\lim_{\omega\to\Omega_{\alpha}^+}{B - F \over 2A}\neq\infty$, i.e. the dispersion surface is continuous.
The switch of sign at $\Omega_{\alpha}$ can be easily explained in a simplified case where $\theta=0$. 
At $\theta=0$, $F=2\sqrt{P^2D^2}=2\mid PD \mid$ and the wave solution \citep{S89}
simplifies into
\begin{equation}
k= \omega\left(S \pm {\mid PD\mid \over P}\right)^{1/2}
= \omega\left(S \pm \textrm{sign}(PD)D\right)^{1/2}
\end{equation}
Since $P<0$ for all $\omega\in[0,1]$, and $D$ swiches sign at $\Omega_{\alpha}$ whereas $(P^2D^2)^{1/2}$ 
does not, the discontinuity is only
introduced by the attempt of writing $k(\omega)$ in an explicit form. As a result, by switching the sign in 
$\displaystyle{B \pm F \over 2A}$ 
at $D=0$, we can follow the continuous dispersion surface.

Whistler branch,
\begin{displaymath}
k=\left\{ \begin{array}{ll} \displaystyle \omega \sqrt{B+F \over 2A} & \omega\in[\omega_1,\Omega_{p})\\
\displaystyle \omega \sqrt{B-F \over 2A} & \omega\in(\Omega_{p},\omega_{e})\\ \end{array} \right.
\end{displaymath}
Due to the strong He cycltron (left-handed) damping around $\Omega_{\alpha}$, $\displaystyle{B + F \over 2A}$
becomes pure imaginary at $\left[\Omega_{\alpha}, \omega_1\right)$, where $\omega_1$ is the smaller root 
of equation $R(\omega)=0$. This $\omega_1$ is the starting point of Whistler branch 
($\omega_1\approx0.585\Omega_p$ very insensitive to density and field strength).
On the other hand, at $\omega_e$ we get $P=0$, the formula reach the pole in parallel direction 
($\theta=0$), and the Whistler branch cuts off at the electron Langmuir oscillation.

\begin{figure}[htb]
\begin{center}
 \includegraphics[width=0.42\textwidth]{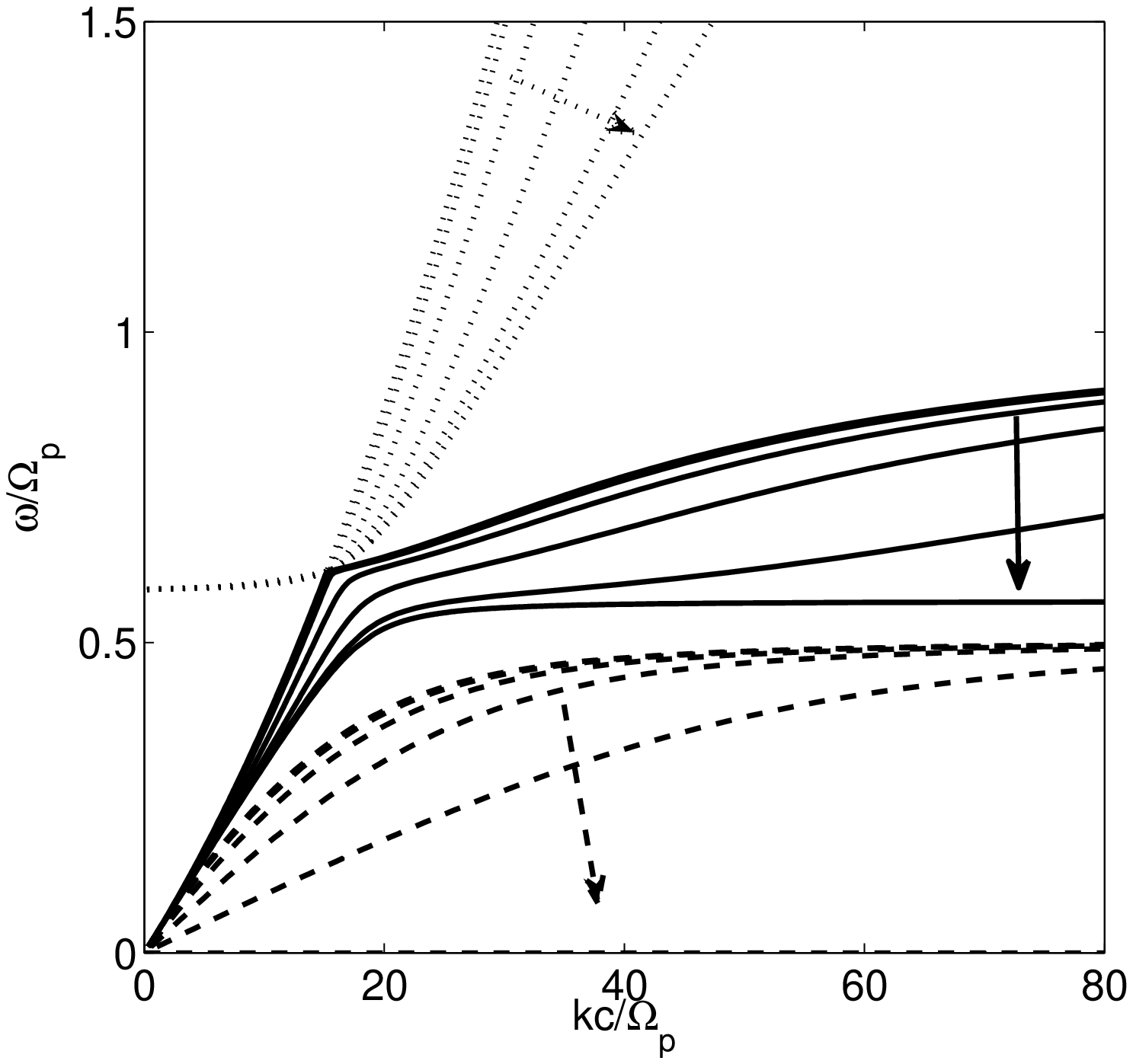}
 \includegraphics[width=0.42\textwidth]{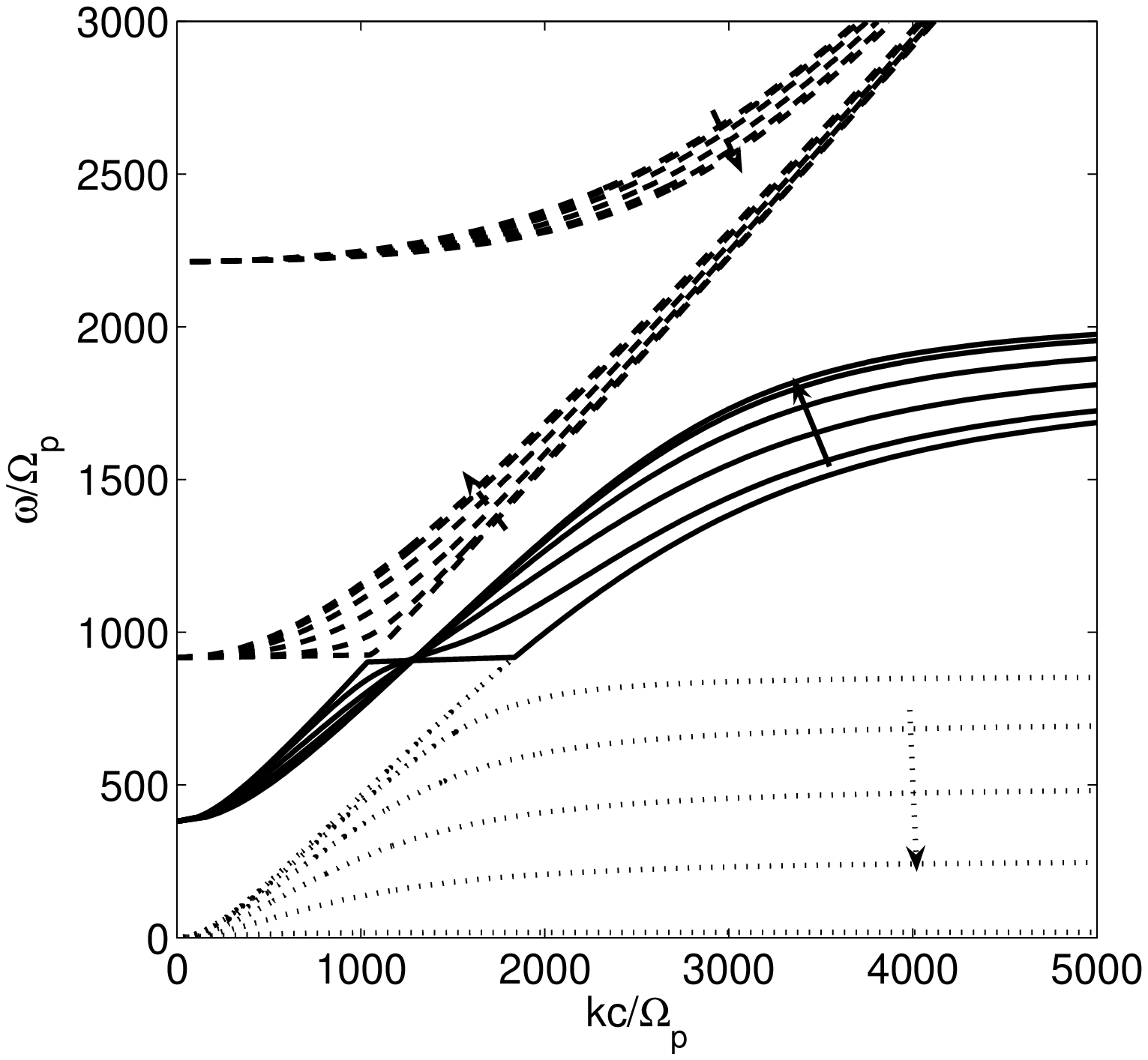}
\end{center}
\caption{
The dispersion relation for a cold fully ionized H and He (10\% by number) plasma 
at different angles ($\theta=0, \pi/10, \pi/5, 3\pi/10, 2\pi/5, \pi/2$
 with arrows pointing to the direction of 
increasing angle). In this example, the only free parameter $\alpha\equiv\omega_{p,e}/\Omega_e=0.5$.
{\it Left}: Aflv\'{e}n (dashed), Fast (solid) and low frequency segment
 of Whistler branch (dotted). 
{\it Right}: High frequency segment of Whistler branch (dotted), Upper
 Hybrid (solid) and the 
two EM branches (dashed).   
}
\label{disp.ps} 
\end{figure}

In the parallel direction, the fast branch starts with electron cyclotron branch and turns into
proton cyclotron branch; the whistler branch, on the other hand, starts
with proton cyclotron branch and turns into electron cyclotron branch
\citep{PL04}. As shown in Figures \ref{disp.ps} and \ref{disps1_prl.ps},
this turning point is also the crossing point of electron cyclotron and
proton cyclotron branches. If we assume turbulence can 
only start at large scales and
low frequency, this would be the only point where whistler branch and
even higher frequency branches 
obtain energy from the turbulence cascade. This crossing wavenumber
$k_\mathrm{c}$ and the efficiency of reverse cascade
determine the low wavenumber cutoff on energy spectrum of proton
cyclotron branch, $k_{\min}$, 
which in turn determines the acceleration efficiency on ions, especially $^3$He due to its special charge to 
mass ratio (see Liu et al. 2006 for a detailed discussion of how $k_{\rm min}$ affects $^3$He 
spectrum). Thereafter, it is worthwhile to calculate this crossing point. By solving $\displaystyle {B + F \over 2A}={B - F \over 2A}$ 
(i.e. $D=0$) or
\begin{equation}
{-\Omega_e\omega_e^2 \over \omega^2-\Omega_e^2} +
{-\Omega_p\omega_p^2 \over \omega^2-\Omega_p^2} +
{-\Omega_{\alpha}\omega_{\alpha}^2 \over \omega^2-\Omega_{\alpha}^2} =0
\end{equation}
we get $\omega_2=0.615\Omega_p$ \footnote{This result depends on the
particle mass and relative 
abundance only, which are approximately constant.} and the corresponding
wavenumber
\begin{eqnarray}
k_\mathrm{c}=\sqrt{460.5\alpha^2-10^{-12}\alpha+0.377}\Omega_p/c 
\end{eqnarray}
where $\alpha=\displaystyle\omega_e/\Omega_e=\sqrt{4\pi n_e m_e} c/B$ is
the only free parameter
for the dispersion relation with the cold plasma approximation.

Upper hybrid branch,
\begin{displaymath}
k=\left\{ \begin{array}{ll} \displaystyle \omega \sqrt{B+F \over 2A} & \omega\in[\omega_3,\Omega_{e})\\
\displaystyle \omega \sqrt{B-F \over 2A} & \omega\in(\Omega_{e},\omega_{4})\\ \end{array} \right.
\end{displaymath}
The upper hybrid branch starts at $\omega_3$, which is the second root of $R(\omega)=0$, and it
becomes electron cyclotron wave in parallel direction. It extends to higher frequencies in other direction
and reaches the resonace frequency $\omega_4$ in perpendicular
direction.\footnote{In the perpendicular direction
$\displaystyle n^2={RL \over S}$, so $\omega_4$ is the root of $S=0$}

At even higher frequency, there are two electromagnetic wave (EM) branches. The lower one is,
\begin{displaymath}
k=\left\{ \begin{array}{ll} \displaystyle \omega \sqrt{B-F \over 2A} & \omega\in[\omega_{p,e},\Omega_{e})\\
\displaystyle \omega \sqrt{B+F \over 2A} & \omega\in(\Omega_{e},\infty)\\ \end{array} \right. ,
\end{displaymath}
and the higher branch is
\begin{equation}
k= \omega \sqrt{B-F \over 2A} \qquad \omega\in[\omega_5,\infty),
\end{equation}
where $\omega_5$ is the root of $L=0$. 

With these segmented functions, one can construct an analytical one-to-one $k(\omega,\theta)$ mapping.
Thereafter, the explicit dispersion relation $\omega({\bf k})$ can be easily solved numerically. 
Figure \ref{disp.ps} shows the calculated dispersion surface for a typical solar flare conditions.
The accuracy and stability of the dispersion relation obtained by solving these equations will facilitate 
the numerical study on wave particle interaction or turbulence cascading.

\section{Damping and Hot Plasma Dispersion Relation}

The cold plasma approximation ignores any thermal motion of charged particles, and therefore, limits 
the resulting dispersion relation to scales larger than the thermal kinetic gyroradius. In the damping 
phase, there are strong couplings between the charged background particles and turbulent motion, 
and the background particles are energized. This process is a major mechanism for heating the solar
flares and other astrophysical plasmas (Petrosian et al. 2006, Bittner et. al. 2007). On the other
hand, the process also significantly damps the plasma turbulence and affects the
dispersion relation. To study the
dispersion relation as well as damping effects (i.e., the imaginary part
of wave frequency $\Im[\omega]$) 
in the damping range, one needs to study the collective motion of thermal
particles. It is easier to use the 
distribution function $f_j(\mathbf{r},\mathbf{v},t)$ to represent the charge and current density,
\begin{eqnarray}
\rho &=& \sum_j q_j \int d^3 v f_j \\ 
\mathbf{J} &=& \sum_j q_j \int d^3 v \mathbf{v} f_j 
\end{eqnarray}
where the subscript $j$ stands for species of particle. The particle distribution function $f_j$ satisfies Vlasov equations,
which includes,
\begin{equation}
{\partial f_j \over \partial t}+\mathbf{v}\cdot\nabla f_j + 
{q_j \over m_j}\left(\mathbf{E}+\mathbf{v}\times\mathbf{B}\right)\cdot\nabla_vf_j = 0
\end{equation}
and the Maxwell equations.

There are many linear, quasilinear and nonlinear approximations to solve
the Vlasov equation set. 
A detailed introduction to these solutions can be found in Swanson's {\it Plasma Waves} (1989). To study turbulence
cascading and particle acceleration,
we only use an analytical approximation for parallel propagating waves and the numerical nonlinear approximation
with the WHAMP (Waves in Homogeneous, Anisotropic, Multicomponent
Plasmas by Ronnmark 1982) code for 
non-parallel propagating waves.


\subsection{Damping of Parallel Propagating Waves}
\label{ppw_damp.sec}

The parallel propagating waves are extensively studied in stochastic acceleration theory due to its
simplicity and efficiency in accelerating particles. The theory is used to explain the electron and ion 
spectra and $^3$He enrichment from solar flares (Petrosian \& Liu 2004; Liu et al. 2004; 2006 on $^3$He 
rich impulsive solar energetic particle events). These studies show that the turbulence energy spectrum and
its high wavenumber cutoff due to damping are important factors that determine the accelerated particle spectra and
relative abundance. In this subsection, we calculate the damping rate
and the cutoff point for parallel propagating waves with the linear approximation.

For parallel and quasi-parallel propagating waves, the Dielectric Tensor can be simplified and provide
the dispersion relation for right-handed cyclotron wave ($R$-wave) and left-handed cyclotron wave ($L$-wave)
(Swanson 1982, p. 158),
\begin{eqnarray}
{k^2 \over \omega^2} &=& 1+
	\sum_j{\omega_{pj}^2 \over \omega k_z v_{j}}
		\left[\left({1\pm\epsilon_j \over 2}\right)Z(\zeta_{1j})+\left({1\mp\epsilon_j \over 2}\right)Z(\zeta_{-1j})\right]
\label{ppwdisp.eq}
\end{eqnarray}
where $\omega=\omega_r+i\omega_i$ is the complex wave frequency 
, $v_j$ here is thermal velocity, $\zeta_{nj}=\displaystyle {\omega+n\Omega_j \over k v_j}$ is a shorthand,
and
\begin{equation}
Z(\zeta_{nj})= {1 \over \sqrt{\pi}}\int_{-\infty}^{\infty}{e^{-\xi^2}d\xi \over \xi-\zeta}
\label{Zfun.eq}
\end{equation}
is the Plasma Dispersion Function that can be approximated with
\begin{equation}
Z(\zeta)=i\sqrt{\pi}e^{-\zeta^2}-
	{1 \over \zeta}\left(1+{1 \over 2\zeta^2}+{3 \over 4 \zeta^4}+\cdots \right)
\label{Zapprox.eq}
\end{equation}
Except for the electromagnetic waves, $\Omega_j\gg kv_j$ for all
particle species and all possible wavenumbers.\footnote{Although cold
plasma approximation suggests that $k$ diverges at particle's
gyrofrequency, the damping rate at these frequencies diverges too, which
prevents the dispersion relation extending to a large wavenumber.} As a result, 
the imaginary part of $Z(\zeta)$ for parallel propagating wave is small
for most frequencies but increase dramatically around particle's
gyrofrequency. For example, despite the summation of all species of
particles in Equation (\ref{ppwdisp.eq}), the contribution from
ions to the $R$-wave (electron cyclotron wave is the only $R$-wave for
solar flare or solar wind) dispersion relation is 
far less than that from electrons. 
Therefore by considering electron cyclotron term only, Swanson (1982) obtained the damping rate for the $R$-wave 
with the first order approximation (Equation \ref{Zapprox.eq}), 
\begin{equation}
{\omega_i \over \omega_r}=
-\frac{\sqrt{\pi}\omega_{pe}^2}
{\omega_r k v_e\left[2+\displaystyle{\omega_{pe}^2\Omega_e \over \omega_r\left(\omega_r-\Omega_e\right)^2}\right]}
\exp\left[-\left({\omega_r-\Omega_e \over k v_e}\right)^2\right]
\label{omgi_swanson.eq}
\end{equation}
For the $L$-wave (for solar flare or solar wind, $L$-wave includes the helium cyclotron and proton
cyclotron wave), the contribution from electron cyclotron term can be ignored. However due to their close gyrofrequency, 
the contributions of proton and helium cyclotron term become comparable. 
Thereafter, we generalized Swanson's (1982) derivation to include all the particle into the calculation and obtain the
damping rate for parallel propagating waves:
\begin{equation}
\frac{\omega_i}{\omega_r}=-\frac
{\displaystyle\sum_{s} \frac{\sqrt{\pi}\omega^2_{p,s}}{\omega_r k v_s}
\exp\biggl[-\left(\frac{\omega_r-\epsilon_s\Omega_s}{k v_s}\right)^2 \biggr]}
{\displaystyle\frac{2k^2}{\omega_r^2}+
\displaystyle\sum_{s} \frac{2\omega_r-\epsilon_s\Omega_s}{\omega_r(\omega_r-\epsilon_s\Omega_s)^2}\omega_{p,s}}
\label{omgi_all.eq}
\end{equation}

\begin{figure}[htb]
\epsscale{0.6}
\centerline{\plotone{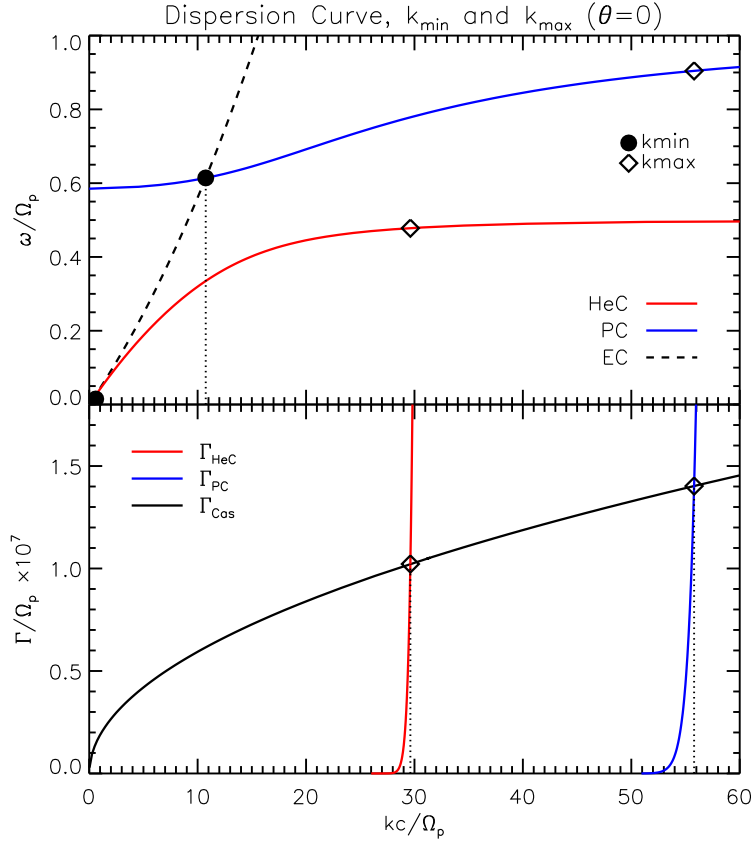}}
\caption{
Dispersion relation for parallel propagating waves
and its high and low wavenumber cutoff, $k_{min}$ and $k_{max}$. {\it Upper panel:} The low wavenumber cutoff 
caused by the topology of dispersion curve i.e., the crossing between EC and PC branch. {\it Lower panel:} 
The high wavenumber cutoff generated by cyclotron damping overwhelming cascading. we assume a simple Kolmogorov
cascading rate for both PC and HeC branch in this case.
}
\label{disps1_prl.ps} 
\end{figure}

When damping rate overwhelms cascading rate from below, the turbulence energy spectrum cuts off, and 
the $k_{max}$ is obtained by solving $\tau_{\mathrm{cas}}^{-1}=\tau_{\mathrm{damp}}^{-1}\equiv2\omega_i$. 
Figure \ref{disps1_prl.ps} illustrates this procedure of finding $k_{max}$ by assuming a Kolmogorov cascading
rate (Zhou \& Matthaeus 1990). From the figure We can also see that the cyclotron damping rate grows extremely fast
(exponential of a square) at particle's cyclotron frequency and thus
$k_{max}$ becomes insensitive to the cascading rate. 
Thereafter, by fitting to the numerical results of $k_{max}$, we am able 
to approximate $k_{max}$ for proton and helium cyclotron branch with plasma temperature and $\alpha$ only,
\begin{eqnarray}
\textrm{PC:} \quad k_{max}=(110\alpha+20)\left({T\over 10^6\mathrm{K}}\right)^{-0.15}-50\alpha+8\\
\textrm{HeC:} \quad k_{max}=(50\alpha+5)\left({T\over 10^6\mathrm{K}}\right)^{-0.16}-10\alpha+6
\end{eqnarray}

Note that although the formula (\ref{omgi_swanson.eq}) and (\ref{omgi_all.eq})
gives the damping rate $\omega_i$ for all $\bf{k}$, the Lauren series approximation of plasma dispersion
function (Equation \ref{Zapprox.eq}) fails at particle's gyrofrequencies 
($\displaystyle\lim_{\omega_r\to\Omega_s}\zeta=0$, where subscript $s$ stands for different
species of particles). At these ranges $\omega_r\simeq\Omega_s$, one can simplify Equation 
(\ref{ppwdisp.eq}) and show that
$\omega_i\ll\omega_r$ (Swanson 1989). Since the accurate number of $\omega_i$ becomes unimportant
at the range $\omega_i\gtrsim\omega_r$, we simply extrapolate the approximated
damping rate with a power law at $\omega_r\simeq\Omega_s$ region for the turbulence cascade and wave-particle
interaction studies.

%% file: ms.bbl
\begin{thebibliography}{}

\bibitem[Andr\'{e} (1985)]{A85} Andr\'{e}, M. 1985, J. Plasma Phys. 33, 1

\bibitem[Bale et al. (2005)]{B05} Bale, S. D., Kellogg, P. J., Mozer, F. S.,
Horbury, T. S., \& Reme, H. 2005, Phys. Rev. Lett., 94, 215002

\bibitem[Beresnyak \& Lazarian (2008)]{BL08}
Beresnyak, A., \& Lazarian, A. 2008, ApJ, 678, 961

\bibitem[Biskamp et al. (1999)]{B99} Biskamp, D., Schwarz, E., Zeiler, A.,
Celani, A., \& Drake, J. F. 1999, Phys. Plasmas, 6, 751

\bibitem[Bittner et al. (2007)]{B07} Bittner, J. M., Liu, S., Fryer, C. L., \&
Petrosian, V. 2007, ApJ, 661, 863

\bibitem[Boldyrev (2002)]{B02} Boldyrev, S. 2002, ApJ, 569, 841

\bibitem[Borovsky \& Funsten (2003)]{BF03} Borovsky, J. E., \& Funsten, H. O.
2003,
JGR, 108(A7), 1284, doi:10.1029/2002JA009625, 2003

\bibitem[Braginskii (1965)]{B65} Braginskii, S. I. 1965, RvPP, 1, 205

\bibitem[Brown \& Melrose (1977)]{BM77} Brown, J. C., \& Melrose, D. B. 1977,
		SoPh, 52, 117

\bibitem[Chandran (2005)]{C05} Chandran, B. D. G. 2005, PRL, 95, 265004

\bibitem[Chen et al. (2007)]{C07} Chen, Y., Reeves, G. D., \& Friedel, R. H. W.
2007, Nature, doi: 10.1038/nphys655

\bibitem[Cho \& Lazarian (2003)]{CL03} Cho, J., \& Lazarian, A. 2003, MNRAS,
345, 325

\bibitem[Cho \& Lazarian (2006)]{CL06} Cho, J., \& Lazarian, A. 2006, ApJ, 638,
811

\bibitem[Cho et al. (2002)]{C02} Cho, J., Lazarian, A., \& Vishniac, E. T.
2002, ApJ, 564, 291

\bibitem[Cho et al. (2003)]{C03} Cho, J., Lazarian, A., \& Vishniac, E. T.
2003, Turbulence and Magnetic Fields in Astrophysics eds. by E. Falgarone, and
T. Passot, Lecture Notes in Physics, 614, 56

\bibitem[Cho \& Vischniac (2000)]{CV00} Cho, J., \& Vishniac, E. T. 2000, ApJ,
538, 217

\bibitem[Cranmer \& van Ballegooijen (2003)]{CV03} Cranmer, S. R., \& van
Ballegooijen, A. A. 2005, ApJS, 156, 265

\bibitem[Cranmer \& van Ballegooijen (2005)]{CV05} Cranmer, S. R., \& van
Ballegooijen, A. A. 2003,
ApJ, 594, 573

\bibitem[Dasso et al. (2005)]{D05} Dasso, S., Milano, L. J., Matthaeus, W. H.,
\& Smith, C. W. 2005, ApJ,
635, L181

\bibitem[Denskat et al. (1983)]{D83} Denskat, K. U., Beinroth, H. J., \&
Neubauer,
F. M. 1983, J. Geophys., 54, 60



\bibitem[Farmer \& Goldreich (2004)]{FG04} Farmer, A. J., \& Goldreich,
		P. 2004, ApJ, 604, 671

\bibitem[Fletcher \& Hudson (2008)]{FH08} Fletcher, L., \& Hudson,
		H. 2008, ApJ, 675, 1645

\bibitem[Galtier et al. (2000)]{G00} Galtier, S. et al., 2000, J. Plasma Phys.
63, 447

\bibitem[Galtier (2006)]{Galtier06} Galtier, S. 2006, J. Plasma
		Phys. 72, 721

\bibitem[Gary \& Borovsky (2004)]{GB04} Gary, S. P., \& Borovsky, E. 2004, JGR,
109, A06105

\bibitem[Gary, Saito, \& Li (2008)]{G08} Gary, S. P., Saito, S., \& Li, H.
2008, Geophys. Res, Lett., 35, L02104,
doi:10.1029/2007GL032327

\bibitem[Ginzburg (1961)]{G61} Ginzburg, V. L. 1961, Propagation of
Electromagnetic Waves in Plasma (New York: Gordon \& Breach)

\bibitem[Ginzburg \& Syrovatskii (1969)]{GS69} Ginzburg, V. L., \& Syrovatskii,
S. I. 1969, The Origin of Cosmic Rays (New York: Gordon \& Breach)




\bibitem[Goldreich \& Sridhar (1995)]{GS95} Goldreich, P., \& Sridhar, S.
1995, ApJ, 438, 763

\bibitem[Hamilton et al. (2008)]{Ham08} Hamilton, K., Smith, C. W., Vasquez, B.
J., \& Leamon, R. J. 2008, JGR, 113, A01106

\bibitem[Howes et al. (2007)]{H07} Howes, G. G. et al. 2007, astro-ph/0707.3149,
astro-ph/0707.3147

\bibitem[Howes et al. (2008)]{H08} Howes, G. G. et al. 2008, PRL, 100, 065004,
doi: 10.1103/PhysRevLett.100.065004

\bibitem[Howes et al. (2008a)]{H08a} Howes, G. G. et al. 2008a, JGR, 113,
A05103,
doi: 10.1029/2007JA012665

\bibitem[Iroshnikov (1963)]{I63} Iroshnikov, P. S. 1963, AZh, 40, 742

\bibitem[Kolmogorov (1941)]{K41} Kolmogorov, A. N. 1941, Dokl. Akad. Nauk SSSR,
30, 301

\bibitem[Kraichnan (1965)]{K65} Kraichnan, R. H. 1965, Phys. Fluids, 8, 1385

\bibitem[Krucker et al. (2008a)]{K08} Krucker, S., Hurford, G. J., MacKinnon, A.
L., Shih, A. Y., \& Lin, R. P. 2008a, ApJ, 678, L63

\bibitem[Krucker et al. (2008b)]{KB08} Krucker, S., et al. 2008b, AA Rev,
		16, 155 DOI 10.1007/s00159-008-0014-9

\bibitem[Lazarian \& Beresnyak (2006)]{LB06} Lazarian, A., \& Beresnyak,
		A. 2006, ApJ, 373, 1195

\bibitem[Leamon et al. (2000)]{L00} Leamon, R. J., Matthaeus, W. H., Smith, C.
W., Zank, G. P.,
 Mullan, D. J., \& Oughton, S. 1998, JGR, 103, 4775

\bibitem[Leamon et al. (1998)]{L98} Leamon, R. J., Smith, C. W., Ness, N. F.,
Matthaeus, W. H., \&
Wong, H. K. 1998, JGR, 103, 4775

\bibitem[Leamon et al. (1999)]{L99} Leamon, R. J., Smith, C. W., Ness, N. F.,
\&
Wong, H. K. 1999, JGR, 104, 22331

\bibitem[Li et al. (2001)]{L01} Li, H., Gary, S. P., \& Stawicki, O. 2001,
Geophys.Res.Letters, 28, 1347




\bibitem[{Liu} et al. (2006a)]{L06a} Liu, S., Melia, F., Petrosian, V., \&
Fatuzzo, M. 2006a,
\apj, 647, 1099.

\bibitem[{Liu} et al. (2004)]{L04} Liu, S., Petrosian, V., \& Mason, G. M.
2004,
\apjl, 613, L81

\bibitem[{Liu} et al. (2006)]{L06} Liu, S., Petrosian, V., \& Mason, G. M.
2006,
\apj, 636, 462

\bibitem[Liu et al. (2008)]{L08} Liu, S., Fan, Z. H., Fryer, C. L., Wang, J. M.,
\& Li, H. 2008, ApJL, in press, astro-ph/0805.2589

\bibitem[Liu et al. (2006b)]{L06b}Liu, W., Liu, S., Jiang, Y., \& Petrosian, V.
2006b, ApJ, 649, 1124

\bibitem[Luo \& Melrose (2006)]{LM06} Luo, Q., \& Melrose, D. 2006, MNRAS, 368,
1151

\bibitem[Markovskii et al. (2008)]{M08} Markovskii, S. A., Vasquez, B., J., \&
Smith, C. W. 2008, ApJ, 675, 1576

\bibitem[Markovskii et al. (2006)]{M06} Markovskii, S. A., Vasquez, B., J.,
Smith, C. W., \& Hollweg, J. V. 2006, ApJ, 639, 1177

\bibitem[Mason et al. (2002)]{M02}Mason, G. M., et al. 2002, ApJ, 574, 1039


\bibitem[Matthaeus et al. (1996)]{M96} Matthaeus, W. M., Ghosh, S., Oughton, S.,
\& Roberts, D. A. 1996, J. Geophys. Res., 101, 7619

\bibitem[Matthaeus et al. (1990)]{M90} Matthaeus, W. M., Goldstein, M. L.,
\& Roberts, D. A. 1990, J. Geophys. Res., 95, 20673

\bibitem[Matthaeus et al. (1998)]{M98} Matthaeus, W. M., Oughton, S., Ghosh, S.,
\& Hossian, M.  1998, Phys. Rev. Lett., 81, 2056

\bibitem[Miller (1997)] {M97} Miller, J. A. 1997, \apj, 491, 939

\bibitem[Miller et al. (1996)] {MLM96} Miller, J. A., LaRosa, T. N., \&
Moore, R. L.,
1996, \apj, 461, 445

\bibitem[Miller \& Roberts (1995)] {MR95} Miller, J. A., \& Roberts, D. A.
1995,
\apj, 452, 912


\bibitem[Montgomery \& Matthaeus (1995)]{MM95} Montgomery, D., \& Matthaeus, W.
H. 1995, ApJ, 447, 706

\bibitem[Montgomery \& Turner (1981)]{MT81} Montgomery, D., \& Turner, L. 1981,
Phys. Fluids 24(5), 825

\bibitem[Ng \& Bhattacharjee (1996)]{NB96} Ng, C. S., \& Bhattacharjee, A. 1996,
ApJ, 465, 845

\bibitem[Ng \& Bhattacharjee (1997)]{NB97} Ng, C. S., \& Bhattacharjee, A. 1997,
Phys. Plasmas, 4(3), 605

\bibitem[Oughton et al. (2006)]{O06} Oughton, S., Dmitruk, P., \& Matthaeus, W.
H. 2006, Phys. Plasmas 13, 042306

\bibitem[Osman \& Horbury (2007)]{OH07} Osman, K. T., \& Horbury, T. S. 2007,
ApJL, 654, L103

\bibitem[Park \& Petrosian (1995)]{PP95} Park, B. T., \& Petrosian, V. 1995,
ApJ, 446, 699.

\bibitem[Parks et al. (2006)]{P06a} Parks, G. K., et al., 2006, Phys. Plasma,
13, 050701

\bibitem[Petrosian \& Liu (2004)] {PL04} Petrosian, V., \& Liu, S.\
2004, \apj,
610, 550

\bibitem[Petrosian et al. (2006)] {P06} Petrosian, V., Yan, H., \&
Lazarian, A.\ 2006, \apj,
644, 603

\bibitem[Porter et al. (1999)]{P99} Porter, D., Pouquet, A., Sytine, I., \&
Woodward, P. 1999, Physica, A, 263, 263

\bibitem[Porter et al. (1998)]{P98} Porter, D. H., Woodward, P. R., \& Pouquet,
A. 1998, Phys. Fluids, 10, 237

\bibitem[Reames et al. (1994)]{R94} Reames, D. V., Meyer, J. P., \& von
		Rosenvinge, T. T. 1994, ApJS, 90, 649

\bibitem[R\"onnmark (1982)]{R82} R\"onnmark, J. 1982, Waves in Homogeneous,
Anisotropic,
Multicomponent Plasmas (Sweden: Kiruna Geophysics Institute)

\bibitem[Saito \& Gary (2007)]{SG07}Saito, S., \& Gary, P. 2007, JGR, 112,
A06116


\bibitem[Shebalin et al. (1983)]{S83} Shebalin, J. V., Matthaeus, W. H., \&
Montgomery, D. 1983, J.
Plasma Phys. 29, 525

\bibitem[Smith et al. (2006)]{S06} Smith, C. W., Hamilton, K., Vasquez, B. J.,
\& Leamon, R. J. 2006, ApJ, 645, L85

\bibitem[Smith et al. (2001)]{Smi01} Smith, C. W., Mulla, D. J., Ness, N. F.,
Skoug, R. M., \& Steinberg, J. 2001, JGR, 106, A9, 18625

\bibitem[Stawicki et al. (2001)]{S01} Stawicki, O., Gary, S. P., \& Li, H.
2001, JGR, 106, 8273

\bibitem[Sridhar \& Goldreich (1994)]{SG94} Sridhar, S., \& Goldreich, P.
1994, ApJ, 432, 612


\bibitem[Stix (1962)]{S62} Stix, T. H. 1962, {\it The Theory of Plasma Waves}
(McGraw-Hill Book
Company, Inc.)

\bibitem[Swanson (1989)]{S89} Swanson, D. G. 1989, Plasma Waves (Academic Press,
Boston)

\bibitem[Tu et al. (2002)]{T02} Tu, C. -Y., Wang, L. -H., \& Marsch, E. 2002,
JGR,  107, A10, 1291

\bibitem[Vestuto et al. (2003)]{V03} Vestuto, J. G., Ostriker, E. C., \& Stone,
J. M. 2003, ApJ, 590, 858

\bibitem[Wu \& Yang (2006)]{WY06} Wu, D., \& Yang, L. 2006, A\&A, 452, L7

\bibitem[Xie \& Ofman (2004)]{XO04} Xie, H., \& Ofman, L. 2004, JGR, 109, A08103

\bibitem[Yan \& Lazarian (2002)]{YL02} Yan, H., \& Lazarian, A. 2002,
Phys.Rev.Letters 89, 281102-1

\bibitem[Yan et al. (2008)]{Y08} Yan, H., Lazarian, A., \& Petrosian, V.  2008,
astro-ph/0801.3786

\bibitem[Yeung \& Zhou (1997)]{YZ97} Yeung, P. K., \& Zhou, Y. 1997, ICASE
Report No. 97-64.

\bibitem[Zakharov \& Kuznetsov (1978)]{ZK78} Zakharov, V. E., \&
		Kuznetsov, E. A. 1978, JETP, 48, 458

\bibitem[Zhang \& Li (2004)]{ZL04} Zhang, T. X., \& Li, B. 2004, Phys. Plasmas,
11, 2172

\bibitem[Zhou \& Matthaeus (1990)]{ZM90} Zhou, Y., \& Matthaeus, W. H. 1990,
JGR, 95, 14881

\end{thebibliography}
